\documentclass[11pt]{article}

\usepackage[margin=1in]{geometry}
\usepackage{amsmath,amssymb,bm}
\usepackage{braket}
\usepackage{graphicx}
\usepackage{tikz}
\usepackage{quantikz}
\usepackage{algorithm}
\usepackage{algpseudocode}
\usepackage{float}
\restylefloat{algorithm}

\usepackage{booktabs}
\usepackage{xcolor}
\usepackage[numbers,sort&compress]{natbib}
\usepackage[colorlinks=true,linkcolor=black,citecolor=blue,urlcolor=blue]{hyperref}
\usepackage{orcidlink}

\newcommand{\Rmat}{\mathbf{R}}
\newcommand{\Qhat}{\hat{Q}}
\newcommand{\ahat}{\hat{a}}
\newcommand{\bhat}{\hat{b}}
\newcommand{\mtil}{\tilde{m}}
\newcommand{\gtil}{\tilde{g}}
\newcommand{\npos}{n_{\mathrm{pos}}}
\DeclareMathOperator{\diag}{diag}
\newcommand{\runin}[1]{\par\smallskip\noindent\textit{#1}\hspace{0.5em}\ignorespaces}

\title{\bfseries Exact quantum circuits for a lattice Boltzmann realization of the Dirac equation}
\author{Nilesh Sawant\,\orcidlink{0000-0001-8403-8943}\thanks{email of the corresponding author: nileshsawant@outlook.com, nsawant@nlr.gov}, Ethan Young\,\orcidlink{0000-0003-1106-7406}, Kevin Patrick Griffin\,\orcidlink{0000-0002-0866-6224}, Michael Martin\,\orcidlink{0000-0002-6526-4408}\\[2pt]
\small Computational Science Center, National Laboratory of the Rockies (NLR), Golden, Colorado 80401, USA}
\date{\today}

\begin{document}
\maketitle

\begin{abstract}
\noindent
The quantum lattice Boltzmann (QLB) scheme of Succi and Dellar advances a
four-component Dirac spinor on a lattice by a fixed sequence of local, exactly
norm-preserving operations: a basis rotation, a collision, a streaming shift, and the
inverse rotation. While this correspondence has been noted qualitatively, it has remained unclear whether the full three-dimensional scheme can be realized exactly and unitarily as a gate-model quantum circuit. This unitarity is a structural property of the scheme rather than an
approximation, which should, in principle, enable an operator-by-operator port onto quantum hardware. Here we make that mapping explicit. We give a gate-level construction of every
operation of the three-dimensional Dirac QLB scheme: the fixed rotation gates, the
collision gate, the streaming shift as a controlled increment on a
position register, the position-dependent potential as a phase oracle, and periodic
and reflecting (bounce-back) boundary conditions as unitary circuits. Together, these constructions provide the first complete demonstration that the Succi--Dellar QLB scheme is exactly representable in the gate model without any approximation beyond the classical method itself. We then compose
them into single-axis, two- and three-dimensional time steps. On a state-vector
emulator the resulting circuits reproduce the classical QLB solver to machine precision
(maximum density deviation between $3.7\times10^{-12}$ and $1.0\times10^{-17}$ across
the one-, two-, and three-dimensional tests), thus verifying our quantum implementation. Our aim is representability and correctness rather than computational speedup; questions of state preparation, asymptotic cost, and hardware noise are identified as open avenues for future work. Appendices restore physical units, derive the scheme's Zitterbewegung in closed form and compare it with a trapped-ion simulation, and execute a reduced two-site instance on a Rigetti superconducting processor, where choosing the qubits from the live calibration lowers the state-preparation error sevenfold. All operators, circuits, tests, and figures are
reproducible from the open-source \texttt{quantumKineticMethods} library.
\end{abstract}

\tableofcontents

\section{Introduction}
\label{sec:intro}

Kinetic formulations replace a velocity transport equation with the transport of a distribution of a
discrete set of velocities, advanced by alternating \emph{stream} and \emph{collide}
updates. For classical hydrodynamics, this is the lattice Boltzmann method
\cite{succi2018}. Succi and Benzi observed that the same construction can be applied to
quantum wave equations \cite{succiBenzi1993}, and Succi later formulated a lattice
Boltzmann scheme for the Dirac equation and, more generally, for quantum field theory
\cite{succi2007}. Dellar, Lapitski, Palpacelli and Succi established the mathematical basis for the three-dimensional
Dirac scheme by identifying the rotations that render the streaming
isotropic \cite{dellar2011}. Palpacelli and co-workers used the scheme to study Klein
tunneling through random impurities \cite{palpacelli2012}.

The feature that distinguishes the Dirac QLB scheme from most discretizations of a
partial differential equation is that \emph{each of its elementary operations is
unitary}: a unitary operation is a reversible linear map that exactly preserves the
Euclidean ($\ell_2$) norm, $\|\psi\|=(\sum_k|\psi_k|^2)^{1/2}$, so it conserves probability.
The rotation is an orthogonal change of spinor basis; the collision is a rotation of pairs
of spinor components; the streaming shift is a permutation of lattice sites. Their product
is therefore unitary, and the discrete $\ell_2$ norm of the field is conserved exactly at
every step. A unitary map on a
$2^n$-dimensional space is precisely what a quantum circuit on $n$ qubits implements.
This correspondence has been noted qualitatively; here it is carried out in full and
checked.

\subsection{Scope and claims}
We restrict attention to a single, well-defined question: \emph{can the Succi--Dellar
Dirac QLB scheme be written, exactly, as a quantum circuit?} We answer it constructively
by exhibiting a circuit for every operation and for their composition in one, two and
three dimensions, and by verifying on a quantum emulator that the circuits reproduce the
classical solver to machine precision. This establishes not only that the scheme is representable, but that each component (rotation, collision, streaming, and boundary handling) is validated independently against its classical reference, yielding a modular set of primitives for future Dirac and lattice-kinetic quantum algorithms. It
does \emph{not} establish any computational advantage: the emulation cost is exponential
in the number of qubits, the encoding of an input state and the read-out of physical
observables are addressed only for the reduced hardware instance of Appendix~\ref{app:hw}, and we make no asymptotic claims. Instead, these limitations delineate what further advances---such as hardware-native compilation, ancilla-optimized streaming, or improved amplitude-encoding techniques---would be required before such QLB-based circuits might be evaluated for practical quantum speedup. These are stated as
open problems in Section~\ref{sec:discussion}.

\subsection{Related work}
Digital quantum algorithms for the Dirac equation based on operator splitting and the
quantum Fourier transform have been given by Fillion-Gourdeau, MacLean and Laflamme
\cite{fillionGourdeau2017}; the kinetic route taken here is different in that it ports an
existing, validated lattice kinetic solver operation by operation, so that each circuit
layer has a classical reference against which it is checked. The gate-synthesis we use is
standard: the Cartan (KAK) decomposition of two-qubit unitaries \cite{shende2006} for the
spinor operators, and multi-controlled-$X$ gates for the streaming shift. Quantum adders
\cite{draper2000,gidney2018} would give a cheaper streaming step but are not
used here (see Section~\ref{sec:resources}).

\subsection{Contributions}
\begin{enumerate}\itemsep2pt
  \item An exact gate-level construction of every operation of the 3D Dirac QLB scheme
    (Sections~\ref{sec:scheme}--\ref{sec:units}): rotations, collision, streaming,
    potential, and periodic and reflecting boundary conditions.
  \item Their composition into single-axis, 2D and 3D time steps by tensoring position
    registers on a shared spinor register (Section~\ref{sec:assemble}).
  \item A numerical verification that the circuits reproduce the classical solver to
    machine precision in 1D, 2D and 3D, including Klein tunneling and a reflecting-box
    bounce (Section~\ref{sec:verify}), together with measured gate counts
    (Section~\ref{sec:resources}).
  \item A small, general ``unitary~$\rightarrow$~verified circuit'' routine that underlies
    all of the above and is reusable for other operators (Section~\ref{sec:assemble}).
\end{enumerate}
All results are reproducible from the \texttt{quantumKineticMethods} library
\cite{qkm2026}; the module and test map is given in Section~\ref{sec:repro}.

\subsection{Overview of the appendices}
Three appendices support the main text. Appendix~\ref{app:phys} restores physical (SI) units to the
dimensionless test parameters of Section~\ref{sec:verify}, expressing the examples as relativistic
electrons at Compton-scale resolution. Appendix~\ref{app:zb} derives the scheme's lattice dispersion
in closed form and reproduces the relativistic \emph{Zitterbewegung}, comparing it term by term with
the trapped-ion analog simulation of Gerritsma et al.~\cite{gerritsma2010}. Appendix~\ref{app:hw}
turns to execution on real quantum hardware: it sets out what a physical processor requires, estimates
the two-qubit gate cost of the scheme, and runs a reduced two-site instance on a superconducting
processor.

\section{From a lattice Boltzmann code to a qubit circuit: a primer}
\label{sec:primer}

This section gives an expository account of the qubit encoding and circuit model and is intended primarily for
readers whose background is in the lattice Boltzmann method rather than in quantum
computing. It may be omitted by readers already familiar with gate-model quantum
computation; the precise formulation is given in Sections~\ref{sec:scheme} and
\ref{sec:encoding}. The following aspects are addressed in
turn: the storage of the field on qubits, the distinction between spinor and position
registers, the mapping of space and dimension onto qubits, the action of a gate circuit
relative to a classical update, time integration, and the KAK decomposition.

\subsection{Amplitude encoding of the field}
In a classical implementation of the Dirac quantum lattice Boltzmann scheme, the Dirac spinor is held as an array in memory: for
a line of $N$ sites carrying a four-component Dirac spinor this is $4N$ complex numbers,
one array element per (site, component). A quantum register stores information differently. A
register of $n$ \emph{qubits} (the quantum counterpart of $n$ classical bits, each a
two-state system) possesses $2^n$ basis states, the binary strings from
$\ket{00\ldots0}$ to $\ket{11\ldots1}$ (the notation $\ket{\cdot}$, read ``ket'', is the
standard label for a basis state), and its state is specified by $2^n$ complex
\emph{amplitudes}, one associated with each pattern. Under \emph{amplitude encoding} these amplitudes serve as
the storage: the $4N$ field values are placed in the $2^n$ amplitudes of a register of
$n=\log_2(4N)$ qubits. Each qubit pattern is interpreted as a binary address, and the
amplitude on that pattern is the field value stored there. In this way, the qubits act as
address lines, while the $2^n$ amplitudes hold the data.

\subsection{Spinor and position registers}
A natural but incorrect picture assigns two qubits to each grid point. In the amplitude
encoding, the two ``spinor'' qubits instead label the four spinor components
$c\in\{0,1,2,3\}$ (two bits provide $2^2=4$ labels) and are \emph{shared across the entire
lattice}. The grid is addressed by a separate \emph{position register} of $\npos=\log_2 N$
qubits per axis, whose $2^{\npos}=N$ patterns enumerate the site index $x$ along that axis.
A line of $32$ sites therefore requires $5$ position qubits rather than $32$, and a line of
$1024$ sites requires $10$. The lattice resides in the amplitudes, and its size enters only
through the number of address qubits, $\log_2 N$, rather than linearly in $N$.
Table~\ref{tab:classicalquantum} summarizes the correspondence.

\subsection{Position, spinor, and dimension}
\begin{figure}[h!]
    \centering
    \includegraphics[width=0.95\linewidth]{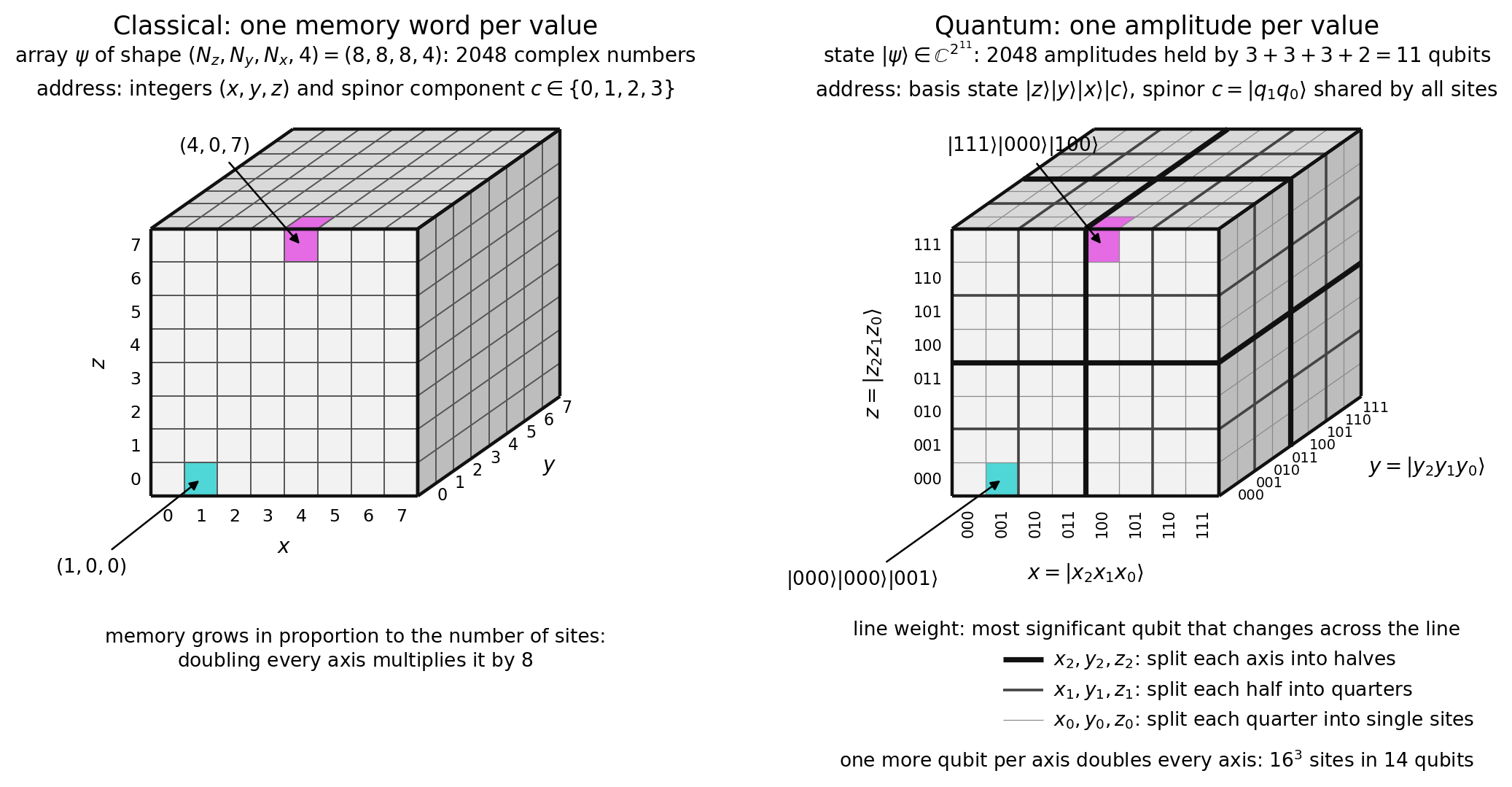}
    \caption{Storage and addressing of the same $8\times8\times8$ lattice with a four-component
      spinor at each site. Left: in a classical array each of the $2{,}048$ complex values is a memory
      word with integer address $(x,y,z)$ and component $c$. Right: in amplitude encoding each value is
      one amplitude of an $11$-qubit state, addressed by the basis state $\ket{z}\ket{y}\ket{x}\ket{c}$.
      The weight of a grid line marks the most significant position qubit that changes across it: the
      first qubit of a register halves its axis, the second halves each half, and the third splits each
      quarter into single sites. The colored cells are the same sites in both pictures, with position
      addresses $(1,0,0)$ or $\ket{000}\ket{000}\ket{001}$ (cyan) and $(4,0,7)$ or
      $\ket{111}\ket{000}\ket{100}$ (magenta).}
    \label{fig:line-classical-vs-quantum}
\end{figure}

\begin{figure}[h!]
    \centering
    \includegraphics[width=\linewidth]{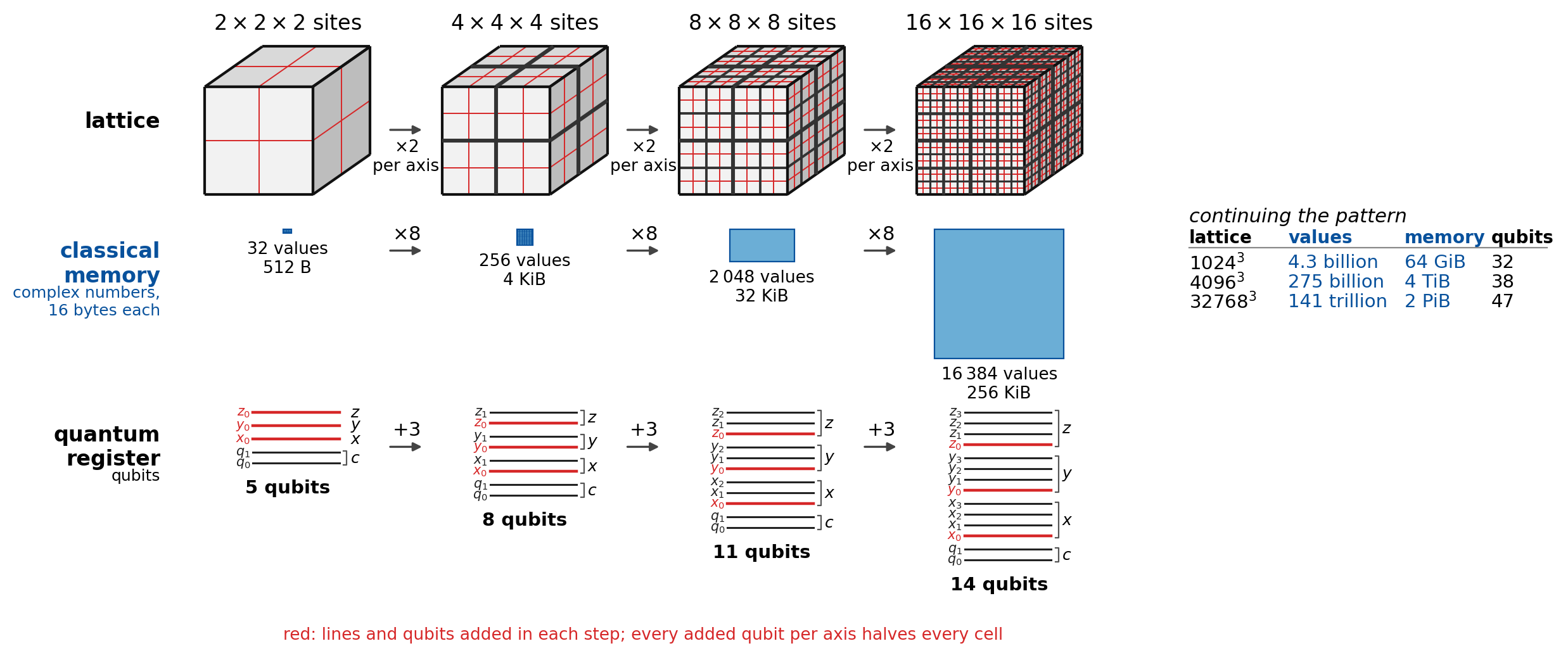}
    \caption{Growth of the storage with the lattice size. Each step doubles the number of sites $N$
      per axis. The classical array of $4N^3$ complex numbers ($16$ bytes each) grows eightfold, while
      the quantum register gains one qubit per axis, $3\log_2 N+2$ qubits in all. Red marks the grid
      lines and qubits added in each step: every added qubit per axis halves every cell. The figure
      refines a fixed domain; doubling the domain at fixed spacing gives the same counts. The table on
      the right continues the pattern to larger lattices. The compression applies to storing and
      updating the state, not to preparing it or reading it out (Section~\ref{sec:primer-circuits}).}
    \label{fig:growth-strip}
\end{figure}

The full register (Section~\ref{sec:encoding}) is two spinor qubits plus one position
register per spatial axis:
\[
  n \;=\; \underbrace{2}_{\text{spinor (4 components)}}
  \;+\; \underbrace{\log_2 N_x + \log_2 N_y + \log_2 N_z}_{\text{position: one register per axis}} .
\]
An $8\times8\times8$ lattice, comprising $512$ sites or $2{,}048$ spinor amplitudes,
is thus represented in $2+3+3+3=11$ qubits. The amplitude of component $c$ at site $(x,y,z)$
sits on the basis state $\ket{z}\ket{y}\ket{x}\ket{c}$, whose binary value is the flat index of
Eq.~\eqref{eq:layout3d}, so that $z$ is the most significant digit and $c$ the least
(Figure~\ref{fig:line-classical-vs-quantum}). The number of stored values is the same as in the
classical array; what is compressed is the number of qubits that hold them. The two spinor qubits are
shared across all three axes; only the position registers grow with the grid, by one qubit each time
the number of sites along an axis doubles. Figure~\ref{fig:growth-strip} illustrates this logarithmic
relationship.

\subsection{Circuits and their relation to a classical update}
\label{sec:primer-circuits}
A classical time step iterates over the $4N$ array, applying the collision and streaming
updates site by site.  As a consequence of amplitude
encoding, a \emph{gate} (a small unitary matrix acting on one or two
qubits) acting on the position qubits acts on the addresses of all sites
simultaneously: a single controlled increment of the position register realizes the
streaming shift for the entire line in one operation, without iteration over sites. This is
the essential structural difference from a classical solver, and it is the reason the
\emph{circuits} (ordered sequence of such gates) constructed below have a size governed by $\log_2 N$ rather than by $N$.

This compactness is subject to two qualifications, which is why no computational advantage
is claimed in the present work. The first concerns input: preparing a general initial field
in the $2^n$ amplitudes (state preparation) is itself a nontrivial operation, with a computational cost 
that scales with the number of amplitudes. The second concerns output: a quantum register
cannot be read out as an array. A \emph{measurement} returns a single address, drawn at
random with probability equal to the squared magnitude of its amplitude, and it collapses
the state. Thus, reconstructing a density profile, or an integrated quantity such as a
transmission coefficient, requires many repeated executions. The compactness of
the storage is genuine, but it does not by itself reduce the cost of the computation. These challenges of quantum state preparation and output are general challenges in quantum computing, not unique to this work.

\subsection{Time integration}
One QLB time step corresponds to a single application of the step circuit
$U_{\mathrm{step}}$, and evolution over $T$ steps corresponds to $T$
applications of the same circuit in series, $U_{\mathrm{step}}^{\,T}$. This is the direct
analogue of the classical time-marching loop, in which one stream-collide update is
repeated; here the loop body is a fixed gate sequence and repetition amounts to
concatenating copies of it. The single-step circuit is constructed once; increasing the
number of steps corresponds to repeated applications of the same gate sequence, so the execution time grows with steps but the circuit ``width'' (qubit count) remains fixed.

\subsection{The KAK decomposition}
The rotation and collision operators are $4\times4$ unitary matrices acting on the two
spinor qubits. Physical hardware does not implement an arbitrary $4\times4$ matrix directly;
it provides a small fixed set of \emph{native} gates, in practice single-qubit rotations
and one two-qubit gate, the controlled-NOT (CX). Quantum compilation is the exact rewriting of a
given matrix as a short sequence of these native gates. The \emph{KAK decomposition}, a
Cartan factorization from the theory of Lie groups named for its $K\,A\,K$ three-factor
form, guarantees that every two-qubit unitary can be rewritten exactly in terms of the native gate set. It also provides the optimal such decomposition: a sequence of single-qubit rotations containing at most
three CX gates. It is the two-qubit
generalization of expressing a planar rotation by a single angle. The transpiler (which rewrites circuits into the device's native gate set) applies
the KAK decomposition to reduce the rotation matrix to one CX gate and the collision matrix
to two (Table~\ref{tab:resources}); the resulting gate counts are minimal because the
decomposition is optimal.

\begin{table}[h]
  \centering
  \caption{Correspondence between a classical lattice Boltzmann implementation and the
    qubit-circuit implementation of the \emph{same} Dirac QLB step, for one axis of $N$
    sites with four spinor components. The circuit size is set by $\log_2 N$, but state
    preparation and measurement (last two rows) carry costs a classical array does not.}
  \label{tab:classicalquantum}
  \renewcommand{\arraystretch}{1.25}
  \begin{tabular}{@{}lll@{}}
    \toprule
    \textbf{Operation} & \textbf{Classical Solver} & \textbf{Qubit Circuit (This Work)} \\
    \midrule
    Field storage      & array of $4N$ complex numbers    & $2+\log_2 N$ qubits (amplitudes) \\
    Grid of $N$ sites  & $N$ cells (per component)         & $\log_2 N$ position qubits \\
    Spinor, 4 comp.    & $4$ numbers per site              & $2$ shared qubits \\
    One time step      & loop over sites: collide, stream  & fixed gate sequence $U_{\mathrm{step}}$ \\
    Streaming shift    & shift the array by one cell       & one controlled increment (all sites at once) \\
    Advance $T$ steps  & repeat the loop $T$ times          & repeat the circuit, $U_{\mathrm{step}}^{\,T}$ \\
    Set initial field  & write the array                   & state preparation (load amplitudes) \\
    Read the result    & read the array                    & measurement (repeated sampling) \\
    \bottomrule
  \end{tabular}
\end{table}

\section{The Dirac quantum lattice Boltzmann scheme}
\label{sec:scheme}

We summarize the scheme in the form implemented in the solver, following
Dellar~et~al.~\cite{dellar2011} and Succi~\cite{succi2007}. The state is a
four-component spinor (wavefunction) $\psi(\mathbf{x},t)$, a function of position
$\mathbf{x}=(x,y,z)$ and time $t$ whose value at each point is a column of four complex
numbers ($\psi\in\mathbb{C}^4$), on a regular lattice. The four components of the Dirac spinor encode two spin states (spin-up and spin-down) and two energy sectors (particle and antiparticle). The spinor obeys the $(3+1)$-dimensional
Dirac equation
\begin{equation}
  i\hbar\,\partial_t\psi
  = \left(-i\hbar c_0\,\bm{\alpha}\cdot\nabla + \beta mc_0^2 - qV\right)\psi ,
  \qquad
  \bm{\alpha}=(\alpha_x,\alpha_y,\alpha_z),
  \label{eq:dirac}
\end{equation}
in which $i=\sqrt{-1}$ is the imaginary unit, $\hbar$ the reduced Planck constant, $c_0$ the
speed of light, $m$ the particle (here electron) rest mass, $q=|e|$ the modulus of the
electron charge, and $V=V(\mathbf{x})$ an applied scalar (electrostatic) potential;
$\partial_t$ denotes the time derivative and $\nabla=(\partial_x,\partial_y,\partial_z)$ the
spatial gradient, so that $\bm{\alpha}\cdot\nabla=\alpha_x\partial_x+\alpha_y\partial_y
+\alpha_z\partial_z$. The four $4\times4$ matrices $\alpha_x,\alpha_y,\alpha_z$ and $\beta$
act on the four spinor components. The equation is advanced with time step $\Delta t$ by
operator (Strang) splitting into per-axis substeps, each substep updating one Cartesian
axis so that a three-dimensional step is a product of one-dimensional updates. The Dirac
matrices are taken in the representation
\begin{equation}
  \alpha_k=\begin{pmatrix}0&\sigma_k\\ \sigma_k&0\end{pmatrix}\ (k=x,y,z),
  \qquad
  \beta=\begin{pmatrix}\mathbb{I}_2&0\\ 0&-\mathbb{I}_2\end{pmatrix},
  \label{eq:diracmats}
\end{equation}
where $\mathbb{I}_2$ is the $2\times2$ identity matrix and $\sigma_x,\sigma_y,\sigma_z$ are
the Pauli matrices,
\begin{equation}
  \sigma_x=\begin{pmatrix}0&1\\1&0\end{pmatrix},\qquad
  \sigma_y=\begin{pmatrix}0&-i\\i&0\end{pmatrix},\qquad
  \sigma_z=\begin{pmatrix}1&0\\0&-1\end{pmatrix}.
  \label{eq:pauli}
\end{equation}
Throughout, $A\otimes B$ denotes the Kronecker (tensor) product of two matrices and
$\mathbb{I}_n$ the $n\times n$ identity matrix. Following Dellar, the scheme is built on the Majorana
form of the Dirac equation, obtained by conjugating Eq.~\eqref{eq:dirac} with the involution
$U=(\alpha_y+\beta)/\sqrt2$ (so that $\psi=U\Psi$), which interchanges the roles of the
$\alpha_y$ and $\beta$ matrices \cite[Eq.~6]{dellar2011}:
\begin{equation}
  \Big[\partial_t + c_0\left(-\alpha_x\,\partial_x + \beta\,\partial_y - \alpha_z\,\partial_z\right)
  + i\,\omega_c\,\alpha_y - i\,g\,\mathbb{I}\Big]\psi = 0 ,
  \qquad \omega_c=\frac{mc_0^2}{\hbar},\quad g=\frac{qV}{\hbar}.
  \label{eq:majorana}
\end{equation}
Its virtue is that the three matrices $\alpha_x=\sigma_x\otimes\sigma_x$,
$\beta=\sigma_z\otimes\mathbb{I}_2$ and $\alpha_z=\sigma_x\otimes\sigma_z$ that multiply the
spatial derivatives are all \emph{real}, so each spatial gradient generates a real
advection that a lattice scheme realizes as a $\pm1$ shift (after a fixed rotation). The
remaining matrix
\begin{equation}
  \alpha_y=\begin{pmatrix}0&\sigma_y\\ \sigma_y&0\end{pmatrix}=\sigma_x\otimes\sigma_y
  \label{eq:alphay}
\end{equation}
is imaginary and has been moved into the algebraic, non-gradient terms: the mass term
$i\omega_c\alpha_y$ (with $\omega_c=mc_0^2/\hbar$ the Compton frequency; real, because
$\alpha_y$ is imaginary) couples spinor components locally,
while the scalar potential enters as the local phase $-ig\,\mathbb{I}$ with $g=qV/\hbar$
(consistent with the $-qV$ of Eq.~\eqref{eq:dirac}, $q$ being the charge modulus). This is
why the three matrices carrying \emph{spatial} streaming are $\{\alpha_x,\beta,\alpha_z\}$,
while the mass acts through $\alpha_y$ and the potential through the identity. The four
matrices $\{\alpha_x,\beta,\alpha_z,\alpha_y\}$ are mutually anticommuting, so each spatial
sweep can be diagonalized independently while the collision (built from $\alpha_y$) is
shared.

\subsection{One substep}
A substep along axis $a\in\{x,y,z\}$ acting on a line of the field is the four-layer
sequence
\begin{equation}
  \psi \;\longmapsto\; \Rmat_a\,\,\mathrm{Stream}_a\!\left[\,\Qhat_a\,\,\Rmat_a^{-1}\psi\,\right],
  \label{eq:substep}
\end{equation}
read right to left: rotate into the frame in which streaming along $a$ is diagonal ($\Rmat_a^{-1}$),
collide ($\Qhat_a$), shift ($\mathrm{Stream}_a$), and rotate back ($\Rmat_a$). These operators will be defined subsequently. Because each factor is unitary, so is the substep.

\runin{Rotations.}
The rotation $\Rmat_z$ that diagonalizes $\alpha_z$ is \cite[Eq.~10]{dellar2011}
\begin{equation}
  \Rmat_z=\frac{1}{\sqrt2}
  \begin{pmatrix} 0&-1&0&1\\ 1&0&-1&0\\ 0&1&0&1\\ 1&0&1&0 \end{pmatrix},
  \qquad
  \Rmat_z^{-1}\alpha_z\,\Rmat_z=\diag(-1,-1,+1,+1).
  \label{eq:Rz}
\end{equation}
The $x$-rotation is obtained by composing $\Rmat_z$ with the spinor rotation that maps
the $z$ streaming axis onto $x$,
\begin{equation}
  S=\exp\!\left(-i\tfrac{\pi}{4}\,\Sigma_y\right),\quad
  \Sigma_y=\mathbb{I}_2\otimes\sigma_y,\qquad
  \Rmat_x=S\,\Rmat_z
  =\frac{1}{\sqrt2}
  \begin{pmatrix} -1&-1&1&1\\ 1&-1&-1&1\\ -1&1&-1&1\\ 1&1&1&1 \end{pmatrix},
  \label{eq:Rx}
\end{equation}
which satisfies $\Rmat_x^{-1}\alpha_x\,\Rmat_x=\diag(-1,-1,+1,+1)$ and, crucially,
leaves the collision generator invariant, $\Rmat_x^{-1}\alpha_y\,\Rmat_x=\alpha_y$, so
that the collision keeps its form in every sweep. The $y$-sweep needs no rotation,
$\Rmat_y=\mathbb{I}_4$, because $\beta$ is already diagonal. (The composite form
Eq.~\eqref{eq:Rx} is the one implemented; a sparser $x$-rotation that diagonalizes
$\alpha_x$ but does \emph{not} preserve $\alpha_y$ produces spurious back-scattering for
generic spinors and is incorrect.)

\runin{Collision.}
With the per-sweep dimensionless mass and potential couplings
\begin{equation}
  \mtil=\frac{mc_0^2\,\Delta t}{s\,\hbar},\qquad
  \gtil=\frac{qV\,\Delta t}{s\,\hbar},\qquad
  \Omega=\mtil^2-\gtil^2,\qquad
  D=1+\tfrac{\Omega}{4}-i\gtil,
  \label{eq:couplings}
\end{equation}
($s$ the number of substeps per time step, $s=3$ in 3D), the collision operator is the
special-unitary ($\mathrm{SU}(2)$) operator
\begin{equation}
  \Qhat=\ahat\,\mathbb{I}_4-i\,\bhat\,\alpha_y,\qquad
  \ahat=\frac{1-\Omega/4}{D},\qquad
  \bhat=\frac{\mtil}{D},\qquad
  |\ahat|^2+|\bhat|^2=1 .
  \label{eq:collision}
\end{equation}
Written out, $\Qhat$ couples the spinor components in the two pairs $(\psi_0,\psi_3)$ and
$(\psi_1,\psi_2)$,
\begin{equation}
  \begin{aligned}
    \psi_0'&=\ahat\,\psi_0-\bhat\,\psi_3, & \psi_3'&=\ahat\,\psi_3+\bhat\,\psi_0,\\
    \psi_1'&=\ahat\,\psi_1+\bhat\,\psi_2, & \psi_2'&=\ahat\,\psi_2-\bhat\,\psi_1,
  \end{aligned}
  \label{eq:collision-comp}
\end{equation}
each a rotation by the same $(\ahat,\bhat)$. In the massless limit $\mtil\to0$ we have
$\bhat=0$ and $\Qhat=\ahat\,\mathbb{I}_4$ with $|\ahat|=1$: the collision is a pure
phase. In a sweep, the collision is applied in the rotated frame,
$\Qhat_a=\Rmat_a^{-1}\Qhat\,\Rmat_a$; since $\Rmat_a^{-1}\alpha_y\Rmat_a=\alpha_y$ for
$a=x,z$ and $\Rmat_y=\mathbb{I}$, one has $\Qhat_a=\Qhat$ up to the sign of $\bhat$ that
distinguishes the $y$-sweep.

\runin{Streaming.}
In the rotated frame the streaming shift moves each spinor component by one site,
forward or backward. The per-component shift signs are fixed by the rotation,
\begin{equation}
  \mathrm{sign}_x=(-,-,+,+),\quad
  \mathrm{sign}_y=(+,+,-,-),\quad
  \mathrm{sign}_z=(-,-,+,+),
  \label{eq:signs}
\end{equation}
i.e.,\ components $0,1$ shift one way and components $2,3$ the other.
Eqs.~\eqref{eq:Rz}--\eqref{eq:signs} define the scheme exactly; everything below is a
gate-level rendering of these objects.

\section{Qubit encoding}
\label{sec:encoding}

The four spinor components are encoded in two qubits, and each lattice axis of
$N=2^{\npos}$ sites in an $\npos$-qubit \emph{position register}. Writing the two spinor
qubits as bits $q_0,q_1\in\{0,1\}$ and the position bits as
$p_0,\dots,p_{\npos-1}\in\{0,1\}$,
\begin{equation}
  c=q_0+2q_1\in\{0,1,2,3\},\qquad
  x=\sum_{j=0}^{\npos-1}2^{\,j}\,p_j,\qquad
  \text{basis index } k=4x+c ,
  \label{eq:layout}
\end{equation}
so $c$ is the spinor-component index, $x\in\{0,\dots,N-1\}$ the site index along the axis,
and $k$ the combined index into the $4N$ amplitudes.
The spinor qubit $q_0$ carries the ``spin'' index and $q_1$ the ``direction'' (upper/lower
bispinor) index; by Eq.~\eqref{eq:signs} the streaming direction depends only on $q_1$.
Table~\ref{tab:spinor-map} lists the four components explicitly, with their two-bit labels, the
spin ($\Sigma_z=\mathbb{I}_2\otimes\sigma_z$) and half ($\beta$) eigenvalues that each bit
carries, and the per-axis streaming signs. Two points there are easy to misread. First, all
four sign tuples of Eq.~\eqref{eq:signs}, like the diagonal of $\Rmat_a^{-1}\alpha_a\Rmat_a$,
are ordered by the component index $c=0,1,2,3$: thus $\mathrm{sign}_x=(-,-,+,+)$ assigns $-$ to
$\psi_0,\psi_1$ and $+$ to $\psi_2,\psi_3$, and not the reverse. Second, $q_1$ is the
upper/lower half of the bispinor, the eigenvalue of $\beta$, and \emph{not} the sign of the
energy: the mass of this scheme acts through $\alpha_y$ (Eq.~\eqref{eq:majorana}), which couples
the pairs $(\psi_0,\psi_3)$ and $(\psi_1,\psi_2)$ (Eq.~\eqref{eq:collision-comp}), so the
positive- and negative-energy eigenstates are superpositions that mix both values of $q_1$; at
the band edge $k=0$ they are $(\psi_0\pm i\,\psi_3)/\sqrt2$, an equal mixture of the two halves.
The identification of $q_1{=}0$ with positive energy holds only for the ordinary Dirac equation,
where the mass sits on $\beta$. A $D$-dimensional lattice uses one position register per axis
and a \emph{shared} spinor register, for a total of $2+\sum_{a}\npos^{(a)}$ qubits.

\begin{table}[H]
  \centering
  \caption{The four spinor components in the two-qubit encoding $c=q_0+2q_1$ of
    Eq.~\eqref{eq:layout}, with $q_0$ the low bit and $q_1$ the high bit. Columns give the
    component $\psi_c$, its bit pattern $\ket{q_1q_0}$, the spin eigenvalue
    $\Sigma_z=\mathbb{I}_2\otimes\sigma_z$ (fixed by $q_0$ alone, $+1\,{=}\,\uparrow$), the half
    eigenvalue $\beta=\diag(+1,+1,-1,-1)$ (fixed by $q_1$ alone, $+1$ the upper half
    $(\psi_0,\psi_1)$), and the per-component streaming signs $\mathrm{sign}_a$ of
    Eq.~\eqref{eq:signs} for the three sweeps. Every row and every sign tuple is ordered by
    $c=0,1,2,3$. The shift direction depends only on $q_1$, while $q_0$ rides along untouched;
    the $\pm E$ energy eigenstates, by contrast, mix the two values of $q_1$.}
  \label{tab:spinor-map}
  \begin{tabular}{@{}cccccccc@{}}
    \toprule
    $c$ & comp. & $\ket{q_1q_0}$ & spin $\Sigma_z$ & half $\beta$ &
      $\mathrm{sign}_x$ & $\mathrm{sign}_y$ & $\mathrm{sign}_z$ \\
    \midrule
    $0$ & $\psi_0$ & $\ket{00}$ & $+1\ (\uparrow)$   & $+1$ (upper) & $-$ & $+$ & $-$ \\
    $1$ & $\psi_1$ & $\ket{01}$ & $-1\ (\downarrow)$ & $+1$ (upper) & $-$ & $+$ & $-$ \\
    $2$ & $\psi_2$ & $\ket{10}$ & $+1\ (\uparrow)$   & $-1$ (lower) & $+$ & $-$ & $+$ \\
    $3$ & $\psi_3$ & $\ket{11}$ & $-1\ (\downarrow)$ & $-1$ (lower) & $+$ & $-$ & $+$ \\
    \bottomrule
  \end{tabular}
\end{table}

The single-axis index of Eq.~\eqref{eq:layout} generalizes to several axes as a
mixed-radix (place-value) number: the spinor is the least-significant digit, of radix $4$
and shared by all axes, while each axis contributes its own position register as a
higher-order block. Ordering the registers as spinor, then $x$, then $y$, then $z$, the
combined index is
\begin{equation}
  k \;=\; c \;+\; 4\bigl(x + N_x\,y + N_x N_y\,z\bigr),
  \label{eq:layout3d}
\end{equation}
with $c\in\{0,\dots,3\}$ the spinor component and $x,y,z$ the site indices along the three
axes ($0\le x<N_x$, $0\le y<N_y$, $0\le z<N_z$, and $N_a=2^{\npos^{(a)}}$). This reduces to
$k=4x+c$ in one dimension and to $k=4(x+N_x\,y)+c$ in two, and is exactly the row-major
(C-order) flattening of an array of shape $(N_z,N_y,N_x,4)$. The qubit ordering is chosen so
that the emulator's state-vector index coincides with this classical flat index, so the
circuit and the reference solver share a single indexing with no intervening permutation. Each
axis sweep acts on the shared spinor qubits together with that axis's position register alone,
so it advances only that axis's position digit (and the shared spinor) while the other axes'
registers ride along unchanged.

This is amplitude encoding of the
field in the sense of the gate model \cite{nielsenChuang2010}: the
$2^{2+\sum_a\npos^{(a)}}$ amplitudes of the qubit register are the components
of $\psi$.

\section{Gate-level construction of the unit operations}
\label{sec:units}

The unit operations are built two different ways, according to their size.

The \emph{spinor operators} (the rotations $\Rmat_a$ and the collision $\Qhat$) are known
as explicit $4\times4$ matrices (Section~\ref{sec:scheme}). We do not design their gate
sequences by hand. Instead we wrap each matrix as a single ``unitary gate'' and pass it to
a \emph{transpiler}: a standard compiler for quantum circuits (here the built-in
\texttt{transpile} routine of Qiskit \cite{qiskit2024}, an in-library function we call
rather than code ourselves) that automatically rewrites a circuit into an equivalent one
using only a chosen set of elementary gates, and optimizes their number. We target the set
$\{R_z,R_y,R_x,\mathrm{CX}\}$, where $R_x(\theta)=e^{-i\theta\sigma_x/2}$,
$R_y(\theta)=e^{-i\theta\sigma_y/2}$ and $R_z(\theta)=e^{-i\theta\sigma_z/2}$ are the
single-qubit rotation gates by angle $\theta$ generated by the Pauli matrices
Eq.~\eqref{eq:pauli}, and $\mathrm{CX}$ (controlled-NOT) is the two-qubit gate that flips a
\emph{target} qubit when the \emph{control} qubit is in the state $\ket1$. For one- and
two-qubit matrices, this matrix-to-circuit synthesis is \emph{exact} and has a known
closed form the transpiler can apply automatically; namely, the Cartan (KAK) decomposition \cite{shende2006}, which expresses any
$4\times4$ unitary as single-qubit rotations interleaved with at most three $\mathrm{CX}$
gates. (A unitary
on more qubits would be handled by the recursive Quantum Shannon decomposition, but every
spinor operator here acts only on the two spinor qubits, so the two-qubit case suffices.)
Our contribution at this step is therefore not the decomposition algorithm, which is supplied by the library's transpile function, but the routine that supplies each target matrix and independently verifies the returned circuit (Section~\ref{sec:assemble}).

The \emph{streaming and boundary operators} are handled the opposite way. As matrices, they
act on the whole position register and are $4N\times4N$, growing with the lattice. Synthesizing a circuit from such a large, structureless matrix is exponentially expensive
and defeats the purpose; instead, we \emph{construct these circuits explicitly, by hand},
from multi-controlled-$X$ (MCX) gates, which flip a target qubit only when several
control qubits are simultaneously $\ket1$. (The single-qubit Pauli bit flip $X=\sigma_x$
and $\mathrm{CX}$ are the zero- and one-control special cases). These hand-built circuits are
passed through the transpiler afterwards, solely to count their gates in the common basis.
Measured gate counts for both kinds of operation are collected in
Table~\ref{tab:resources}.

\subsection{Rotations}
Each $\Rmat_a$ is a fixed $4\times4$ unitary acting on the two spinor qubits, with no
free parameters and no dependence on grid size or physics. Transpiling Eqs.~\eqref{eq:Rz} or
\eqref{eq:Rx} to the standard gate set gives a constant-size block: one CX gate and
depth $4$, reproducing the target to a phase-invariant gate fidelity of $1$ to twelve
digits (Table~\ref{tab:resources}). Figure~\ref{fig:transpiled}(top) shows the
transpiled $\Rmat_x$.

\subsection{Collision}
\begin{figure}[H]
  \centering
  \includegraphics[height=3.1cm]{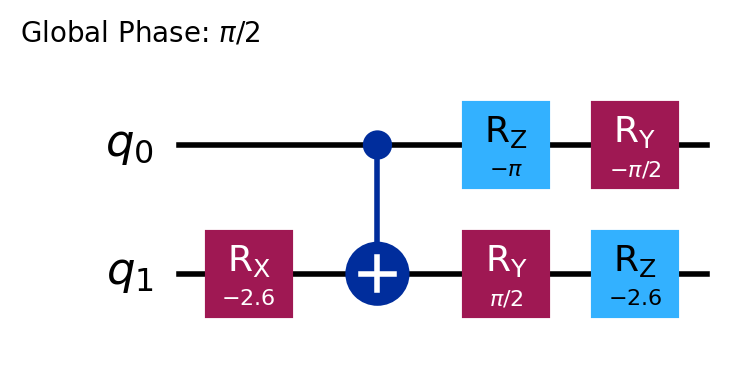}\hfill
  \includegraphics[height=3.1cm]{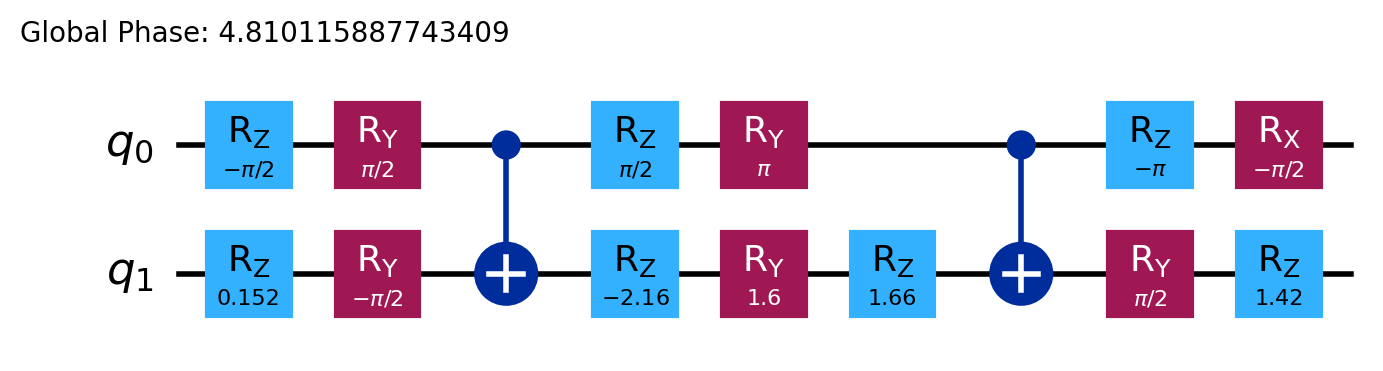}
  \caption{Transpiled unit gates on the two spinor qubits, in the basis
    $\{R_z,R_y,R_x,\mathrm{CX}\}$. Top: the fixed $x$-rotation $\Rmat_x$ of
    Eq.~\eqref{eq:Rx} (one CX, depth $4$). Bottom: the collision $\Qhat(\mtil{=}0.3,
    \gtil{=}0.1)$ of Eq.~\eqref{eq:collision} (two CX, depth $9$). Both reproduce their
    target $4\times4$ unitary to gate fidelity $1$ (twelve digits).}
  \label{fig:transpiled}
\end{figure}

When the mass and potential are uniform, the collision operator $\Qhat_a$ reduces to a single fixed two-qubit gate. Transpiling the collision operator yields an exact circuit with two CX gates and depth $9$
(Figure~\ref{fig:transpiled}, bottom). The massless collision
$\Qhat=\ahat\,\mathbb{I}$ is a global phase and compiles to a single-qubit phase (zero
CX gates). A position-dependent potential is handled in Section~\ref{sec:potential}. By Eq.~\eqref{eq:collision-comp}, the gate factors into two independent $\mathrm{SU}(2)$ rotations. The
two-body $\mathrm{SU}(2)$ structure is the Dirac analogue of the parametric collision
circuits used for classical lattice Boltzmann equilibria \cite{lbmqc2025,sawant_mean_2025}, with the real
scalar amplitudes there replaced by the complex spinor couplings $(\ahat,\bhat)$.

\subsection{Streaming as a controlled increment}

In the rotated frame, streaming is a shift of the position register: it moves each spinor
component rigidly by one lattice site, forward or backward along the axis according to the signs of Eq.~\eqref{eq:signs}.
Under the amplitude encoding of Eq.~\eqref{eq:layout} the amplitude of a given component at
site $x$ lives on the position-register basis state $\ket{x}$, so advancing that component
from site $x$ to site $x+1$ for \emph{every} $x$ at once is the map
$\ket{x}\mapsto\ket{x+1}$ on the $N=2^{\npos}$ basis states of the register. That map only
relabels basis states, so it is a permutation and therefore unitary; it is the
\emph{increment} operator, ordinary binary ``add one'' applied to the integer
$x=\sum_j 2^{\,j}p_j$ stored in the register. A backward shift $\ket{x}\mapsto\ket{x-1}$ is
the \emph{decrement}, its inverse. No amplitude is created or destroyed, so streaming needs
no collision-type gate and reduces entirely to arithmetic on the address bits.

The increment must be controlled, because the two halves of the spinor stream in opposite
directions; by Eq.~\eqref{eq:signs} the
direction is set entirely by the single spinor qubit $q_1$: components $0,1$ (which have
$q_1=0$) move one way and components $2,3$ (which have $q_1=1$) the other. Streaming is
therefore a \emph{controlled} increment, with $q_1$ as the control. For the $x$- and
$z$-sweeps the register is incremented ($x\mapsto x+1$) on the $q_1=1$ components and
decremented ($x\mapsto x-1$) on the $q_1=0$ components; the $y$-sweep exchanges the two
directions. The spin qubit $q_0$ takes no part in streaming and is left untouched.

\begin{figure}[h]
  \centering
  \begin{minipage}[c]{0.30\textwidth}
    \centering
    \begin{quantikz}[column sep=6pt,row sep=10pt]
      \lstick{$p_0$} & \ctrl{1} & \ctrl{1} & \gate{X} & \qw \\
      \lstick{$p_1$} & \ctrl{1} & \targ{}   & \qw      & \qw \\
      \lstick{$p_2$} & \targ{}  & \qw       & \qw      & \qw
    \end{quantikz}
  \end{minipage}\hfill
  \begin{minipage}[c]{0.66\textwidth}
    \centering
    \includegraphics[width=\linewidth]{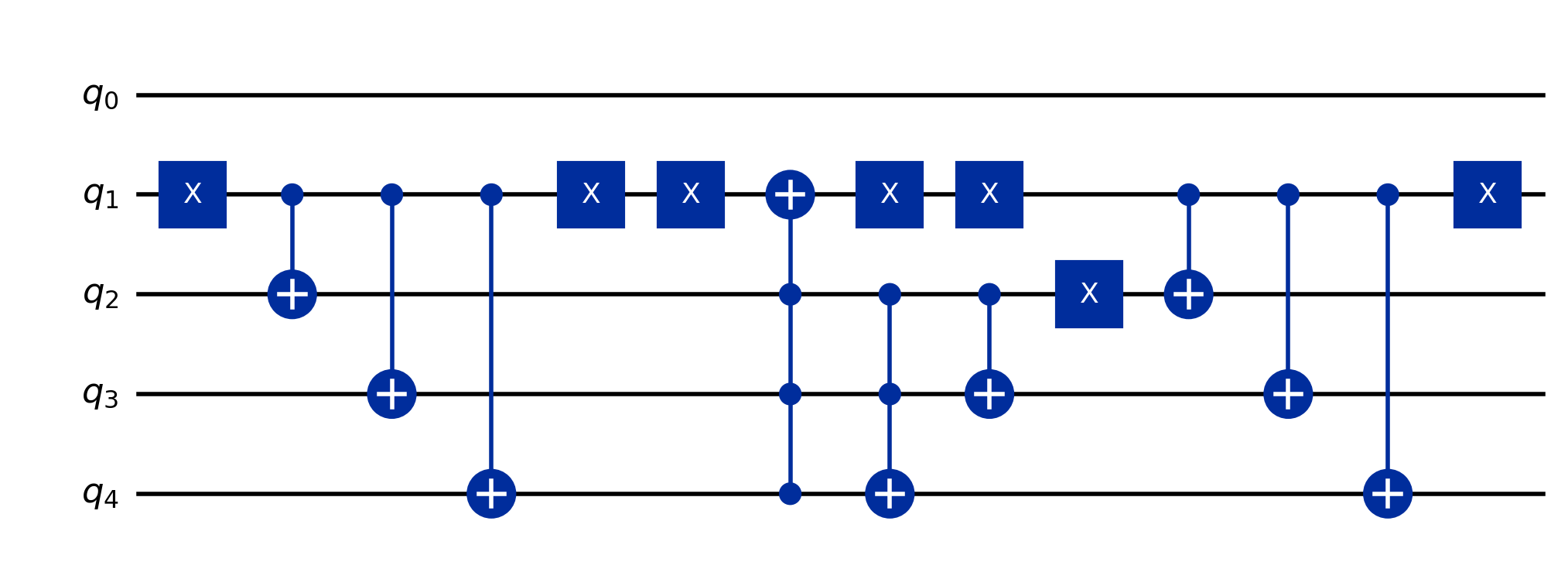}
  \end{minipage}
  \caption{Streaming circuits. Left: the increment $+1\bmod 2^{\npos}$ of
    Eq.~\eqref{eq:incr} for $\npos=3$ (a ripple of multi-controlled-$X$ gates). Right:
    the reflecting (bounce-back) streaming of Section~\ref{sec:bc} for $\npos=3$: the
    direction qubit $q_1$ is folded into the position register (controlled complement),
    a single $(\npos{+}1)$-bit increment is applied, and the register is unfolded. A
    mover reaching a wall crosses the fold and returns with its direction reversed. Each
    adjacent pair of $X$ gates on the $q_1$ line is the identity ($X^2=\mathbb{I}$): the
    panel shows the unoptimized, hand-built circuit, in which those $X$s separately bracket
    the controlled complement and set $q_1$ as the increment's most significant bit, and the
    transpiler cancels them.}
  \label{fig:stream}
\end{figure}

The increment itself is binary addition, which follows the schoolbook carry rule. Label the address bits so
that $p_0$ is the least significant bit (LSB) and $p_{\npos-1}$ the most significant.
Incrementing flips $p_0$ unconditionally; a carry then ripples upward, flipping $p_1$ only
if $p_0$ was $1$, flipping $p_2$ only if $p_0$ and $p_1$ were both $1$, and in general
flipping $p_k$ only if all lower bits $p_0,\dots,p_{k-1}$ were simultaneously $1$. The
instruction ``flip $p_k$ when $p_0,\dots,p_{k-1}$ are all $1$'' is precisely one
multi-controlled-$X$ gate with target $p_k$ and controls $p_0,\dots,p_{k-1}$, so the whole
increment is a ripple of multi-controlled-$X$ gates,
\begin{equation}
  \mathrm{Incr}=\Bigl(\textstyle\prod_{k=\npos-1}^{1}\!
  \mathrm{MCX}(p_0,\dots,p_{k-1}\!\to p_k)\Bigr)\,X(p_0),
  \label{eq:incr}
\end{equation}
where the product runs over the target bits. Because every carry condition tests the bit
values \emph{before} the increment, the gates are applied from the most significant target
downward: first flip $p_{\npos-1}$ (controlled on all lower bits), then $p_{\npos-2}$, and so
on, with the unconditional flip of $p_0$ coming last, precisely because $p_0$ is a control of
every other gate. Figure~\ref{fig:stream} (left) shows the resulting circuit for $\npos=3$,
whose three gates are, in application order, $\mathrm{MCX}(p_0,p_1\!\to\!p_2)$, then
$\mathrm{CX}(p_0\!\to\!p_1)$, then $X(p_0)$. Writing a register state as $\ket{p_2p_1p_0}$ (so
that $x=p_0+2p_1+4p_2$) and labeling each arrow by the bit its gate flips, or ``no flip''
when that gate's controls are not all set, three representative inputs evolve as
\begin{align*}
  x=1:\quad \ket{001}&\xrightarrow{\ \text{no flip}\ }\ket{001}
    \xrightarrow{\ p_1\ }\ket{011}\xrightarrow{\ p_0\ }\ket{010}\ \ (x{=}2),\\
  x=3:\quad \ket{011}&\xrightarrow{\ \ p_2\ \ }\ket{111}
    \xrightarrow{\ p_1\ }\ket{101}\xrightarrow{\ p_0\ }\ket{100}\ \ (x{=}4),\\
  x=7:\quad \ket{111}&\xrightarrow{\ \ p_2\ \ }\ket{011}
    \xrightarrow{\ p_1\ }\ket{001}\xrightarrow{\ p_0\ }\ket{000}\ \ (x{=}0),
\end{align*}
the last line showing the modular wrap $7\mapsto0$ that makes the shift periodic.

Eq.~\eqref{eq:incr} holds for any lattice size, and the circuit that implements it is a single loop over the target bits: for $k$ from
$\npos-1$ down to $1$ it appends one multi-controlled-$X$ with target $p_k$ and the running
list of lower bits $p_0,\dots,p_{k-1}$ as controls, and finally the unconditional $X(p_0)$.
The operator matrix, which is $2^{\npos}\times2^{\npos}$ and grows with the lattice, is never
formed; only its known carry structure is used. Because the loop is parameterized by the
register size $\npos$, the identical few lines generate the correct circuit for any number of
grid points $N=2^{\npos}$, placing just $\npos$ gates. This is the sense in which the streaming
circuit is built ``by hand'': the \emph{algorithm} is fixed once and for all, not a separate
circuit for each $N$. At this stage each multi-controlled-$X$ is a single symbolic
instruction; only the subsequent transpilation to the elementary set
$\{R_z,R_y,R_x,\mathrm{CX}\}$ expands it into two-qubit gates, which is where the counts of
Table~\ref{tab:resources} arise. The reflecting boundary of Section~\ref{sec:bc} is produced by
the very same loop, applied to the position register extended by the direction qubit $q_1$
(Algorithm~\ref{alg:refl}), so no new construction is needed there either.

For streaming, the controlled increment adds $q_1$ as one extra control on every gate of
Eq.~\eqref{eq:incr}, so the ripple fires only on the $q_1=1$ components; the decrement is the
same circuit executed in reverse order (each gate being its own inverse), which subtracts one
and carries the $q_1=0$ components the other way (Figure~\ref{fig:stream}, left). Because
the arithmetic is taken modulo $2^{\npos}$, the site
past the last one wraps to the first, so Eq.~\eqref{eq:incr} already realizes \emph{periodic}
streaming; this matches the periodic $y$- and $z$-sweeps and the bulk of the $x$-sweep.
With the ancilla-free multi-controlled-$X$ synthesis used here the gate count grows as
reported in Table~\ref{tab:resources}; a ripple-carry adder with clean ancilla qubits
\cite{gidney2018} would reduce a controlled increment to $O(\npos)$ gates, and an ancilla-free
Fourier adder \cite{draper2000} to $O(\npos^2)$ gates, optimizations we do not pursue here.

Nothing new is required to stream in two or three dimensions. Each axis carries its own
position register $P_a$ (Section~\ref{sec:encoding}), while the two spinor qubits are shared
by all axes. The $a$-sweep applies the controlled increment above to $P_a$ alone and leaves
the registers of the other axes untouched, so they ride along as spectators. The only
per-axis difference is the fixed spinor rotation $\Rmat_a$ of
Eqs.~\eqref{eq:Rz}--\eqref{eq:Rx}: it selects, through the direction qubit $q_1$, \emph{which}
pair of spinor components are the forward- and backward-movers along that axis, after which
the shift gates are identical from axis to axis, merely rewired to a different register.
Motion along a diagonal is produced by composition: within one time step the $x$-sweep
advances a packet in $x$ and the $y$-sweep advances it in $y$, for a net displacement of
$(\pm1,\pm1)$ lattice units. The complete time step is the ordered product of the single-axis
sweeps (Section~\ref{sec:assemble}, Eq.~\eqref{eq:step}), which is the operator (dimensional)
splitting used in any classical lattice Boltzmann code that sweeps one lattice direction at a
time. Because the position registers are disjoint, the per-axis streaming circuits act on
separate qubits and share no gates, so the cost of $D$-dimensional streaming is simply the
sum of the $D$ one-dimensional costs.

\subsection{Position-dependent potential}
\label{sec:potential}
A scalar potential that varies from site to site, $V=V(x)$, enters the scheme entirely
through the collision. The dimensionless coupling of Eq.~\eqref{eq:couplings} becomes the
local value $\gtil(x)=qV(x)\Delta t/(s\hbar)$, so the collision operator $\Qhat$ of
Eq.~\eqref{eq:collision} carries a site-dependent argument $\gtil(x)$. Only the collision
layer of the substep changes; the fixed rotations and the streaming shift are untouched. How
the site dependence is realized as a gate divides into two cases.

\begin{figure}[h]
  \centering
  \includegraphics[width=0.9\linewidth]{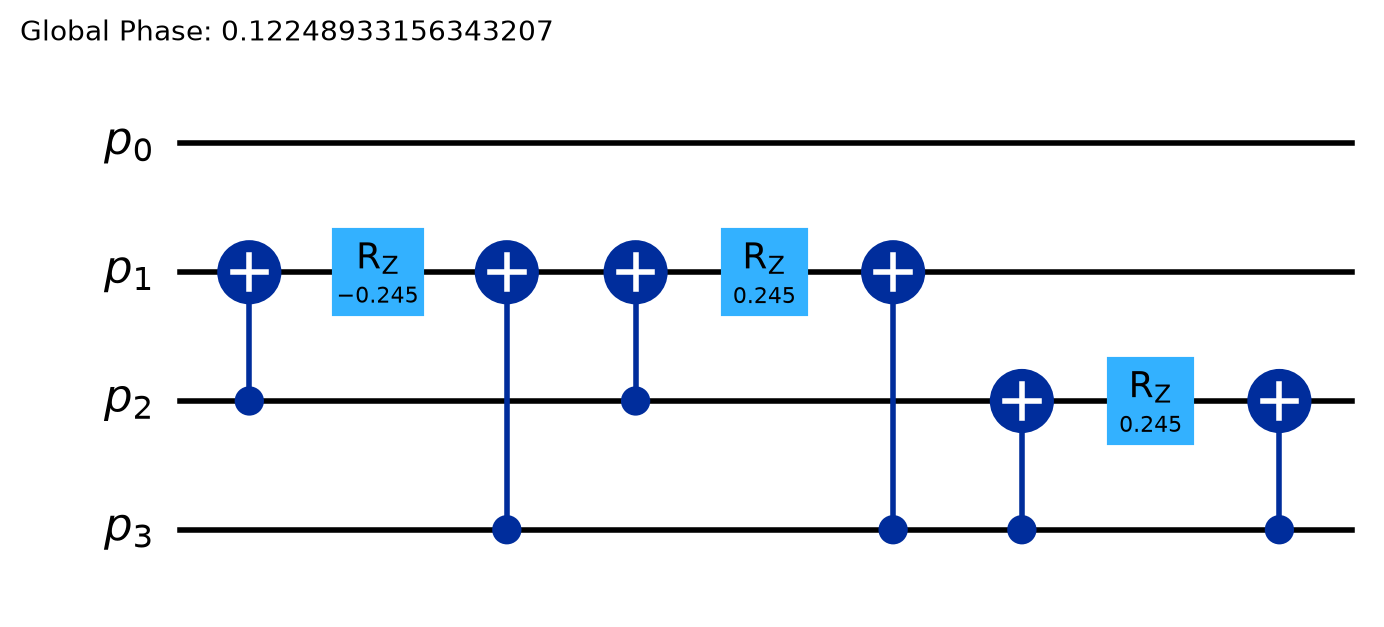}
  \caption{The transpiled massless potential phase oracle $\mathcal{O}_V$ of
    Eq.~\eqref{eq:oracle} for the $\npos=4$ barrier of Table~\ref{tab:resources}:
    $\gtil=0.5$ on the four sites $x=6,7,8,9$ of a $16$-site line and $0$ elsewhere,
    compiled to $\{R_z,R_y,R_x,\mathrm{CX}\}$ (six $\mathrm{CX}$, three $R_z(\pm\theta/2)$,
    depth $9$, and a global phase $\theta/4$, with $\theta=2\arctan(\tfrac14)\approx0.49$).
    The lowest position bit $p_0$ carries no gate because the barrier phase does not depend
    on it; the $\mathrm{CX}$ ladders compute the bit parities of the phase polynomial
    $\theta\,(p_3+p_1p_2-p_1p_3-p_2p_3)$ and the $R_z$ rotations stamp its phases
    (Section~\ref{sec:potential}).\protect\footnotemark[1]}
  \label{fig:oracle}
\end{figure}

\runin{Massless case: a phase oracle.}
In the massless limit $\mtil\to0$ the mass amplitude $\bhat$ of Eq.~\eqref{eq:collision}
vanishes and the collision is a pure phase, $\Qhat=\ahat\,\mathbb{I}_4$ with $|\ahat|=1$.
Setting $\mtil=0$ in Eq.~\eqref{eq:couplings} gives $\Omega=-\gtil^2$, so the numerator and
denominator of $\ahat=(1-\Omega/4)/D$ become $1-\Omega/4=1+\gtil^2/4$ and
$D=1+\Omega/4-i\gtil=1-\gtil^2/4-i\gtil$. Both are controlled by the single complex number
$z=1-\tfrac{i}{2}\gtil$, whose square returns the denominator and whose squared modulus
returns the numerator,
\begin{equation}
  z^2=\bigl(1-\tfrac{i}{2}\gtil\bigr)^2=1-\tfrac{\gtil^2}{4}-i\gtil=D,\qquad
  |z|^2=1+\tfrac{\gtil^2}{4}=1-\tfrac{\Omega}{4}.
  \label{eq:massless-z}
\end{equation}
Writing $\bar z$ for the complex conjugate of $z$ (the overbar denotes complex conjugation,
so that $|z|^2=z\bar z$), the collision amplitude is therefore the ratio of a modulus to a
square, which collapses to a pure phase,
\begin{equation}
  \ahat=\frac{1-\Omega/4}{D}=\frac{|z|^2}{z^2}=\frac{z\bar z}{z^2}=\frac{\bar z}{z}=e^{-2i\arg z},
  \label{eq:massless-ratio}
\end{equation}
making $|\ahat|=1$ manifest for every real $\gtil$. Since
$\arg z=\arg\!\bigl(1-\tfrac{i}{2}\gtil\bigr)=-\arctan\!\bigl(\tfrac{1}{2}\gtil\bigr)$, the
phase $\theta=-2\arg z$ is the closed form
\begin{equation}
  \ahat(x)=e^{\,i\theta(x)},\qquad
  \theta(x)=2\arctan\!\bigl(\tfrac{1}{2}\gtil(x)\bigr)\;\xrightarrow[\ \gtil\to0\ ]{}\;\gtil(x),
  \label{eq:massless-phase}
\end{equation}
a known real phase per substep, the small-coupling limit following from $\arctan u\to u$. Since $\Qhat$ is now proportional to the identity on the
spinor, it multiplies all four spinor components by the \emph{same} factor $\ahat(x)$ and so
acts on the position register alone. The result is a \emph{phase oracle}, a diagonal unitary
that stamps each position basis state $\ket{x}$ with its own phase,
\begin{equation}
  \mathcal{O}_V=\diag\bigl(\ahat(0),\dots,\ahat(N-1)\bigr),\qquad
  \mathcal{O}_V\ket{x}=e^{\,i\theta(x)}\ket{x},
  \label{eq:oracle}
\end{equation}
with $\ahat(x)\equiv\ahat(\gtil(x))$. A general phase profile is synthesized as a diagonal
unitary on the $\npos$ position qubits at a cost of $O(N)$ elementary gates ($6$ CX and depth
$9$ for the $\npos=4$ barrier of Table~\ref{tab:resources}, drawn in
Figure~\ref{fig:oracle}), but physically structured
potentials are much cheaper: a linear ramp $V(x)\propto x$ makes
$\theta(x)=\sum_j(\alpha\,2^{\,j})p_j$ a sum over the address bits $p_j$, so the oracle is
just $\npos$ independent single-qubit $R_z$ rotations with no entangling gates, and a
piecewise-constant barrier over an address-aligned block of sites is a single
multi-controlled phase. Figure~\ref{fig:oracle} works through the small example tabulated,
a four-site barrier ($\gtil=0.5$ on sites $x=6,\dots,9$ of a $16$-site line): those sites
form two adjacent pairs that differ only in the lowest bit $p_0$, so the phase does not
depend on $p_0$ and that qubit carries no gate, while the remaining pattern is the phase
polynomial $\theta\,(p_3+p_1p_2-p_1p_3-p_2p_3)$ with $\theta=2\arctan(\tfrac14)$, which the
transpiler renders as three $R_z(\pm\theta/2)$ rotations wrapped in $\mathrm{CX}$ ladders
(six $\mathrm{CX}$, depth $9$), well below the $2^{\npos}-1$ two-qubit gates a generic
diagonal would need. Because the collision contributes no spinor mixing here
($|\ahat|=1$), the potential acts purely by imprinting phase: a massless particle at normal
incidence is perfectly transmitted through an electrostatic step (Klein tunneling), while at
oblique incidence the phase gradient still turns part of the packet around, as in the
two-dimensional benchmark of Section~\ref{sec:verify} (reflected fraction $R\approx0.38$).

\footnotetext[1]{The global phase $\theta/4$ is physically unobservable, since it multiplies
every amplitude equally, and is dropped on hardware; it is recorded only so that the compiled
circuit equals $\mathcal{O}_V$ exactly rather than up to an overall phase.}

\begin{figure}[h]
  \centering
  \includegraphics[width=0.40\linewidth]{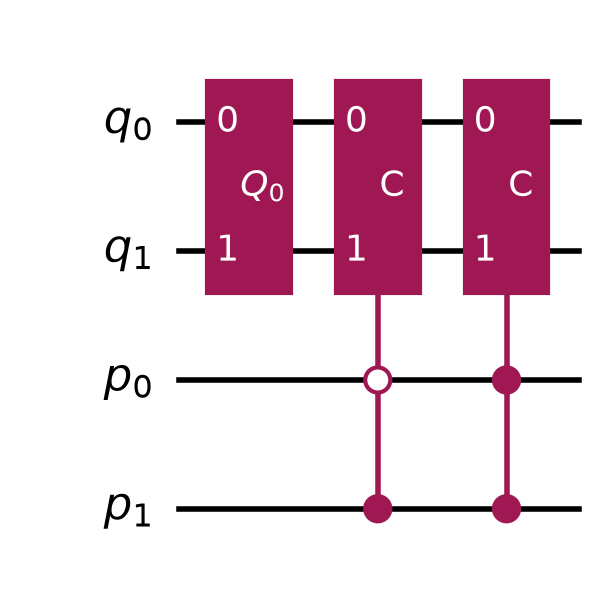}
  \caption{Position-multiplexed collision for a massive potential
    (Eq.~\eqref{eq:multiplexed}), on a $\npos=2$ line with a two-site barrier ($\gtil=0.5$ on
    sites $x=2,3$, mass $\mtil=0.3$). The vacuum collision $\Qhat_0=\Qhat(\mtil,0)$ acts on the
    two spinor qubits $(q_0,q_1)$ unconditionally; the correction
    $C=\Qhat(\mtil,\gtil)\,\Qhat(\mtil,0)^{-1}$ is applied to the spinor under control of the
    position register $(p_0,p_1)$, once per barrier site (filled circle: control on $1$; open
    circle: control on $0$). Off the barrier $C$ is the identity, so no gate is placed there.
    The two corrections shown share the same $C$ and $p_1{=}1$, hence reduce to a single $C$
    controlled on $p_1{=}1$.}
  \label{fig:mux}
\end{figure}
\runin{Massive case: a position-multiplexed collision.}
When $\mtil\neq0$ the mass amplitude $\bhat$ of Eq.~\eqref{eq:collision} is nonzero, so the
collision $\Qhat(\mtil,\gtil(x))$ is no longer a phase but a genuine two-qubit
$\mathrm{SU}(2)$ gate that mixes the spinor components in the pattern of
Eq.~\eqref{eq:collision-comp}, and its coupling $\gtil(x)$ still varies from site to site. The
circuit must therefore apply a \emph{different} two-qubit spinor gate according to the value
stored in the position register. Such an operation is a \emph{multiplexed} gate, also called a
uniformly controlled gate: the position register selects, like the address of a lookup table,
which spinor gate the two spinor qubits receive, namely $\Qhat(\mtil,\gtil(x))$ at site $x$. It
is the non-diagonal counterpart of the phase oracle, a full spinor rotation in place of a
scalar phase at each address.

A separate gate at all $N$ sites is unnecessary because the potential is nonzero only on part
of the lattice. We split each collision into a site-independent part and a correction,
\begin{equation}
  \Qhat(\mtil,\gtil(x))=\underbrace{\Qhat(\mtil,\gtil(x))\,\Qhat(\mtil,0)^{-1}}_{C(x)}\;\Qhat(\mtil,0),
  \label{eq:multiplexed}
\end{equation}
apply the \emph{vacuum} collision $\Qhat(\mtil,0)$ once to the two spinor qubits, the
mass-only gate common to every site, and then apply the correction $C(x)$, which is the
identity wherever $\gtil(x)=0$ and so acts only on the support of the potential. Each
nontrivial $C(x)$ is a two-qubit spinor gate conditioned on the position register, that is an
$\npos$-controlled gate; a barrier covering $O(N)$ sites costs $O(N)$ such controlled gates,
and each is far more expensive than a phase because it entangles the spinor and position
registers. This is the price of the mass: the potential now rotates the spinor components
(Eq.~\eqref{eq:collision-comp}) rather than merely stamping a phase on them.

Figure~\ref{fig:mux} shows the smallest instructive case, a four-site line ($\npos=2$) with a
two-site barrier ($\gtil=0.5$ on sites $x=2,3$) and mass $\mtil=0.3$. The vacuum collision
$\Qhat_0=\Qhat(\mtil,0)$ acts on the two spinor qubits everywhere; the correction
$C=\Qhat(\mtil,\gtil)\,\Qhat(\mtil,0)^{-1}$ is then applied to the spinor qubits under control of
the position register, once for site $2$ (controls $p_0{=}0,\,p_1{=}1$) and once for site $3$
(controls $p_0{=}1,\,p_1{=}1$). Because the two barrier sites carry the same height, $C$ is the
\emph{same} gate at both and the two controls together cover both values of $p_0$ with
$p_1{=}1$, so the pair is equivalent to a single $C$ controlled only on $p_1{=}1$, the
address-aligned block $\{2,3\}$. This is the concrete form of the statement that a
piecewise-constant barrier on an aligned block reduces to one fixed spinor gate under a
multi-controlled selection of that block. Because $\bhat\neq0$ makes $C$ mix the spinor
components, the barrier now reflects the gapped particle even at normal incidence, as in the
one-dimensional massive-barrier benchmark of Section~\ref{sec:verify}, in contrast to the
massless (Klein) transmission above.

\runin{Placement in the sweep and limiting checks.}
Either form simply replaces the collision layer $\Qhat_a$ in the substep of
Figure~\ref{fig:sweep}, and, like the collision, is applied
once per substep, so its coupling carries the same $1/s$ factor as Eq.~\eqref{eq:couplings}.
The massless oracle commutes with the fixed rotations, because it is proportional to the
identity on the spinor and hence $\Rmat_a^{-1}\mathcal{O}_V\Rmat_a=\mathcal{O}_V$, so it may
be inserted directly on the position register; the massive correction $C(x)$ is applied in
the rotated frame alongside $\Qhat_a$. Two limits fix ideas: a spatially \emph{uniform}
potential reduces Eq.~\eqref{eq:oracle} to a global phase $\ahat\,\mathbb{I}_N$, which is
physically unobservable and compiles to a single-qubit phase with no CX gates; and switching
the potential off, $\gtil\equiv0$, returns $\ahat=1$ and the free scheme of
Section~\ref{sec:scheme}.

\subsection{Boundary conditions}
\label{sec:bc}
Boundaries are treated per axis. Two boundary conditions are unitary and therefore admit
exact circuits.
\begin{itemize}\itemsep2pt
  \item \emph{Periodic}: the modular increment of Eq.~\eqref{eq:incr} already wraps
    around, so periodic streaming is the default.
  \item \emph{Reflecting (bounce-back)}: a hard wall at each end of the axis, at which a
    mover reverses its direction and stays inside the domain, so that no probability is lost
    and the operation is unitary. It is realized by a single \emph{folded} increment,
    described in detail below (Figure~\ref{fig:stream}, right; Algorithm~\ref{alg:refl}).
\end{itemize}

\runin{Bounce-back as periodic streaming on a folded ring.}
A plain periodic increment implies that a forward mover leaving the last site would
reappear at the first, teleporting across the domain. A reflecting wall must instead turn
the mover around while keeping it inside the domain. The construction that does this exactly
is the method of images, familiar from classical lattice Boltzmann bounce-back, namely to
double the line into a ring. Label the $2N$ states by (direction, site), writing $R_x$ for
the forward mover at site $x$ and $L_x$ for the backward mover, and arrange them in the
cyclic order
\begin{equation}
  \underbrace{R_0\to R_1\to\cdots\to R_{N-1}}_{\text{forward arc}}\;\to\;
  \underbrace{L_{N-1}\to\cdots\to L_1\to L_0}_{\text{backward arc}}\;\to\;R_0 ,
  \label{eq:ring}
\end{equation}
where the final arrow closes the ring. Following the arrows, a forward mover advances one
site at a time until it reaches the right wall at $R_{N-1}$, where the next step turns it
into the backward mover $L_{N-1}$ at the \emph{same} site; the backward movers then walk
back to the left wall at $L_0$, where the ring closes by turning $L_0$ into $R_0$. Every
arrow is a single physical hop, and the two turning points fall exactly on the two walls, so
one step around this ring \emph{is} reflecting streaming. Being a single $2N$-cycle, that
step is a plain increment modulo $2N$ of a coordinate that runs once around the ring, and the
modular wrap $2N{-}1\mapsto0$ supplies one of the two wall crossings.

For $N=4$ the ring Eq.~\eqref{eq:ring} is
$R_0\to R_1\to R_2\to R_3\to L_3\to L_2\to L_1\to L_0\to R_0$. A forward mover approaching the
right wall advances $R_2\to R_3$, reflects $R_3\to L_3$ (its direction reversed, its site held
at the wall), and returns $L_3\to L_2$; a backward mover reaching the left wall reflects
$L_0\to R_0$. No amplitude is ever carried across the domain.

The ring is addressed by an $(\npos{+}1)$-bit register formed from the $\npos$ position bits
together with the direction qubit $q_1$, which supplies the doubling to $2N=2^{\npos+1}$
slots. To lay the forward and backward arcs head to tail, the ``unfold'' step of
Algorithm~\ref{alg:refl} complements the position bits of the backward movers, so that their
sites run from $N-1$ down to $0$ as the ring requires, while the forward movers are left
unchanged; $q_1$ is the most significant bit of the register. A single plain increment
$p\mapsto p+1\bmod 2^{\npos+1}$ then walks the ring, and the ``fold'' step undoes the
complement. In the interior of either arc the increment leaves the top bit alone, so the
mover simply advances one site; at a wall all the lower bits are set, the carry reaches the
top bit, and $q_1$ flips, which is precisely the reversal of direction. The spin qubit $q_0$
is never touched, so the reflection reverses the direction and preserves the spin, as a hard
wall must (Figure~\ref{fig:stream}, right). For the $y$-sweep the forward and backward roles
of $q_1$ are exchanged, because the signs in Eq.~\eqref{eq:signs} are reversed there, so $q_1$
is flipped once before and once after the folded increment in order to reuse the identical
circuit.

An \emph{open/absorbing} boundary is not unitary (it removes probability from the
domain) and therefore cannot be a pure gate circuit; it would require mid-circuit
measurement and reset. This is a genuine limitation of the method, and we make no
attempt to force it into a unitary form.

\begin{algorithm}[H]
\caption{Reflecting (bounce-back) streaming circuit for axis $a$}
\label{alg:refl}
\begin{algorithmic}[1]
\Require spinor qubits $(q_0,q_1)$ with $q_1$ the direction, position register $P$
\If{$a=y$} $X(q_1)$ \EndIf \Comment{make $q_1{=}1$ the right mover}
\State $X(q_1)$;\ \ \textbf{for} $p\in P$: $\mathrm{CX}(q_1\!\to p)$;\ \ $X(q_1)$
  \Comment{unfold: complement $P$ when $q_1{=}0$}
\State apply $\mathrm{Incr}$ to the extended register $(p_0,\dots,p_{\npos-1},q_1)$
  \Comment{$+1 \bmod 2^{\npos+1}$}
\State $X(q_1)$;\ \ \textbf{for} $p\in P$: $\mathrm{CX}(q_1\!\to p)$;\ \ $X(q_1)$
  \Comment{fold back}
\If{$a=y$} $X(q_1)$ \EndIf
\end{algorithmic}
\end{algorithm}

\section{Assembling a time step}
\label{sec:assemble}

\subsection{Single axis}
A single-axis sweep $\mathrm{Sweep}_a$ is the substep of Eq.~\eqref{eq:substep}: the fixed
rotation $\Rmat_a^{-1}$, the collision $\Qhat_a$, the (periodic or reflecting) streaming
circuit, and $\Rmat_a$. Figure~\ref{fig:sweep} shows the circuit. Its unitary equals the exact
classical substep operator $\Rmat_a\,\mathrm{Stream}_a\,\Qhat_a\,\Rmat_a^{-1}$ on the
$4N$-dimensional space, to machine precision.

\begin{figure}[H]
  \centering
  \begin{quantikz}[column sep=8pt,row sep=14pt]
    \lstick{$q_0$ (spin)} & \qw & \gate[2]{\Rmat_a^{-1}} & \gate[2]{\Qhat_a} & \qw & \qw & \gate[2]{\Rmat_a} & \qw \\
    \lstick{$q_1$ (dir.)} & \qw &                        &                   & \ctrl{1} & \octrl{1} &          & \qw \\
    \lstick{$\ket{x}$}    & \qwbundle{\npos} & \qw       & \qw               & \gate{+1} & \gate{-1} & \qw     & \qw
  \end{quantikz}
  \caption{One QLB substep along axis $a$ (Eq.~\eqref{eq:substep}): rotate into the
    characteristic frame, collide on the two
    spinor qubits, stream by a $q_1$-controlled increment/decrement of the position
    register, and rotate back. The filled/open control on $q_1$ selects
    increment/decrement; for reflecting walls the two controlled shifts are replaced by
    the single folded increment of Algorithm~\ref{alg:refl}. For a massless potential the
    collision is replaced by the phase oracle Eq.~\eqref{eq:oracle} on the position register.}
  \label{fig:sweep}
\end{figure}

\subsection{Two and three dimensions}
A $D$-dimensional time step is the product of the single-axis sweeps, each acting on
its own position register and on the shared spinor register:
\begin{equation}
  U_{\mathrm{step}}
  = \mathrm{Sweep}_z\,\mathrm{Sweep}_y\,\mathrm{Sweep}_x ,
  \label{eq:step}
\end{equation}
composed on qubits $(q_0,q_1)\cup P_x$, $(q_0,q_1)\cup P_y$, $(q_0,q_1)\cup P_z$
respectively. Because the spinor register is shared, the 3D
circuit inherits the correctness of the 1D sweep. Per-axis boundary conditions are
selected independently, so a one- or two-dimensional problem is a three-dimensional
lattice with thin periodic transverse directions, exactly as in a classical lattice
solver.

\subsection{A general porting and verification routine}
All of the above is produced and checked by a single routine (Algorithm~\ref{alg:port}):
given a target unitary $U$, it compiles $U$ to the chosen gate set and then \emph{verifies}
that the compiled circuit reproduces $U$ on the state-vector emulator, returning a fidelity
together with the two-qubit gate count and the depth.

The fidelity compares the target $U$ with the unitary $U_{\mathrm{circ}}$ that the compiled
circuit actually implements,
\begin{equation}
  F=\frac{\bigl|\mathrm{tr}(U^\dagger U_{\mathrm{circ}})\bigr|}{\dim},\qquad \dim=2^n ,
  \label{eq:fidelity}
\end{equation}
in which $U^\dagger$ is the conjugate transpose (adjoint) of $U$ and $\mathrm{tr}$ the matrix
trace. The numerator $\mathrm{tr}(U^\dagger U_{\mathrm{circ}})$ is the Hilbert--Schmidt
(Frobenius) inner product of the two operators, the natural overlap between matrices; dividing
by $\dim$ normalizes it so that identical operators give $F=1$, since
$\mathrm{tr}(U^\dagger U)=\mathrm{tr}(\mathbb{I})=\dim$. We use this operator overlap, rather
than the overlap $|\langle\psi|U^\dagger U_{\mathrm{circ}}|\psi\rangle|$ of a single evolved
state, for two reasons.

First, it is \emph{invariant under an overall phase}. Taking the modulus discards a factor
$e^{i\varphi}$: if $U_{\mathrm{circ}}=e^{i\varphi}U$ then
$\mathrm{tr}(U^\dagger U_{\mathrm{circ}})=e^{i\varphi}\dim$ and $F=1$. An overall phase is
physically unobservable and is routinely introduced by the compiler (for instance the global
phase of Figure~\ref{fig:oracle}), so a fidelity that penalized it would report spurious
failures.

Second, and more importantly, $F=1$ certifies an \emph{exact} match on \emph{every} input at
once. By the Cauchy--Schwarz inequality in the Hilbert--Schmidt inner product,
$|\mathrm{tr}(U^\dagger U_{\mathrm{circ}})|\le\dim$ for two unitaries, with equality if and
only if $U_{\mathrm{circ}}=e^{i\varphi}U$. Thus $F=1$ is not the agreement of one test vector,
which two different operators can share by coincidence, but a guarantee that the compiled
circuit equals the target on the whole $2^n$-dimensional space up to an irrelevant phase.
Reaching $F=1$ to twelve digits is therefore a strong, input-independent statement that the
port is exact to floating-point round-off. Alongside $F$ the routine returns the depth, the
number of gate layers executed in sequence, and the two-qubit ($\mathrm{CX}$) gate count, the
dominant hardware cost, so the single call that certifies an operator also measures it.

The check has three deliberate limitations. First, it is an \emph{emulator} check: forming
the $\dim\times\dim$ matrix $U_{\mathrm{circ}}$ and its trace costs $O(4^n)$ time and memory,
so $F$ can be evaluated only at the modest sizes reached here, and it is not a quantity that a real
device could return, since hardware yields samples rather than the operator itself; the routine
certifies the \emph{construction}, it is not a scalable test. Second, it verifies
\emph{compilation}, that the circuit equals its target matrix, and not the \emph{physics}, that
the target matrix is the intended operator; the latter is the separate density comparison of
Section~\ref{sec:verify} against the classical solver, so the two checks are complementary.
Third, it is noiseless, and says nothing about behavior under hardware noise, which
Appendix~\ref{app:hw} examines for a reduced instance. The phase blindness of $F$ is
harmless here because the compiled circuit is itself retained exactly, its global phase
included, and matters only when a block is used as a \emph{controlled} sub-operation, where
that phase becomes relative; it is then carried in the circuit even though $F$ does not score
it.

The rotations and collision are ported by feeding their $4\times4$ matrices to this routine;
the streaming and boundary operators are checked against their classical permutation matrices.
The same routine is what one would apply to any future operator (Section~\ref{sec:discussion}).

\begin{algorithm}[H]
\caption{\textsc{PortAndVerify}$(U,\text{basis})$}
\label{alg:port}
\begin{algorithmic}[1]
\State $\mathrm{qc}\gets\textsc{Transpile}(\textsc{UnitaryGate}(U),\text{basis})$ \Comment{KAK / Shannon}
\State $U_{\mathrm{circ}}\gets\textsc{Unitary}(\mathrm{qc})$ \Comment{state-vector emulator}
\State $F\gets |\mathrm{tr}(U^\dagger U_{\mathrm{circ}})|/\dim$ \Comment{phase-invariant fidelity}
\State \Return $(\mathrm{qc},\,F,\,\textsc{CXcount}(\mathrm{qc}),\,\textsc{Depth}(\mathrm{qc}))$
\end{algorithmic}
\end{algorithm}

\section{Numerical verification}
\label{sec:verify}

\subsection{Setup}
Circuits are executed on the Qiskit Aer \cite{qiskit2024} state-vector emulator on a
graphics processing unit (GPU). For each test we build the initial spinor field $\psi_0$, evolve it both with the
classical solver and with the circuit of Eq.~\eqref{eq:step} applied for the same number
of steps, and compare the resulting probability densities
$\rho(\mathbf{x})=\sum_c|\psi_c(\mathbf{x})|^2$, where the sum runs over the four spinor
components $\psi_c$ and $\rho(\mathbf{x})$ is the probability of finding the particle at
lattice site $\mathbf{x}$. In every test the initial field is a Gaussian wave packet
modulated by a plane-wave carrier and carried by a fixed four-spinor,
\begin{equation}
  \psi_0(\mathbf{x}) = A\,
  \exp\!\Bigl(-\frac{\lVert\mathbf{x}-\mathbf{x}_0\rVert^2}{2\sigma^2}\Bigr)\,
  e^{\,i\,\mathbf{k}\cdot\mathbf{x}}\; u ,
  \label{eq:initcond}
\end{equation}
where $\mathbf{x}=(x,y,z)$ ranges over the integer lattice sites and the ingredients are:
$\mathbf{x}_0$, the site at which the packet is centered at $t=0$; $\sigma$, the width
(standard deviation, in lattice sites) of the Gaussian envelope, a larger $\sigma$ giving a
broader and better-collimated packet that disperses less; $\mathbf{k}=(k_x,k_y,k_z)$, the
carrier wavevector in radians per site, whose winding $e^{i\mathbf{k}\cdot\mathbf{x}}$ sets
the packet's mean momentum and hence a group velocity \cite{mulloth_high_2015} along $\mathbf{k}$; $u$, a constant
four-component spinor fixing the internal Dirac state, taken as a positive-velocity
eigenvector of the lattice streaming (velocity) operator so that the packet moves as a single
coherent lobe rather than splitting into counter-propagating branches; and $A$, the constant
that normalizes $\lVert\psi_0\rVert=1$. The spinor is $u=\Rmat_x\,(0,0,1,1)^{\!\top}/\sqrt2$
for the axis-aligned movers of the 1D and 2D tests, and the positive-velocity eigenvector of
the appropriate velocity operator for the diagonal and oblique movers described below.
Eq.~\eqref{eq:initcond} with the per-test parameters listed below fully specifies each
initial state. The figure of merit is the maximum absolute density deviation
$\max|\Delta\rho|$, the largest difference between the circuit and classical densities
over all sites and recorded times; the state fidelity
$|\langle\psi_{\mathrm{circ}}|\psi_{\mathrm{class}}\rangle|$, the overlap of the two
normalized state vectors (equal to $1$ for identical states), is $1$ to twelve digits in
every case. We distinguish two errors: (i) the \emph{circuit-vs-classical}
deviation reported here, which
measures the fidelity of the \emph{port} and is at the level of floating-point round-off;
and (ii) the scheme's own discretization error relative to the continuum Dirac equation
(dispersion, etc.), which is a property of the Succi--Dellar method and is validated
separately in the library's solver tests. Only (i) is addressed here.

\subsection{Results}
\begin{figure}[htb]
  \centering
  \includegraphics[width=0.96\linewidth]{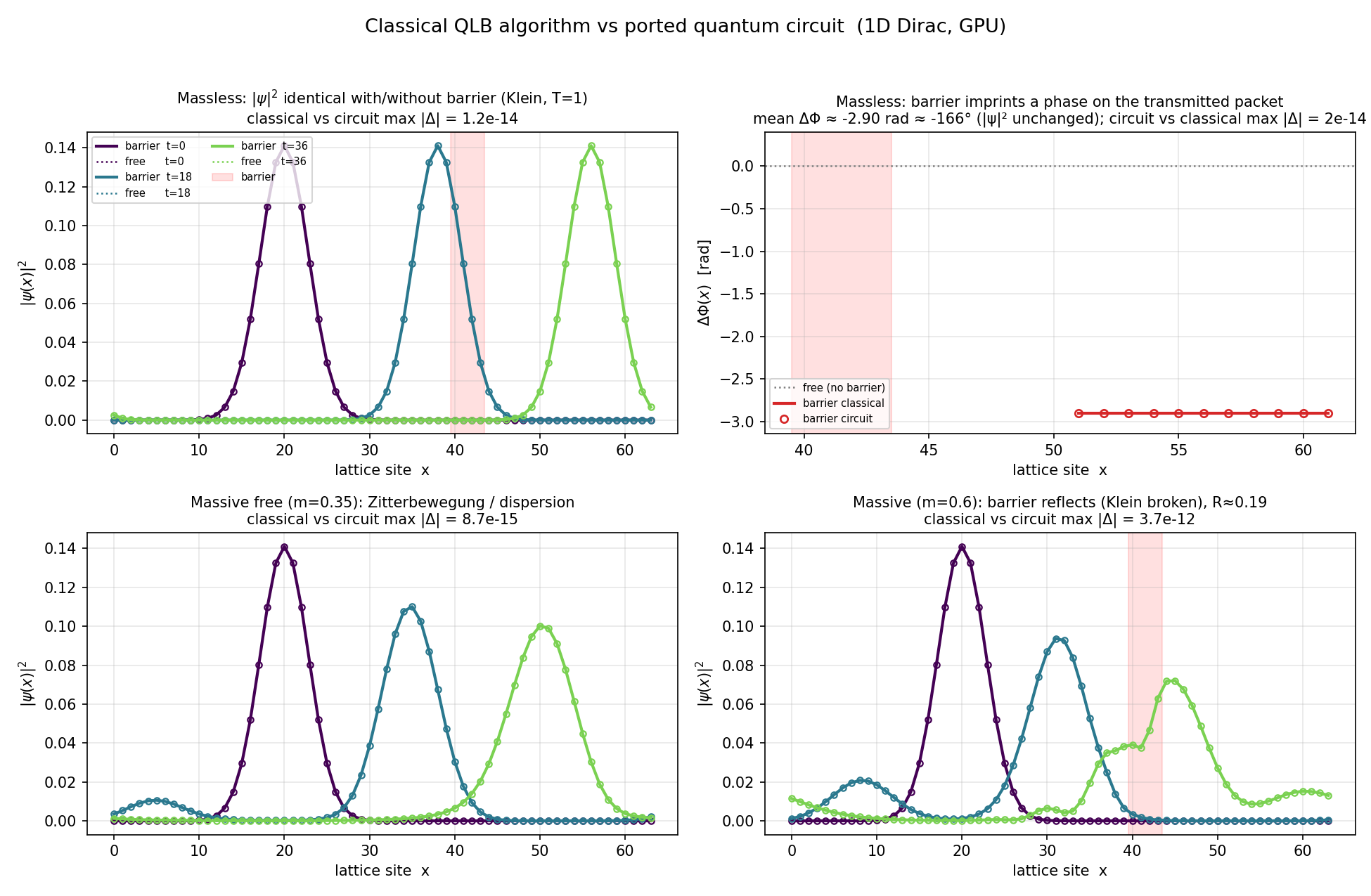}
  \caption{One-dimensional validation on a $64$-site line (classical QLB as lines, circuit
    as open markers), packet width $\sigma=4$, momentum $k=0.6$, at $t=0,18,36$. Top row: a
    massless packet is perfectly transmitted through a barrier (Klein, $T=1$, left), which
    nonetheless imprints a phase $\approx-166^\circ$ on it while leaving $|\psi|^2$ unchanged
    (right). Bottom row: a massive free packet ($\mtil=0.35$) disperses with Zitterbewegung
    (left), and a massive packet ($\mtil=0.6$) is partly reflected by the barrier
    ($R\approx0.19$, right) once the mass gap breaks the Klein protection. Circuit and
    classical densities agree from $8.7\times10^{-15}$ to $3.7\times10^{-12}$.}
  \label{fig:v1d}
\end{figure}

\begin{figure}[htb]
  \centering
  \includegraphics[width=0.96\linewidth]{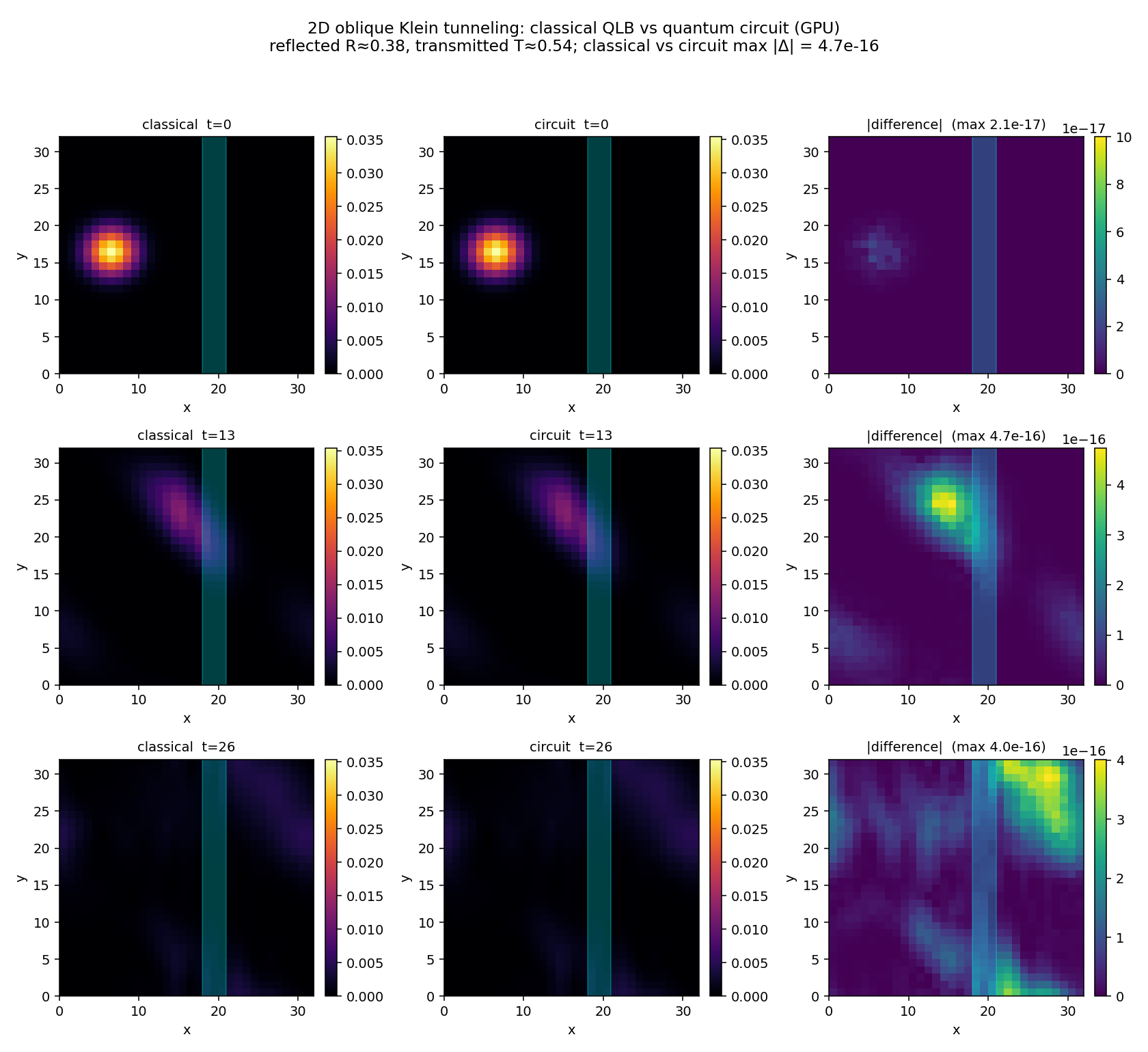}
  \caption{Two-dimensional validation on a $32\times32$ lattice: a massless packet at oblique
    incidence on a vertical barrier (cyan strip). Columns: classical density, circuit density,
    and their absolute difference; rows: $t=0,13,26$. The packet splits into reflected and
    transmitted lobes ($R\approx0.38$, $T\approx0.54$). $\max|\Delta\rho|=4.7\times10^{-16}$.}
  \label{fig:v2d}
\end{figure}

\begin{figure}[htb]
  \centering
  \includegraphics[width=0.9\linewidth]{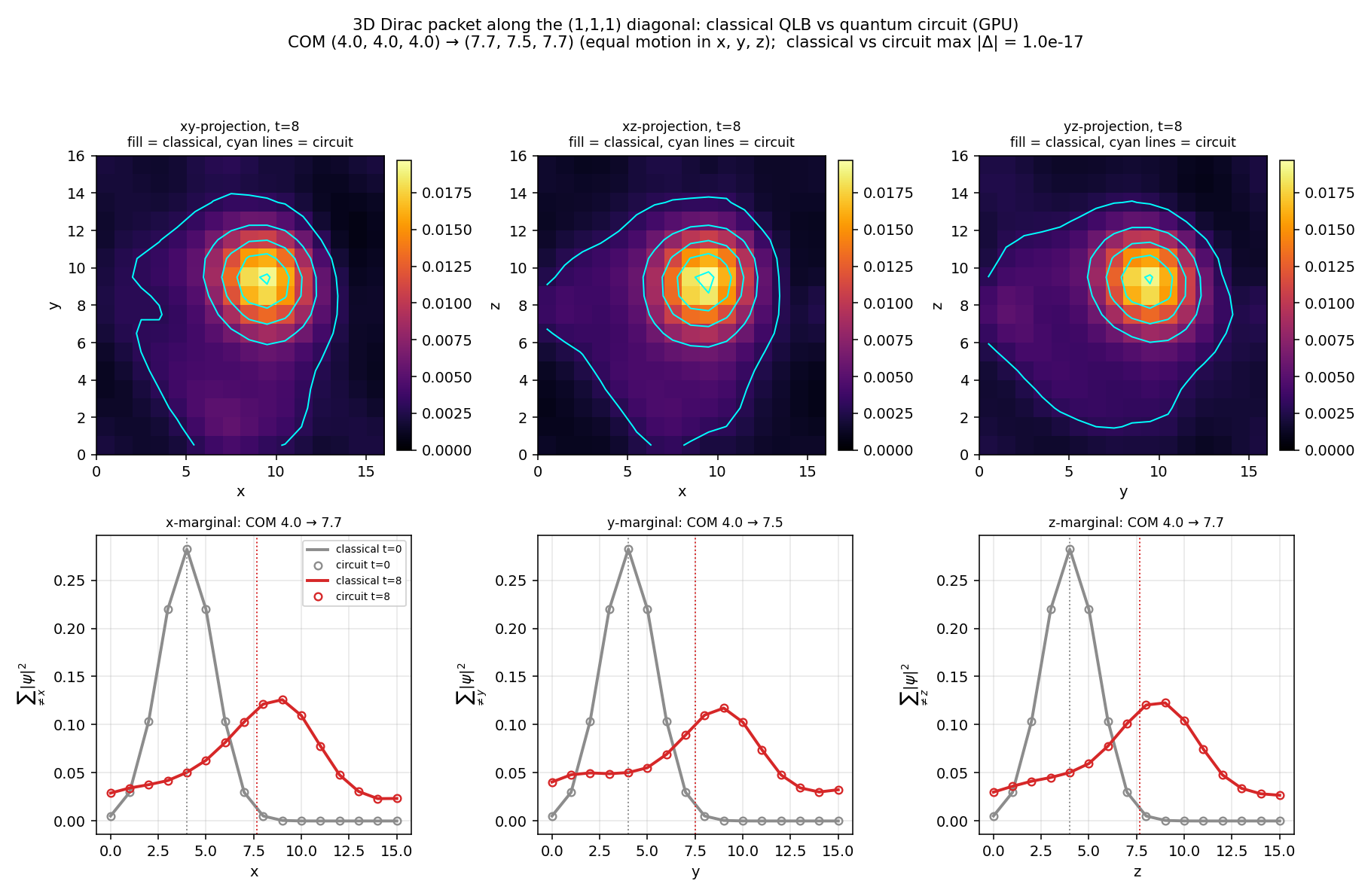}
  \caption{Three-dimensional validation on a $16^3$ lattice: a massless packet launched along
    the body diagonal with equal group velocity on all three axes, so every sweep is
    exercised. Top row: the $xy$, $xz$ and $yz$ density projections at $t=8$ (classical filled,
    circuit as cyan contours); bottom row: the $x$, $y$ and $z$ marginals at $t=0$ and $t=8$
    (classical lines, circuit markers). The center of mass moves diagonally from $(4,4,4)$ to
    $\approx(7.7,7.5,7.7)$. $\max|\Delta\rho|=1.0\times10^{-17}$.}
  \label{fig:v3d}
\end{figure}

\begin{figure}[htb]
  \centering
  \includegraphics[width=0.98\linewidth]{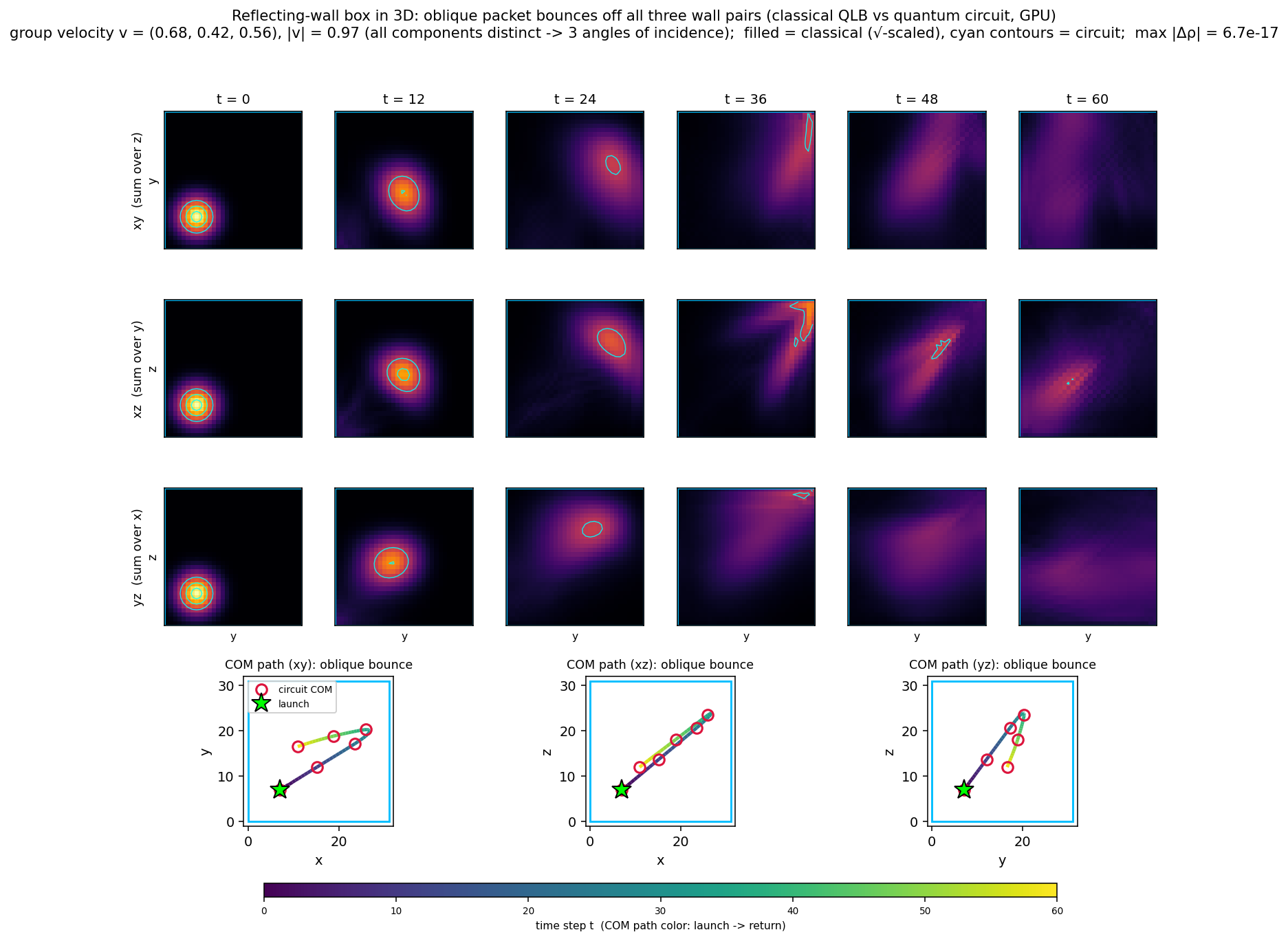}
  \caption{Reflecting (bounce-back) boundary conditions in 3D. A massless packet launched
    from the corner $(7,7,7)$ of a $32^3$ box with group velocity $v\approx(0.68,0.42,0.56)$
    strikes the $x$-, $z$- and $y$-wall pairs at three distinct angles
    ($\approx45^\circ,55^\circ,65^\circ$) and three different times, and returns. Rows:
    density projected onto the $xy$, $xz$ and $yz$ planes; columns: time steps
    $t=0,\dots,60$; filled color is the classical scheme, cyan contours the circuit. Bottom:
    the center-of-mass trajectory in each plane, colored by time
    (launch~$\rightarrow$~return), with circuit markers on the classical path.
    $\max|\Delta\rho|=6.7\times10^{-17}$; probability conserved.}
  \label{fig:vbc}
\end{figure}

Table~\ref{tab:validation} collects the deviations; Figures~\ref{fig:v1d}--\ref{fig:vbc}
show representative runs.

\begin{table}[H]
  \centering
  \caption{Circuit-vs-classical maximum density deviation $\max|\Delta\rho|$ over all
    sites and recorded times; the state fidelity is $1$ to twelve digits throughout. All
    values are at or near floating-point round-off; the $1$D entry is the largest, being
    round-off accumulated through the massive per-site collision over its $36$-step run.}
  \label{tab:validation}
  \begin{tabular}{@{}llll@{}}
    \toprule
    Test & Lattice (qubits) & Physics & $\max|\Delta\rho|$ \\
    \midrule
    1D free / barrier / phase & $2^6$ line ($8$) & massless \& massive & $3.7\times10^{-12}$ \\
    2D oblique Klein         & $2^5\!\times\!2^5$ ($12$) & massless barrier & $4.7\times10^{-16}$ \\
    3D diagonal mover        & $2^4\!\times\!2^4\!\times\!2^4$ ($14$) & massless free & $1.0\times10^{-17}$ \\
    3D reflecting box        & $2^5\!\times\!2^5\!\times\!2^5$ ($17$) & massless, bounce-back & $6.7\times10^{-17}$ \\
    \bottomrule
  \end{tabular}
\end{table}

\runin{1D (Figure~\ref{fig:v1d}).} A line of $N=2^6=64$ sites ($8$ qubits). A packet
of width $\sigma=4$ centered at $x_0=20$ with carrier momentum $k=0.6$ is evolved for $36$
steps (shown at $t=0,18,36$) in four regimes. (a) \emph{Massless, free versus barrier}: a
four-site barrier ($\gtil=0.9$ on sites $40$--$43$) leaves the density indistinguishable
from the free packet, the perfect transmission ($T=1$) of Klein tunneling at normal
incidence; agreement $1.2\times10^{-14}$. (b) \emph{Massless phase}: the same barrier
leaves $|\psi|^2$ unchanged but imprints a phase on the transmitted packet, here a mean
$\Delta\Phi\approx-2.9$~rad ($\approx-166^\circ$); agreement $2\times10^{-14}$. (c)
\emph{Massive free} ($\mtil=0.35$): the packet broadens and grows the trembling secondary
lobe of \emph{Zitterbewegung} (the rapid quivering of a Dirac wave packet that arises from
interference between its positive- and negative-energy components, which the mass term
couples together, and which disappears in the massless limit), exercising the collision
gate; agreement $8.7\times10^{-15}$. (d)
\emph{Massive barrier} ($\mtil=0.6$, $\gtil=2.0$): the mass gap breaks the Klein
protection, so the barrier now reflects part of the packet, reflected fraction
$R\approx0.19$; agreement $3.7\times10^{-12}$, the largest deviation in the paper, being
round-off accumulated through the per-site massive collision over the run.

\runin{2D (Figure~\ref{fig:v2d}).} A $32\times32$ lattice ($12$ qubits). A massless
packet ($\sigma=3$, center $(6,16)$, momentum $(k_x,k_y)=(0.6,0.5)$) is launched obliquely
at a vertical barrier ($\gtil=0.9$ on the columns $x=18$--$20$) and evolved for $26$ steps
($t=0,13,26$). At oblique incidence the Klein protection is imperfect, so the packet splits
into a reflected and a transmitted lobe, $R\approx0.38$ and $T\approx0.54$; agreement
$4.7\times10^{-16}$.

\runin{3D (Figure~\ref{fig:v3d}).} A $16\times16\times16$ lattice ($14$ qubits). A
massless packet ($\sigma=2$, center $(4,4,4)$, carrier momentum $0.5$ along the body
diagonal), whose spinor is the positive-velocity eigenstate of
$(\alpha_x+\beta+\alpha_z)/\sqrt3$, has equal group velocity on all three axes, so one free
run exercises the $x$-, $y$- and $z$-sweeps together; over $8$ steps its center of mass
moves from $(4,4,4)$ to $\approx(7.7,7.5,7.7)$; agreement $1.0\times10^{-17}$.

\runin{3D reflecting box (Figure~\ref{fig:vbc}).} A $32^3$ box ($17$ qubits) with
reflecting walls on all three axes. A massless packet ($\sigma=3.2$) launched from the low
corner $(7,7,7)$ with carrier momentum $(1.00,0.68,0.15)$ has oblique group velocity
$v\approx(0.68,0.42,0.56)$ ($|v|\approx0.97$) with three unequal components, so it strikes
the $x$-, $z$- and $y$-wall pairs at three distinct angles ($\approx45^\circ,55^\circ,
65^\circ$) and three different times; one $60$-step run therefore tests every reflecting
plane independently. The packet bounces and returns, its center of mass tracing the oblique
path with probability conserved; agreement $6.7\times10^{-17}$.

\noindent In every case the ported circuit reproduces the classical scheme to machine
precision: \emph{the circuit is the scheme.}

\section{Gate counts}
\label{sec:resources}

Table~\ref{tab:resources} lists the measured two-qubit (CX) gate count and circuit depth of
each building block, after transpilation to $\{R_z,R_y,R_x,\mathrm{CX}\}$, as a function of the
per-axis register size $\npos=\log_2 N$. The spinor operators (the rotations, the collision,
and the massless potential oracle) act only on the two shared spinor qubits, so they are
constant-size and independent of the lattice. Only streaming grows with $\npos$, and it
dominates the cost. A single-axis sweep is one streaming circuit plus the two fixed rotations
($1$ CX each) and the collision ($2$ CX if massive, none if massless), and a $D$-dimensional
time step is the sum of the $D$ single-axis sweeps (Eq.~\eqref{eq:step}); a free $32^3$ step
($\npos=5$ per axis) costs about $2700$ CX, nearly all of it streaming.

The streaming counts are an un-optimized upper bound: we build the increment without helper
qubits, from bare multi-controlled-$X$ gates, which is simple but grows quickly with $\npos$.
Adder-based increments would bring streaming down to $O(\npos)$ gates with clean ancilla
qubits \cite{gidney2018}, or $O(\npos^2)$ gates without them \cite{draper2000}, and reflecting streaming is already
much cheaper than periodic because it uses one increment rather than two. The spinor gates are
minimal (the KAK decomposition gives one CX for a rotation and two for the collision). We
report these numbers only to characterize the primitives, and make no claim that they are the
minimal cost of the scheme.

\begin{table}[H]
  \centering
  \caption{Measured CX gate count and depth of each building block after transpilation to
    $\{R_z,R_y,R_x,\mathrm{CX}\}$ (optimization level $3$), for a per-axis register of
    $\npos=\log_2 N$ qubits. The spinor operators are constant-size; streaming grows with
    $\npos$ and dominates a time step (Eq.~\eqref{eq:step}).}
  \label{tab:resources}
  \begin{tabular}{@{}lrr@{}}
    \toprule
    Building block & CX & depth \\
    \midrule
    Rotation $\Rmat_a$ (fixed)          & $1$ & $4$ \\
    Collision $\Qhat$ (massive)         & $2$ & $9$ \\
    Collision $\Qhat$ (massless)        & $0$ & $0$ \\
    Potential phase oracle, $\npos=4$   & $6$ & $9$ \\
    \midrule
    Periodic streaming, $\npos=3,4,5,6$   & $106,331,900,2203$ & $215,647,1831,4494$ \\
    Reflecting streaming, $\npos=3,4,5,6$ & $22,45,79,123$     & $39,81,140,233$     \\
    \bottomrule
  \end{tabular}
\end{table}

\section{Discussion and outlook}
\label{sec:discussion}

We give an exact, operation-by-operation quantum-circuit realization of the
three-dimensional Succi--Dellar Dirac QLB scheme, with the tools and evidence to trust it.
Specifically, we provide (i) explicit gate-level constructions for every unit operation, the
fixed rotations, the two-qubit collision, streaming as a controlled increment on the position
register, the position-dependent potential as a phase oracle in the massless case and a
multiplexed collision in the massive case, and periodic and reflecting boundaries as unitary
circuits; (ii) their composition into single-axis, two- and three-dimensional time steps by
tensoring one position register per axis on a shared spinor register; (iii) a numerical
demonstration that the assembled circuits reproduce the classical solver to machine precision
in 1D, 2D and 3D, including Klein tunneling and a three-wall reflecting box; (iv) measured
gate counts that characterize the cost of each primitive; and (v) a small, reusable
``unitary~$\rightarrow$~verified circuit'' routine (Algorithm~\ref{alg:port}) that ported and
independently checked every operator here and applies unchanged to future ones. Because each
circuit layer is validated against its classical reference, the port is trustworthy layer by
layer, not only end to end. Two aspects are meant to be reusable beyond this scheme: the
methodology, porting a validated lattice-kinetic solver to circuits one operation at a time
with a classical check at each step; and the primitives themselves, the increment-based
streaming, the folded-ring bounce-back, and the phase-oracle potential, for others building
Dirac or lattice-Boltzmann circuits.

The result is one of exact representability, and we claim no more; delineating what a full
assessment of computational advantage would still require is itself part of the contribution.
Three costs remain open. \emph{Input}: preparing a general spinor field as an
amplitude-encoded state is a state-preparation problem in its own right, which the emulator
side-steps here by direct state loading. \emph{Output}: physical observables (densities,
currents, transmission coefficients) must be estimated by measurement with the usual sampling
overhead, whereas we compare full state vectors, which hardware cannot provide.
\emph{Asymptotics}: we report measured gate counts at accessible sizes and note the standard
adder-based improvements, but make no scaling claim. Naming these boundaries precisely is what
lets a subsequent study attack them directly.

The construction suggests several continuations, in increasing order of difficulty:
(i) hardware-native compilation and error mitigation for instances larger than the reduced
run of Appendix~\ref{app:hw}, to gauge what present devices can execute;
(ii) ancilla-based
adders to bring streaming to $O(\npos)$ depth; (iii) the potential landscapes of the
Klein-tunneling benchmark \cite{palpacelli2012} at scale; and (iv) the many-body
quantum-field-theory extension of the scheme \cite{succi2007}, where the collision no
longer has the two-body $\mathrm{SU}(2)$ form and must satisfy the equal-time
commutation relations in addition to unitarity, a genuinely open problem for which the
verification routine of Algorithm~\ref{alg:port} provides a ready target function.

\section{Reproducibility}
\label{sec:repro}

All operators, circuits, tests and figures are in the public
\texttt{quantumKineticMethods} library \cite{qkm2026}. The relevant modules are: \texttt{operators.py} (the matrices of
Section~\ref{sec:scheme}), \texttt{streaming.py} (the increment of Eq.~\eqref{eq:incr} and
the bounce-back of Algorithm~\ref{alg:refl}), \texttt{sweep.py} (the substep of
Eq.~\eqref{eq:substep}), \texttt{twod.py} and \texttt{threed.py} (the time step of
Eq.~\eqref{eq:step}), \texttt{potential.py}
(Section~\ref{sec:potential}), \texttt{port.py} (Algorithm~\ref{alg:port}), and
\texttt{backend.py} (the emulator interface). The validation suite \texttt{test\_port.py}
checks every operator and composition against its classical reference; the plotting
scripts \texttt{plot\_validation*.py} regenerate Figures~\ref{fig:v1d}--\ref{fig:vbc}, with
\texttt{plot\_validation\_zitterbewegung.py} regenerating Figure~\ref{fig:zb} and
Table~\ref{tab:zb}, and \texttt{make\_figures.py} in this paper's directory regenerates the
circuit diagrams and Table~\ref{tab:resources}. The classical solver that provides the reference is
\texttt{dirac\_qlb\_solver.py}. The hardware study of Appendix~\ref{app:hw} is reproduced by
\texttt{run\_hardware\_zitterbewegung.py} (the hardware primitives, native-gate compilation, and the
state-preparation, readout, and emulator cross-check of Section~\ref{app:hw-demo}),
\texttt{run\_2site\_density.py} and \texttt{run\_2site\_trembling.py} (the position- and
velocity-versus-time runs of Section~\ref{app:hw-results}), and \texttt{plot\_2site\_results.py}
(Figure~\ref{fig:hw-results} and its tables), with \texttt{export\_quil.py} and
\texttt{qcs\_quil/run\_qcs.py} for the pinned run of Section~\ref{app:hw-pinned}; the raw device counts returned by the processor are
stored with them.

\section*{Author contributions}
N.S. developed the quantum-circuit constructions and the open-source
\texttt{quantumKineticMethods} library, performed the numerical verification, and wrote the
manuscript. E.Y. initiated the group's research direction in quantum computing in a
supervisory role. E.Y. and K.G. contributed to the conceptual development through discussions, took part in
verifying the circuit constructions, and reviewed and edited the manuscript. M.M.
contributed to the conceptual development through brainstorming discussions during the proposal
stage of the project. A large-language-model based agent assisted with copy-editing and reformatting of the
manuscript, with the \LaTeX{} typesetting, and with writing and refactoring code in the
library; all constructions, results, and their verification were devised and checked by the
authors, who take full responsibility for the work.

\section*{Acknowledgements}
This work was authored by the National Laboratory of the Rockies (NLR) for the U.S. Department of Energy (DOE) under Contract No. DE-AC36-08GO28308. The views expressed in the article do not necessarily represent the views of the DOE or the U.S. Government. The U.S. Government retains and the publisher, by accepting the article for publication, acknowledges that the U.S. Government retains a nonexclusive, paid-up, irrevocable, worldwide license to publish or reproduce the published form of this work, or allow others to do so, for U.S. Government purposes.

This work builds on the quantum lattice Boltzmann formulation of S.~Succi and the three-dimensional isotropy analysis of P.~Dellar and co-workers. 
We thank Wesley Jones for initiating and coordinating the quantum computing activities at the center. Thanks to Obadiah Reid and Ross Larsen for participating in discussions 
during the proposal stage of the project. A portion of the research was performed using computational resources sponsored by the Department of Energy's Office of Critical Minerals and Energy Innovation and located at the National Laboratory of the Rockies.
Time for the project was partially covered by NLR internal funds Scalable Parallel Accelerated Computing (SPAC) and HPC User \& Application Support (HPCAPPS).

The hardware demonstration of Appendix~\ref{app:hw-results} was performed on the Rigetti Cepheus-1 processor through the Open Quantum cloud platform~\cite{openquantum}, whose access we gratefully acknowledge. 
We thank Rigetti for providing access to the Cepheus-1-108Q quantum processor through their Quantum Cloud Services (QCS), which enabled the fine-grained experiments of Appendix~\ref{app:hw-pinned}.

\bibliographystyle{unsrtnat}
\bibliography{references}

\appendix

\section{Physical units}
\label{app:phys}

\subsection{Unit conversion}
\label{app:units}
The scheme is formulated in lattice units, so the examples of Section~\ref{sec:verify} quote
only dimensionless ratios. Physical values follow once a single dimensional scale, the lattice
spacing $\Delta x$ (a length, in meters, m), is fixed; the time step $\Delta t$ (in seconds, s)
is then set by the light-cone rule
\begin{equation}
  c_0\,\Delta t = \Delta x ,
  \label{eq:lightcone}
\end{equation}
with $c_0\approx2.998\times10^{8}~\mathrm{m\,s^{-1}}$ the speed of light, expressing that streaming
advances one lattice site per substep at speed $c_0$. The remaining constants are the reduced
Planck constant $\hbar\approx1.055\times10^{-34}~\mathrm{J\,s}$ and the particle rest mass $m$ (in
kilograms, kg), which fix the reduced Compton wavelength $\bar\lambda_{\mathrm C}=\hbar/(mc_0)$ (a
length, m). Writing $N$ for the number of sites along an axis and $s$ for the number of substeps
per time step ($s=1,2,3$ in one, two and three dimensions; both dimensionless counts), the
dimensionless quantities of Section~\ref{sec:verify} convert as follows (SI units in brackets).
\begin{itemize}\itemsep2pt
  \item \emph{Lengths} (site index $x$, packet width $\sigma$, center $x_0$; in sites) are
    multiplied by $\Delta x$ to give meters [m], and a box of $N$ sites spans $N\Delta x$.
  \item \emph{Times} (step counts $t,T$) are multiplied by $\Delta t=\Delta x/c_0$ to give
    seconds [s].
  \item \emph{The carrier wavevector} $k$ (radians per site) becomes
    $k_{\mathrm{phys}}=k/\Delta x$ [rad\,m$^{-1}$], with momentum $p=\hbar k/\Delta x$
    [kg\,m\,s$^{-1}$]; group velocities, reported with $|v|\le1$, are in units of $c_0$
    [m\,s$^{-1}$].
  \item \emph{Mass and potential} invert the couplings of Eq.~\eqref{eq:couplings},
    \begin{equation}
      m=\frac{s\hbar\,\mtil}{c_0^2\Delta t}=\frac{s\hbar\,\mtil}{c_0\,\Delta x},\qquad
      qV=\frac{s\hbar\,\gtil}{\Delta t}=\frac{s\hbar\,\gtil\,c_0}{\Delta x},
      \label{eq:phys-mV}
    \end{equation}
    giving the rest mass $m$ [kg] (equivalently the rest energy $mc_0^2$ [J]) and the potential
    energy $qV$ [J]; division by the elementary charge $q=|e|$ re-expresses these energies in
    electronvolts (eV), the unit used in Table~\ref{tab:physunits}.
\end{itemize}
Eq.~\eqref{eq:couplings} also shows that the mass coupling is the lattice spacing measured
in reduced Compton wavelengths,
\begin{equation}
  \mtil=\frac{\Delta x}{s\,\bar\lambda_{\mathrm C}},
  \label{eq:mtil-compton}
\end{equation}
a dimensionless ratio, so resolving the Compton scale requires $\mtil\ll1$. Introducing the
energy scale
\begin{equation}
  E_0 \equiv \frac{\hbar}{\Delta t} = \frac{\hbar c_0}{\Delta x},
  \label{eq:E0}
\end{equation}
an energy [J], most naturally quoted in MeV, every energy and momentum is a simple multiple of a
dimensionless input: $mc_0^2=s\,\mtil\,E_0$, $qV=s\,\gtil\,E_0$ and $pc_0=k\,E_0$ (each an energy,
[J] or MeV). Because the reported physics (the Klein transmission, the
\emph{Zitterbewegung}, the reflected fractions) depends only on the dimensionless ratios, the
tests are scale-free: one choice, such as the particle species or the grid resolution, fixes
$\Delta x$ and hence every entry in Appendix~\ref{app:physunits}.

\subsection{Physical units for the test parameters}
\label{app:physunits}
To attach concrete SI values we fix the single free scale by taking the massive, free
one-dimensional packet ($\mtil=0.35$, $s=1$) to be an electron. By Eq.~\eqref{eq:mtil-compton}
this gives
\begin{equation}
  \Delta x = 0.35\,\frac{\hbar}{m_{\mathrm e}c}\approx1.35\times10^{-13}~\mathrm{m},\qquad
  \Delta t = \frac{\Delta x}{c}\approx4.51\times10^{-22}~\mathrm{s},\qquad
  E_0=\frac{\hbar}{\Delta t}\approx1.46~\mathrm{MeV}.
  \label{eq:refscale}
\end{equation}
Table~\ref{tab:physunits} then expresses every test of Section~\ref{sec:verify} in these units
through $mc_0^2=s\,\mtil\,E_0$, $qV=s\,\gtil\,E_0$ and $pc_0=k\,E_0$. The tests thus describe
relativistic Dirac fermions at Compton-scale resolution, with sub-MeV to few-MeV masses,
momenta and potentials, on picometer boxes evolved over tens of zeptoseconds
($1~\mathrm{zs}=10^{-21}$~s). A different choice of $\Delta x$ rescales the table through
Appendix~\ref{app:units}: energies and momenta as $1/\Delta x$, lengths as $\Delta x$, and times
as $\Delta x$.

\begin{table}[htb]
  \centering
  \caption{The verification tests of Section~\ref{sec:verify} in physical units, for the
    reference scale of Eq.~\eqref{eq:refscale}. Dimensionless values are given in parentheses.
    The one-dimensional column lists its two massive cases ($\mtil=0.35$ and $0.6$) and its two
    barrier heights ($\gtil=0.9$ and $2.0$); the remaining tests are massless ($\mtil=0$), and
    only the two-dimensional test carries a potential ($\gtil=0.9$, applied over $s=2$
    substeps). The group velocity $|v|/c_0$ is the packet's center-of-mass drift per step in
    units of $c_0=\Delta x/\Delta t$ ($|v|=\sqrt{\sum_a v_a^2}$ over the swept axes); its three
    one-dimensional entries are the massless case and the two massive cases.}
  \label{tab:physunits}
  \begin{tabular}{@{}lcccc@{}}
    \toprule
     & 1D & 2D & 3D & 3D box \\
    \midrule
    Lattice spacing $\Delta x$           & \multicolumn{4}{c}{$1.35\times10^{-13}$~m} \\
    Time step $\Delta t$                 & \multicolumn{4}{c}{$4.51\times10^{-22}$~s} \\
    Velocity $\Delta x/\Delta t=c$       & \multicolumn{4}{c}{$3.00\times10^{8}$~m\,s$^{-1}$} \\
    \midrule
    Sites per axis $N$      & $64$ & $32$ & $16$ & $32$ \\
    Substeps $s$            & $1$  & $2$  & $3$  & $3$  \\
    Box size $N\Delta x$    & $8.7$~pm & $4.3$~pm & $2.2$~pm & $4.3$~pm \\
    Packet width $\sigma$   & $0.54$~pm $(4)$ & $0.41$~pm $(3)$ & $0.27$~pm $(2)$ & $0.43$~pm $(3.2)$ \\
    Momentum $|p|c$         & $0.88$~MeV $(0.6)$ & $1.14$~MeV $(0.78)$ & $0.73$~MeV $(0.5)$ & $1.78$~MeV $(1.22)$ \\
    Group velocity $|v|/c_0$  & $1.00,\,0.68,\,0.40$ & $0.70$ & $0.79$ & $0.97$ \\
    Mass $mc_0^2$             & $0.51,\,0.88$~MeV & $0$ & $0$ & $0$ \\
    Potential $qV$          & $1.31,\,2.92$~MeV & $2.63$~MeV & $0$ & $0$ \\
    Duration $T\Delta t$    & $16$~zs $(36)$ & $12$~zs $(26)$ & $3.6$~zs $(8)$ & $27$~zs $(60)$ \\
    \bottomrule
  \end{tabular}
\end{table}

\subsection{Group velocities of the test packets}
\label{app:vg}
The group-velocity row of Table~\ref{tab:physunits} rewards a closer look, because the packets
drift at visibly different fractions of $c_0$ for two unrelated reasons: the one-dimensional
spread is set by \emph{mass}, whereas the differences among the massless two- and
three-dimensional movers are a purely \emph{lattice} (dimensional-splitting) effect. It is
worth separating the two.

Consider mass first. The one-dimensional entries $1.00,\,0.68,\,0.40$ share a
single carrier momentum ($k=0.6$, $|p|c_0=0.88$~MeV) but carry three different masses,
$\mtil=0,\,0.35,\,0.6$. At fixed momentum a heavier packet has more energy and therefore moves
more slowly, $v_g=pc_0^2/E$ with $E=\sqrt{(pc_0)^2+(mc_0^2)^2}$: the massless packet rides the light
cone at $v=c$, while the two massive packets fall to $0.68\,c_0$ and $0.40\,c_0$. (These measured
center-of-mass drifts lie below the continuum estimates $0.86$ and $0.71$ because at
$k=0.6$~rad per site the lattice dispersion is already appreciable; the ordering with mass is
unchanged.) The two massless one-dimensional runs, free and barrier, propagate identically, so
the four sub-tests of Figure~\ref{fig:v1d} collapse to the three velocities listed.

The lattice effect explains why the oblique massless movers are slower. In one dimension a massless packet is an exact
right-moving eigenstate of the streaming step and advances one site per step, exactly $c_0$. In
two and three dimensions the step is factored into per-axis sweeps
(Section~\ref{sec:assemble}), and two effects then pull the drift below the light cone. First,
the per-axis Dirac velocity operators $\alpha_x$, $\beta$ and $\alpha_z$ mutually anticommute
and square to the identity, so every spinor obeys
$\langle\alpha_x\rangle^2+\langle\beta\rangle^2+\langle\alpha_z\rangle^2\le1$: the mean speed of
a Dirac state is bounded by $c_0$, and an oblique mover must \emph{divide} that single unit of
speed among the axes, leaving no component at $c_0$. Second, because the sweeps do not commute, an
obliquely launched packet is not a single rigid eigenmode of the composite step; it disperses
slightly, and its center of mass advances by less than one site per sweep. Together these give
$|v|=0.70$ in two dimensions and $0.79$ in three. A direct fingerprint is the two-dimensional
mover, which reuses the axis-aligned one-dimensional spinor
$u=\Rmat_x\,(0,0,1,1)^{\!\top}/\sqrt2$ but is launched with $k=(0.6,0.5)$: although $k_x\approx
k_y$, its measured drift $v\approx(0.62,0.33)$ is strongly biased toward $x$, precisely because
the spinor is aimed along $x$ and is only an approximate eigenstate of the two-dimensional step.

The reflecting-box packet is faster than the other oblique movers: it reaches $0.97\,c$, close to the
light cone, for two reinforcing reasons. Its carrier momentum $k\approx(1.00,0.68,0.15)$ is
strongly anisotropic and dominated by a single axis, so it behaves almost like a fast
axis-aligned mover that keeps most of its speed budget on one sweep, rather than sharing it
evenly as the body-diagonal three-dimensional mover ($k=(0.5,0.5,0.5)$, hence the smaller
$0.79$) must. In addition, its spinor is constructed as the exact eigenvector of the discrete
one-step operator $M(k)=M_z(k_z)\,M_y(k_y)\,M_x(k_x)$, the Fourier symbol of one full step,
taking the branch with all $\langle\alpha_a\rangle>0$; being a genuine Bloch eigenmode of the
scheme, it propagates coherently with little dispersion and stays near $c_0$, whereas the two- and
three-dimensional movers use continuum or one-dimensional spinors that only approximate the
lattice eigenstate. In short, obliquity that is shared evenly across axes and a spinor that is
not an exact lattice eigenmode both slow a massless packet, while a near-axis carrier ridden by
the exact eigenmode keeps it close to light speed.

\section{Zitterbewegung and comparison with a trapped-ion simulation}
\label{app:zb}

Gerritsma and co-workers realized the one-dimensional Dirac dynamics on a single trapped ion
and observed \emph{Zitterbewegung}, the trembling motion of a relativistic wave packet, together
with the crossover from relativistic to non-relativistic behavior \cite{gerritsma2010}. Theirs
is an \emph{analog} simulation: the ion's own Hamiltonian is tuned to equal the Dirac
Hamiltonian, with position encoded in a continuous motional quadrature. The construction of this
paper reproduces the same physics \emph{digitally}, and because the ported circuit equals the
classical solver exactly (Section~\ref{sec:verify}), we run the experiment both ways and confirm
that they coincide.

\subsection{The 1+1D Dirac limit of the scheme}
This subsection derives, step by step, the $4\times4$ matrix $U(k)$ that governs the scheme's
dispersion. The method is a Fourier (von Neumann) analysis \cite{mulloth_high_2015} of a single sub-step, of the kind used
to study the stability and dispersion of a classical lattice Boltzmann model; it uses nothing
beyond the sub-step of Section~\ref{sec:scheme}.

The state of the scheme is a four-component complex spinor
$\psi(x)=\big(\psi_0(x),\psi_1(x),\psi_2(x),\psi_3(x)\big)$ carried at each integer lattice site
$x=0,1,\dots,N-1$, so the field is an array of $N\times4$ complex numbers (the four spinor
components play the role that the discrete populations $f_i$ play in a lattice Boltzmann model).
These four components are a Dirac bispinor: with the component index written $c=q_0+2q_1$
(Eq.~\eqref{eq:layout}), the low bit $q_0\in\{0,1\}$ is the spin label and the high bit
$q_1\in\{0,1\}$ the upper or lower half of the bispinor, so $\psi_0,\psi_1$ are the two spin states of
the upper half and $\psi_2,\psi_3$ those of the lower; in the rotated frame of Eq.~\eqref{eq:signs}
the same bit $q_1$ is the streaming direction, components with $q_1=0$ moving one way and those with
$q_1=1$ the other. The collision mixes the components within the pairs $(\psi_0,\psi_3)$ and
$(\psi_1,\psi_2)$ (Eq.~\eqref{eq:collision-comp}); at nonzero mass this mixing couples the
positive- and negative-energy branches found below, whose interference is the Zitterbewegung.
One sub-step along $x$ maps $\psi$ to a new field $\psi'$ through four operations, applied in
order: (i)~\emph{rotate}, multiply the four-vector at every site by the fixed $4\times4$ matrix
$\Rmat_x^{-1}$; (ii)~\emph{collide}, multiply the four-vector at every site by the fixed
$4\times4$ matrix $\Qhat$; (iii)~\emph{stream}, displace spinor component $c$ by $s_c\in\{+1,-1\}$
sites, so that after streaming the value of component $c$ at site $x$ equals its pre-stream value
at the neighboring site $x-s_c$; and (iv)~\emph{rotate back}, multiply the four-vector at every
site by $\Rmat_x$. Operations (i), (ii) and (iv) are \emph{local}: each acts on the four components
at a single site, with the same matrix at every site, and none couples different sites. Only
operation (iii), streaming, couples different sites.

Rather than track all $N\times4$ amplitudes, we apply the
sub-step to one particularly simple field, a \emph{lattice plane wave}. By this we mean the field
consisting of the same fixed four-component vector $\hat u$ at every site, multiplied on site $x$
by a phase factor that advances steadily with $x$,
\begin{equation}
  \psi(x)=\hat u\;e^{ikx},\qquad x=0,1,\dots,N-1 .
  \label{eq:zb-planewave}
\end{equation}
Here $\hat u$ is a constant four-component complex vector, identical at all sites, and $k$, the
\emph{wavenumber}, is a real number measured in radians per site. The factor $e^{ikx}$ is a
complex number of unit modulus whose argument advances by $k$ from one site to the next: it equals
$1$ at $x=0$, $e^{ik}$ at $x=1$, $e^{2ik}$ at $x=2$, and so on around the periodic ring of $N$
sites. Equation~\eqref{eq:zb-planewave} is the continuum plane wave $e^{ikx}$ sampled at the
integer sites, and it is the same trial field $f(x)=\hat f\,e^{ikx}$ that one substitutes into a
lattice Boltzmann update in a von Neumann analysis. It is the natural test field for two reasons.
First, any field can be written as a sum of such plane waves, one for each allowed $k$ (its
discrete Fourier transform), so it suffices to follow a single plane wave. Second, as verified
below by direct substitution, the sub-step maps a plane wave of wavenumber $k$ to another plane
wave of the same $k$, altering only the constant vector $\hat u$; the $N\times4$ update therefore
reduces, for each $k$ independently, to a single $4\times4$ matrix acting on $\hat u$.

To see how each operation acts on the plane wave, substitute Eq.~\eqref{eq:zb-planewave} into the
sub-step and follow the amplitude $\hat u$ through the four operations.
\begin{itemize}\itemsep2pt
  \item \emph{Rotation and collision.} The three matrices $\Rmat_x^{-1}$, $\Qhat$ and $\Rmat_x$
    are identical at every site and act only on the four components at a given site, without
    reference to neighboring sites, so they leave the phase factor $e^{ikx}$ unchanged. Applying
    $\Rmat_x^{-1}$, for instance, gives
    $\Rmat_x^{-1}\big(\hat u\,e^{ikx}\big)=\big(\Rmat_x^{-1}\hat u\big)\,e^{ikx}$, again a plane
    wave of the same $k$, whose amplitude vector is now $\Rmat_x^{-1}\hat u$. Each local operation
    therefore multiplies the amplitude vector by its $4\times4$ matrix and leaves the phase
    unchanged.
  \item \emph{Streaming.} This is the only operation that involves the wavenumber. Streaming sets
    the new value of component $c$ at site $x$ equal to its old value at site $x-s_c$; on the plane
    wave that old value is $\hat u_c\,e^{ik(x-s_c)}$, and factoring out the common $e^{ikx}$,
    \begin{equation}
      \hat u_c\,e^{ik(x-s_c)}=\big(e^{-iks_c}\big)\,\hat u_c\,e^{ikx}.
      \label{eq:zb-shift}
    \end{equation}
    Streaming thus multiplies component $c$ of the amplitude by the constant phase $e^{-iks_c}$ and
    again leaves $e^{ikx}$ unchanged. Collecting the four components, streaming multiplies the
    amplitude vector by the diagonal $4\times4$ matrix
    $S(k)=\operatorname{diag}\big(e^{-iks_0},e^{-iks_1},e^{-iks_2},e^{-iks_3}\big)$. This is the
    central point of the analysis: a displacement in real space (streaming) becomes multiplication
    by a phase in $k$-space, as in the von Neumann analysis of any stream-and-collide scheme.
\end{itemize}

Applying the four operations to $\hat u$ in order gives the chain
\[
  \hat u\;\xrightarrow{\ \Rmat_x^{-1}\ }\;\Rmat_x^{-1}\hat u
  \;\xrightarrow{\ \Qhat\ }\;\Qhat\,\Rmat_x^{-1}\hat u
  \;\xrightarrow{\ S(k)\ }\;S(k)\,\Qhat\,\Rmat_x^{-1}\hat u
  \;\xrightarrow{\ \Rmat_x\ }\;U(k)\,\hat u ,
\]
so one sub-step multiplies the amplitude vector by the single $4\times4$ matrix
\begin{equation}
  U(k)=\Rmat_x\,S(k)\,\Qhat\,\Rmat_x^{-1}
      =\Rmat_x\,\operatorname{diag}\big(e^{-ik s_c}\big)\,\Qhat\,\Rmat_x^{-1}.
  \label{eq:zb-symbol}
\end{equation}
The matrix $U(k)$ is the representation of the sub-step on a single Fourier mode, that is, its
\emph{Fourier symbol}; the full field evolves as the superposition of these independent $4\times4$
evolutions, one for each $k$.

The evolution is governed by the eigenvalues of $U(k)$. An \emph{eigenvalue} of the $4\times4$ matrix $U(k)$ is a
number $\lambda$ for which some nonzero vector $v$ (an \emph{eigenvector}) satisfies
$U(k)\,v=\lambda v$: under one sub-step that particular amplitude is simply rescaled by $\lambda$.
$U(k)$ has four eigenvalues, and each has modulus one. This is because $U(k)$ is a product of the
unitary factors $\Rmat_x,\Rmat_x^{-1},\Qhat$ and the streaming phases $e^{-iks_c}$ (a
\emph{unitary} matrix is one that preserves the length $\|w\|=\sqrt{\sum_i|w_i|^2}$ of every
vector), so from $U(k)\,v=\lambda v$ and $\|U(k)\,v\|=\|v\|$ one gets $|\lambda|\,\|v\|=\|v\|$,
hence $|\lambda|=1$. A complex number of modulus one can always be written $\lambda=e^{i\theta}$
with $\theta$ a real angle; this angle is called the \emph{eigenphase}. All of the dispersion
information is carried by the four eigenphases, computed next.

The eigenvalues are easiest to find after reducing $U(k)$ to a $2\times2$ problem. Conjugating a matrix by an invertible matrix does
not change its eigenvalues, because the characteristic polynomial of $U(k)$
(Eq.~\eqref{eq:zb-symbol}) is unchanged:
\[
  \det\!\big(U(k)-\lambda\mathbb I\big)
   =\det\!\big(\Rmat_x(S(k)\Qhat-\lambda\mathbb I)\Rmat_x^{-1}\big)
   =\det(\Rmat_x)\,\det\!\big(S(k)\Qhat-\lambda\mathbb I\big)\,\det(\Rmat_x^{-1})
   =\det\!\big(S(k)\Qhat-\lambda\mathbb I\big),
\]
using $\det(\Rmat_x)\det(\Rmat_x^{-1})=1$. So $U(k)$ has the same four eigenvalues as
$M(k)\equiv S(k)\,\Qhat$, and we may discard $\Rmat_x$. Two ingredients are now written out
explicitly. The streaming signs of the $x$-sweep are $s=(-1,-1,+1,+1)$ (Eq.~\eqref{eq:signs}), so
the streaming factor is $S(k)=\operatorname{diag}(e^{ik},e^{ik},e^{-ik},e^{-ik})$; and the
collision (Eq.~\eqref{eq:collision-comp}) mixes the four components only within the two pairs
$(0,3)$ and $(1,2)$,
\[
  \Qhat=\begin{pmatrix} \ahat&0&0&-\bhat\\[1pt] 0&\ahat&\bhat&0\\[1pt] 0&-\bhat&\ahat&0\\[1pt]
    \bhat&0&0&\ahat\end{pmatrix},
\]
where for the free particle ($\gtil=0$) the amplitudes $\ahat,\bhat$ are real and satisfy
$\ahat^2+\bhat^2=1$ (Eq.~\eqref{eq:collision}). Multiplying $M(k)=S(k)\Qhat$ (each row of $\Qhat$
is scaled by the corresponding diagonal entry of $S(k)$),
\[
  M(k)=\begin{pmatrix}
   \ahat e^{ik}&0&0&-\bhat e^{ik}\\[1pt]
   0&\ahat e^{ik}&\bhat e^{ik}&0\\[1pt]
   0&-\bhat e^{-ik}&\ahat e^{-ik}&0\\[1pt]
   \bhat e^{-ik}&0&0&\ahat e^{-ik}\end{pmatrix},
\]
which is block-diagonal: the pair $(0,3)$ does not mix with the pair $(1,2)$. It therefore
suffices to diagonalize one $2\times2$ block, say the one on components $(0,3)$,
\begin{equation}
  B(k)=\begin{pmatrix} \ahat e^{ik} & -\bhat e^{ik}\\[3pt] \bhat e^{-ik} & \ahat e^{-ik}\end{pmatrix}.
  \label{eq:zb-block}
\end{equation}

The two eigenvalues of a $2\times2$ matrix $B$ solve the
quadratic $\lambda^2-(\operatorname{tr}B)\,\lambda+\det B=0$, with $\operatorname{tr}B$ the sum of
the diagonal entries and $\det B$ the determinant. For Eq.~\eqref{eq:zb-block},
\[
  \operatorname{tr}B=\ahat e^{ik}+\ahat e^{-ik}=2\ahat\cos k,\qquad
  \det B=(\ahat e^{ik})(\ahat e^{-ik})-(-\bhat e^{ik})(\bhat e^{-ik})=\ahat^2+\bhat^2=1,
\]
using $e^{ik}e^{-ik}=1$, $\cos k=\tfrac12(e^{ik}+e^{-ik})$, and $\ahat^2+\bhat^2=1$. The quadratic
is thus $\lambda^2-2\ahat\cos k\,\lambda+1=0$, whose solutions are
\[
  \lambda=\ahat\cos k\pm\sqrt{(\ahat\cos k)^2-1}
         =\ahat\cos k\pm i\sqrt{1-(\ahat\cos k)^2},
\]
the square root being imaginary since $|\ahat\cos k|\le1$.

Each solution has modulus
$\sqrt{(\ahat\cos k)^2+\big(1-(\ahat\cos k)^2\big)}=1$, and the two are complex conjugates of one
another. Writing them as $e^{\mp iE(k)}$ identifies
\begin{equation}
  \cos E(k)=\ahat\cos k,\qquad \sin E(k)=\sqrt{1-(\ahat\cos k)^2}\ge0 .
  \label{eq:zb-eig}
\end{equation}
This is why the eigenvalues take the form $e^{-iE(k)}$ and $e^{+iE(k)}$: the determinant of the
block is $1$, which forces the product of its two eigenvalues to be $1$, and their modulus is $1$,
which forces them to be complex conjugates; two numbers on the unit circle that are conjugate and
multiply to one are exactly $e^{+iE}$ and $e^{-iE}$ for a single real angle $E(k)\ge0$. (The other
block, on $(1,2)$, has the same $\operatorname{tr}$ and $\det$ and hence the same pair, so the four
eigenphases of $U(k)$ are $\pm E(k)$, each occurring twice.)

\runin{The dispersion relation.} The nonnegative angle $E(k)$, the size of the phase each branch
turns through in one sub-step (the two carrying $e^{\mp iE}$), is the \emph{dispersion relation}
(energy versus wavenumber). Substituting the
free-particle amplitude $\ahat=(1-\mtil^2/4)/(1+\mtil^2/4)$
(Eqs.~\eqref{eq:couplings}--\eqref{eq:collision} with $\gtil=0$) into Eq.~\eqref{eq:zb-eig} gives
it in closed form,
\begin{equation}
  \cos E(k)=\frac{1-\mtil^2/4}{1+\mtil^2/4}\,\cos k .
  \label{eq:zb-disp}
\end{equation}
This is \emph{exact}; it is the relation used for the $E$ (and $2E$) columns of
Table~\ref{tab:zb} and the solid curve of Figure~\ref{fig:zb}. For a well-resolved packet $k$ and
$\mtil$ are both small, and expanding each factor to second order,
$\cos E=1-\tfrac12E^2+\cdots$, $\cos k=1-\tfrac12k^2+\cdots$ and
$\ahat=(1-\tfrac14\mtil^2)(1+\tfrac14\mtil^2)^{-1}=1-\tfrac12\mtil^2+\cdots$, turns
Eq.~\eqref{eq:zb-disp} into $1-\tfrac12E^2\simeq1-\tfrac12k^2-\tfrac12\mtil^2$, that is
\begin{equation}
  E(k)\simeq\sqrt{k^2+\mtil^{\,2}} ,
  \label{eq:zb-cont}
\end{equation}
the continuum Dirac relation $E=\sqrt{(pc)^2+(mc^2)^2}$ with momentum $pc\leftrightarrow k$
and mass $mc^2\leftrightarrow\mtil$ (lattice units
$\hbar=c=\Delta x=\Delta t=1$). Table~\ref{tab:zb} lists both: the exact Eq.~\eqref{eq:zb-disp}
tracks the measured trembling, while the approximation Eq.~\eqref{eq:zb-cont} departs from it once
$\mtil$ is no longer small.

\runin{Zitterbewegung: a beat between the two branches.} Let $u_+$ and $u_-$ be the normalized
eigenvectors of the block with eigenvalues $e^{-iE}$ and $e^{+iE}$. The subscript records the sign
of the energy: one sub-step is the discrete time-evolution operator, and a state of energy
$\varepsilon$ turns by the phase $e^{-i\varepsilon}$ each step (the lattice analogue of the continuum
$e^{-i\varepsilon t}$, with $\hbar=\Delta t=1$), so the positive-energy branch $u_+$ (energy $+E$)
carries the eigenvalue $e^{-iE}$ and the negative-energy branch $u_-$ (energy $-E$) carries
$e^{+iE}$. Being eigenvectors of the unitary $U(k)$ for distinct eigenvalues, they are orthogonal,
$u_+^\dagger u_-=0$. Applying $U(k)$
(equivalently the block) $T$ times multiplies each by its eigenvalue $T$ times,
$u_\pm\mapsto e^{\mp iET}u_\pm$. A packet built on a single branch, $\psi=u_+$, is then only
multiplied by the overall phase $e^{-iET}$ and moves rigidly. A packet built on \emph{both}, here
the equal superposition $\psi=\tfrac{1}{\sqrt2}(u_++u_-)$ used in the runs below, evolves after $T$
steps to $\psi(T)=\tfrac{1}{\sqrt2}\big(e^{-iET}u_++e^{+iET}u_-\big)$. The mean of a physical
observable, represented by a Hermitian matrix $O$ (one equal to its own conjugate transpose,
$O^\dagger=O$, as every measurable quantity is), is
$\langle O\rangle(T)=\psi(T)^\dagger O\,\psi(T)$, where the conjugate transpose
$\psi(T)^\dagger=\tfrac{1}{\sqrt2}\big(e^{+iET}u_+^\dagger+e^{-iET}u_-^\dagger\big)$ turns the column
into a row and conjugates its phases. Multiplying the row, the matrix $O$, and the column expands
into four terms, one for each pairing of a branch from $\psi(T)^\dagger$ with a branch from
$\psi(T)$,
\begin{align*}
  \langle O\rangle(T)
  &=\tfrac12\big(e^{+iET}u_+^\dagger+e^{-iET}u_-^\dagger\big)\,O\,
     \big(e^{-iET}u_++e^{+iET}u_-\big)\\
  &=\tfrac12\big(u_+^\dagger O\,u_++u_-^\dagger O\,u_-
     +e^{+2iET}\,u_+^\dagger O\,u_-+e^{-2iET}\,u_-^\dagger O\,u_+\big),
\end{align*}
the phase of each term being the product of the phases of the two branches that meet in it. In the
first two (``diagonal'') terms the branch phases are opposite and cancel, $e^{+iET}e^{-iET}=1$,
leaving a constant; in the last two (``cross'') terms they are equal and \emph{add}, to
$e^{\pm2iET}$. Because $O$ is Hermitian the two cross terms are complex conjugates of one another,
$u_-^\dagger O\,u_+=\big(u_+^\dagger O\,u_-\big)^{*}$ (the star denotes the complex conjugate); so,
writing $z=e^{+2iET}u_+^\dagger O\,u_-$ for the first of them, their sum is
$z+z^{*}=2\operatorname{Re}z$, in which $\operatorname{Re}$ denotes the \emph{real part} of a
complex number ($\operatorname{Re}(a+ib)=a$ for real $a,b$, so $w+w^{*}=2\operatorname{Re}\,w$ for
every $w$). The factor $2$ cancels the $\tfrac12$, and the four terms collapse to
\begin{equation}
  \langle O\rangle(T)
  =\underbrace{\tfrac12\big(u_+^\dagger O\,u_++u_-^\dagger O\,u_-\big)}_{\text{steady}}
   +\underbrace{\operatorname{Re}\!\big[e^{+2iET}\,u_+^\dagger O\,u_-\big]}_{\text{oscillating}} .
  \label{eq:zb-split}
\end{equation}
The steady part does not depend on $T$; the oscillating part is a cosine, since
$\operatorname{Re}\big[e^{+2iET}z_0\big]=|z_0|\cos(2ET+\arg z_0)$ for any constant complex $z_0$.
Every observable therefore trembles at the same angular frequency $2E$: this beat between the two
branches is the Zitterbewegung, of frequency
\begin{equation}
  \omega_{\mathrm{ZB}}=2E(k_0).
  \label{eq:zb-freq}
\end{equation}

\runin{The trembling amplitude in closed form.} The physical observable is the velocity, which in
the scheme is the matrix $\alpha_x$; its mean $\langle\alpha_x\rangle$ is the quantity whose
spectral peak defines $\omega_{\mathrm{ZB}}$ and whose swing is plotted in Figure~\ref{fig:zb}. By
the identity~\eqref{eq:zb-split} with $O=\alpha_x$, the velocity oscillates with amplitude
$A_v=|u_+^\dagger\alpha_x\,u_-|$ (for normalized $u_\pm$), the modulus of the cross matrix element,
which we now evaluate exactly. It is simplest in the characteristic frame, where $\alpha_x$ is
diagonal, $\Rmat_x^{-1}\alpha_x\Rmat_x=\operatorname{diag}(-1,-1,+1,+1)$ (Eq.~\eqref{eq:Rx}), so on
the $(0,3)$ block it is $\Lambda=\operatorname{diag}(-1,+1)$. In this frame the block eigenvectors
follow from $B(k)$ (Eq.~\eqref{eq:zb-block}) by solving $B(k)\,u=e^{\mp iE}u$. The first row of
$\big(B(k)-e^{\mp iE}\mathbb I\big)u=0$ reads $(\ahat e^{ik}-e^{\mp iE})\,u_0=\bhat e^{ik}\,u_1$;
taking $u_0=\bhat e^{ik}$ to clear the factor makes $u_1=\ahat e^{ik}-e^{\mp iE}$, and inserting
$e^{\mp iE}=\cos E\mp i\sin E=\ahat\cos k\mp i\sin E$ (Eq.~\eqref{eq:zb-eig}) together with
$\ahat e^{ik}=\ahat\cos k+i\,\ahat\sin k$ gives
\[
  u_1=\ahat e^{ik}-\big(\ahat\cos k\mp i\sin E\big)=i\,\ahat\sin k\pm i\sin E=i(\ahat\sin k\pm\sin E).
\]
Hence, before normalization,
\begin{equation}
  B(k)\,u_\pm=e^{\mp iE}u_\pm,\qquad
  u_\pm=\big(\,\bhat\,e^{ik},\ \ i(\ahat\sin k\pm\sin E)\,\big) .
  \label{eq:zb-evec}
\end{equation}
Their overlap through $\Lambda$ is, using $(i\rho)^{*}(i\tau)=\rho\tau$ for real $\rho,\tau$ and
then $\sin^2 E=1-\ahat^2\cos^2 k$ together with $\ahat^2+\bhat^2=1$,
\[
  u_+^\dagger\Lambda\,u_-
   =-|\bhat|^2+(\ahat\sin k+\sin E)(\ahat\sin k-\sin E)
   =-\bhat^2+\ahat^2\sin^2 k-\sin^2 E=-2\bhat^2 ,
\]
while the same two identities give the norms $\|u_+\|\,\|u_-\|=2\bhat\sin E$. The velocity
amplitude is their ratio,
\begin{equation}
  A_v=\frac{|u_+^\dagger\alpha_x\,u_-|}{\|u_+\|\,\|u_-\|}=\frac{\bhat}{\sin E},
  \label{eq:zb-Av}
\end{equation}
and, since a sinusoid of angular frequency $\omega_{\mathrm{ZB}}=2E$ has position amplitude equal
to its velocity amplitude divided by that frequency, the packet's position swings by
\begin{equation}
  R_{\mathrm{ZB}}=\frac{A_v}{\omega_{\mathrm{ZB}}}=\frac{\bhat}{2E\,\sin E} .
  \label{eq:zb-amp}
\end{equation}
Equations~\eqref{eq:zb-Av}--\eqref{eq:zb-amp} are \emph{exact}. The mass enters only through
$\bhat=\mtil/(1+\mtil^2/4)$: at $\mtil=0$ it vanishes, so $\Qhat=\mathbb I$, $M(k)=S(k)$ is already
diagonal, the $\pm$ branches are decoupled basis states that never interfere, and there is no
trembling. For a well-resolved packet ($\mtil\ll1$) one has $\bhat\to\mtil$ and $\sin E\to E$, so
$A_v\to\mtil/E$ and $R_{\mathrm{ZB}}\to\mtil/2E^2$, the leading-order forms; the exact
$R_{\mathrm{ZB}}$ is what the measured trembling follows, and it departs from $\mtil/2E^2$ once
$\mtil$ leaves the Compton-resolved regime (at $\mtil=2$, $R_{\mathrm{ZB}}=0.32$ versus
$\mtil/2E^2=0.41$). As a function of mass, $R_{\mathrm{ZB}}$ grows from zero at $\mtil=0$, reaches a
maximum near $\mtil\approx0.7$, and decreases again as $E$ grows with the mass. This maximum is
essentially the relativistic-to-non-relativistic crossover: the leading-order amplitude
$R_{\mathrm{ZB}}\simeq\mtil/2E^2\simeq\mtil/\big[2(k^2+\mtil^2)\big]$ has zero derivative at
$\mtil=k$, exactly where the rest energy equals the momentum energy ($mc^2=pc$); the exact
$R_{\mathrm{ZB}}$ only nudges the peak slightly upward, to $\mtil\approx0.7$ at the carrier $k_0=0.6$
of the runs below, i.e.\ $mc^2/pc\approx1.2$, just above unity.

\subsection{Correspondence with the trapped-ion parameters}
The trapped-ion experiment and the present scheme realize the \emph{same} 1+1D Dirac operator, in
different units and on different hardware. This subsection makes the correspondence explicit and 
summarizes it in Table~\ref{tab:zb-map}.

\runin{The trapped-ion analog simulator.} The apparatus of Gerritsma et al.\ \cite{gerritsma2010} is
a single ion held in an electromagnetic (Paul) trap. Two of the ion's internal electronic levels
serve as the two Dirac spinor components, while the ion's motion in the trap, a quantum harmonic
oscillator, stores the particle's position and momentum in a continuous motional coordinate (a
\emph{quadrature}, the oscillator's counterpart of position). Laser beams couple the internal levels
to the motion so that the ion's own Hamiltonian is tuned to \emph{equal} the one-dimensional Dirac
Hamiltonian
\begin{equation}
  H_{\mathrm{ion}}=c\,p\,\sigma_x+mc^2\,\sigma_z ,
  \label{eq:zb-Hion}
\end{equation}
where $p$ is the motional momentum, $\sigma_x,\sigma_z$ are the Pauli matrices \eqref{eq:pauli} acting
on the two internal levels, and the coefficients are fixed by the laser rather than by nature: the
effective speed of light is $c=2\eta\tilde\Omega\Delta$, with $\eta$ the Lamb--Dicke parameter (a
dimensionless measure of how strongly the light couples to the ion's motion), $\tilde\Omega$ a Rabi
frequency (the rate at which the beam drives the internal transition), and $\Delta$ the spatial width
of the motional ground state; the effective rest energy $mc^2=\hbar\Omega$ is set by a separate
carrier coupling of Rabi frequency $\Omega$. The simulation is \emph{analog}: the apparatus applies
no gates but lets the ion evolve continuously under \eqref{eq:zb-Hion}, reading out the mean position
$\langle x\rangle$ against time, its trembling being the Zitterbewegung. Because $c$ and $mc^2$ are
laboratory settings, both can be dialed freely, which is precisely what lets a slow tabletop system
imitate the fast relativistic regime.

The present scheme reaches that operator digitally and directly,
because each factor of a sub-step is already an exact rotation generated by a Dirac matrix. Take the
one-step symbol $U(k)=\Rmat_x\,S(k)\,\Qhat\,\Rmat_x^{-1}$ of Eq.~\eqref{eq:zb-symbol} and read its
two nontrivial factors in turn.

\emph{The collision is a rotation about $\alpha_y$.} Since $\alpha_y^2=\mathbb I$ and the
free-particle amplitudes satisfy $\ahat^2+\bhat^2=1$ (Eq.~\eqref{eq:collision}), writing
$\ahat=\cos\theta$ and $\bhat=\sin\theta$ matches $\Qhat$ to the matrix Euler identity
\begin{equation}
  \Qhat=\ahat\,\mathbb I-i\bhat\,\alpha_y=\cos\theta\,\mathbb I-i\sin\theta\,\alpha_y=e^{-i\theta\alpha_y},\qquad
  \theta=\arcsin\bhat=\arcsin\frac{\mtil}{1+\mtil^2/4}\simeq\mtil .
  \label{eq:zb-Qrot}
\end{equation}
The exponential collapses to these two terms because $\alpha_y^2=\mathbb I$: in the series
$e^{-i\theta\alpha_y}=\sum_n(-i\theta)^n\alpha_y^{\,n}/n!$ the even powers give $\alpha_y^{2m}=\mathbb I$ and
sum to $\cos\theta\,\mathbb I$, while the odd powers give $\alpha_y^{2m+1}=\alpha_y$ and sum to
$-i\sin\theta\,\alpha_y$ (the matrix form of $e^{-i\theta}=\cos\theta-i\sin\theta$). It is a rotation by
the small angle $\theta\simeq\mtil$, the only approximation made below.

\emph{The streaming is a rotation about $\alpha_x$.} In the characteristic frame $S(k)$ multiplies
component $c$ by the phase $e^{-iks_c}$ with signs $s=(-1,-1,+1,+1)$
(Eqs.~\eqref{eq:signs},~\eqref{eq:zb-shift}), and those signs are exactly the eigenvalues of
$\alpha_x$ in that frame, $\Rmat_x^{-1}\alpha_x\Rmat_x=\operatorname{diag}(-1,-1,+1,+1)$
(Eq.~\eqref{eq:Rx}), so
\[
  S(k)=\operatorname{diag}\big(e^{-iks_c}\big)=\exp\!\big(-ik\,\Rmat_x^{-1}\alpha_x\Rmat_x\big),
\]
an exact rotation by the angle $k$.

\emph{One sub-step is their product.} Insert $\Rmat_x^{-1}\Rmat_x=\mathbb I$ between the streaming
and collision factors, so that each is conjugated by $\Rmat_x$ on its own,
\[
  U(k)=\Rmat_x\,S(k)\,\big(\Rmat_x^{-1}\Rmat_x\big)\,\Qhat\,\Rmat_x^{-1}
   =\big(\Rmat_x\,S(k)\,\Rmat_x^{-1}\big)\big(\Rmat_x\,\Qhat\,\Rmat_x^{-1}\big).
\]
Conjugation passes through a matrix exponential, $\Rmat\,e^{M}\Rmat^{-1}=e^{\Rmat M\Rmat^{-1}}$, so the
first factor undoes the frame rotation carried by the streaming generator (the inner
$\Rmat_x\Rmat_x^{-1}$ pairs cancel around $\alpha_x$),
\[
  \Rmat_x\,S(k)\,\Rmat_x^{-1}=\Rmat_x\,e^{-ik\,\Rmat_x^{-1}\alpha_x\Rmat_x}\,\Rmat_x^{-1}=e^{-ik\alpha_x},
\]
while the second leaves the collision untouched, $\Rmat_x\,\Qhat\,\Rmat_x^{-1}=\Qhat=e^{-i\theta\alpha_y}$,
because $\Qhat$ is built from $\mathbb I$ and $\alpha_y$ and the rotation preserves it,
$\Rmat_x\alpha_y\Rmat_x^{-1}=\alpha_y$ (Eq.~\eqref{eq:Rx}). Their product is the exact factorization
\begin{equation}
  U(k)=e^{-ik\alpha_x}\,e^{-i\theta\alpha_y}.
  \label{eq:zb-factor}
\end{equation}
Combining the two exponentials into one is the Baker--Campbell--Hausdorff expansion: for matrices
$A,B$ that need not commute,
\[
  e^{A}e^{B}=\exp\!\Big(A+B+\tfrac12[A,B]+\cdots\Big),\qquad [A,B]\equiv AB-BA ,
\]
the omitted terms being higher nested commutators. With $A=-ik\alpha_x$ and $B=-i\theta\alpha_y$ the
leading exponent $A+B=-i(k\alpha_x+\theta\alpha_y)$ is simply the sum of the two generators, and the
first correction is a single commutator. Since $\alpha_x$ and $\alpha_y$ anticommute
($\alpha_x\alpha_y=-\alpha_y\alpha_x$, hence $[\alpha_x,\alpha_y]=2\alpha_x\alpha_y$),
\[
  \tfrac12[A,B]=\tfrac12(-ik)(-i\theta)\,[\alpha_x,\alpha_y]=-k\theta\,\alpha_x\alpha_y=O(k\theta),
\]
a second-order term in the small angles $k,\theta$ that is dropped together with the higher
commutators (each smaller still); this neglect is precisely the operator (Strang) splitting error of
applying streaming and collision in sequence rather than together, familiar from split
lattice-kinetic and advection-reaction solvers. To leading order the two rotations therefore combine
into one by their summed generators, $U(k)\simeq e^{-iH_{\mathrm{QLB}}}$ with
\begin{equation}
  H_{\mathrm{QLB}}=k\,\alpha_x+\mtil\,\alpha_y .
  \label{eq:zb-Hqlb}
\end{equation}
This is the 1+1D Dirac Hamiltonian outright: a momentum term $k\,\alpha_x$ and a mass term
$\mtil\,\alpha_y$, built from the two anticommuting Dirac matrices. The momentum term is \emph{exact}:
at zero mass $\theta=0$, so $U(k)=e^{-ik\alpha_x}$ has eigenphases $\pm k$ and the massless dispersion
is $E=k$, a single linear light cone at unit speed for every $k$ (with no lattice fermion doubling);
only the mass term is approximate, through $\theta\simeq\mtil$.

Equations~\eqref{eq:zb-Hion} and \eqref{eq:zb-Hqlb} are the \emph{same} Hamiltonian: each is a
momentum term along one matrix and a mass term along a second that anticommutes with it (the pair
$\sigma_x,\sigma_z$ for the ion; $\alpha_x,\alpha_y$ here), a choice of basis with no physical effect.
Both therefore carry the relativistic spectrum $\pm\sqrt{(\text{momentum})^2+(\text{mass})^2}$,
matched term by term by the dictionary
\begin{equation}
  pc\;\longleftrightarrow\;k,\qquad mc^2\;\longleftrightarrow\;\mtil .
  \label{eq:zb-dict}
\end{equation}
Gerritsma et al.\ encode the momentum in the motional quadrature and the mass in a laser detuning;
here the momentum is the lattice wavenumber $k$ (radians per site) and the mass is the collision
coefficient $\mtil$.

Every feature of the trembling is a function of the
spectrum alone, and hence only of the single dimensionless ratio $mc^2/pc$. The trembling frequency is
$\omega_{\mathrm{ZB}}=2E/\hbar$ with $E=\sqrt{(pc)^2+(mc^2)^2}$, and the
relativistic-to-non-relativistic crossover sits where that ratio is of order unity. Unity is the
natural scale because $E$ is built from a momentum energy $pc$ and a rest energy $mc^2$: the packet is
ultrarelativistic when the momentum energy dominates ($mc^2\ll pc$, $E\simeq pc$) and
non-relativistic when the rest energy dominates ($mc^2\gg pc$, $E\simeq mc^2+p^2/2m$). The group
velocity makes this quantitative. It depends only on the ratio,
$v/c=pc/E=1/\sqrt{1+(mc^2/pc)^2}$,\footnote{Both forms follow
from $E=\sqrt{(pc)^2+(mc^2)^2}$: factoring out the rest energy,
$E=mc^2\sqrt{1+(pc/mc^2)^2}\simeq mc^2+p^2/2m$ when $pc\ll mc^2$; and the group velocity of a wave
packet is $v=d\omega/dk=dE/dp$ (using $E=\hbar\omega$ and $p=\hbar k$, so the $\hbar$'s cancel), which
for this $E$ differentiates to $dE/dp=pc^2/E$, i.e.\ $v/c=pc/E$.} and as $mc^2/pc$ grows from zero to
infinity the speed falls steadily from $c$ (an ultrarelativistic packet moving at the speed of light)
toward $0$ (a heavy, non-relativistic packet moving far below it). It equals $c/\sqrt2$ precisely
when $mc^2=pc$, the point where the momentum and rest energies are equal; this equal-energy point,
$mc^2/pc=1$, is the balance point, and because the speed varies smoothly with the ratio the crossover
is a broad region around unity rather than a sharp threshold. None of this depends on the laboratory
constants $c$, $\hbar$ or the trap frequency separately. Reproducing the
experiment therefore does not mean importing its kilohertz frequencies or micrometer trap size, but
scanning the same range of $mc^2/pc$. By the dictionary \eqref{eq:zb-dict} the two knobs coincide:
the ion sweeps the ratio through its laser detuning, the scheme through the single coefficient
$\mtil$.

\runin{Reproducing the crossover.} We fix the carrier at $k_0=0.6$ (so $pc=k_0=0.6$) and sweep
$\mtil$ from $0$ to $2$, carrying $mc^2$ from well below to well above $pc$ and through the crossover
$mc^2=pc$ at $\mtil=k_0$. This is the same crossover Gerritsma et al.\ traverse by tuning the
laser detuning at fixed momentum. Absolute numbers (frequencies in hertz, lengths in meters) enter
only once a single physical scale $\Delta x$ is fixed, as in Appendix~\ref{app:phys}; the
dimensionless sweep already contains all of the physics.

\begin{table}[htb]
  \centering
  \caption{The same 1+1D Dirac operator on two platforms: $H=c\,p\,\sigma_x+mc^2\sigma_z$ in the
    trapped ion, and $H=k\,\alpha_x+\mtil\,\alpha_y$ here. The trembling frequency and the crossover
    depend only on the dimensionless ratio
    $mc^2/pc$, so the two platforms reproduce the same physics without sharing units.}
  \label{tab:zb-map}
  \begin{tabular}{@{}lll@{}}
    \toprule
    Dirac ingredient   & Trapped ion \cite{gerritsma2010}         & QLB scheme (this work) \\
    \midrule
    spinor             & two internal ion levels                  & $2$ spinor qubits ($4$ components) \\
    momentum $p$       & motional quadrature $\hat p$             & wavenumber $k$; $pc\leftrightarrow k$ \\
    mass $mc^2$        & laser detuning $\hbar\Omega$             & collision coefficient $\mtil$ (units of $E_0$) \\
    light speed $c$    & laser coupling $2\eta\tilde\Omega\Delta$ & $\Delta x/\Delta t=1$ (one site per step) \\
    scanned quantity   & detuning (mass) at fixed $p,c$           & $\mtil$ at fixed $k_0$ \\
    \bottomrule
  \end{tabular}
\end{table}

\subsection{Setup}
\runin{The initial state.} We take a line of $N=2^8=256$ sites ($10$ qubits), carrier $k_0=0.6$ (so
$pc=k_0=0.6$), and a Gaussian envelope of width $\sigma=14$ centered on the lattice. The carried
spinor is the equal superposition $\tfrac{1}{\sqrt2}(u_++u_-)$ of two eigenmodes $u_\pm$ of the
one-step symbol $U(k_0)$ (Eq.~\eqref{eq:zb-symbol}) with opposite energies $\pm E(k_0)$, the
trembling state of Eq.~\eqref{eq:zb-split} whose two branches beat against each other to give the
Zitterbewegung. Picking $u_+$ and $u_-$ takes a little care, because each energy $\pm E$ occurs
\emph{twice}: the collision splits the four spinor components into two independent pairs (blocks),
$(0,3)$ and $(1,2)$ (Eq.~\eqref{eq:collision-comp}), and each block carries its own $+E$ and $-E$
mode. The trembling we measure is that of the velocity $\alpha_x$, whose swing, by
Eq.~\eqref{eq:zb-split}, has amplitude equal to the cross matrix element $|u_+^\dagger\alpha_x u_-|$,
the ``velocity coupling'' that says how strongly $\alpha_x$ links the two energy branches. In the
frame where $\alpha_x$ is diagonal (the characteristic frame of Eq.~\eqref{eq:Rx}, in which the modes
are the block eigenvectors of Eq.~\eqref{eq:zb-evec}) the velocity keeps each block to itself, so a
$+E$ and a $-E$ taken from \emph{different} blocks have $u_+^\dagger\alpha_x u_-=0$ and do not tremble
at all; we therefore take both from the \emph{same} block, where the coupling is nonzero. On that
block the two modes are $u_\pm=\big(\bhat\,e^{ik_0},\ i(\ahat\sin k_0\pm\sin E)\big)$
(Eq.~\eqref{eq:zb-evec}), and in their equal sum the $\pm\sin E$ parts cancel, leaving the carried
four-spinor (components ordered $0,1,2,3$)
\[
  u=\Rmat_x\,\big(\bhat\,e^{ik_0},\ 0,\ 0,\ i\,\ahat\sin k_0\big)^{\!\top},
\]
the simple block vector $(\bhat e^{ik_0},0,0,i\ahat\sin k_0)$ rotated back to the computational basis
by $\Rmat_x$. The initial field is this spinor at every site, times the Gaussian carrier $e^{ik_0x}$
of Eq.~\eqref{eq:initcond} and normalized to $\lVert\psi_0\rVert=1$.

\runin{Evolution.} The state is advanced for $90$ steps in two independent ways: with the exact
classical one-step operator, the $4N\times4N$ matrix
$\Rmat_x\,\mathrm{Stream}_x\,\Qhat\,\Rmat_x^{-1}$ of Eq.~\eqref{eq:substep} applied by dense
matrix-vector multiplication, and with the ported circuit of Eq.~\eqref{eq:step} (a single
$x$-sweep here) run on the qiskit-aer state-vector emulator on a GPU. After every step we record,
from each, the mean position $\langle x(t)\rangle=\sum_x x\,\rho(x)$, where
$\rho(x)=\sum_c|\psi_c(x)|^2$ is the probability density of Section~\ref{sec:verify}; the mean velocity
$\langle\alpha_x(t)\rangle=\psi(t)^\dagger\alpha_x\psi(t)$; and the density profile $\rho$ itself (for
the $\max|\Delta\rho|$ column of Table~\ref{tab:zb}).

\runin{Measuring the trembling frequency.} The frequency $\omega_{\mathrm{ZB}}$ of
Table~\ref{tab:zb} is extracted from the velocity signal $\langle\alpha_x(t)\rangle$, sampled once per
step over the $90$-step run, by a windowed Fourier transform in four stages. First, subtract the
best-fit straight line, removing the steady baseline---the constant term of Eq.~\eqref{eq:zb-split}
together with any slow drift as the two branches separate---so that only the oscillation remains.
Second, multiply by a \emph{Hann window} $w(t)=\tfrac12\bigl(1-\cos(2\pi t/T)\bigr)=\sin^2(\pi t/T)$, a
smooth bell equal to $1$ at the middle of the record and tapering to $0$ at both ends $t=0,T$. The
window is needed because a discrete Fourier transform treats the finite record as one period of an
indefinitely repeating signal: unless the signal happens to start and end at the same value, the jump
at that wrap-around radiates spurious frequencies (\emph{spectral leakage}) that smear the peak;
tapering both ends to zero removes the jump, so the true line stands out cleanly, at the mild cost of
broadening it slightly. Third, \emph{zero-pad} by a factor of eight, appending zeros to the windowed
signal until it is eight times its original length before transforming. The transform samples the
spectrum only at discrete frequencies spaced by $2\pi/N$ (radians per step), and with $N\approx90$
these bins are coarse enough that the true peak can fall between two of them; padding to $8N$ points
adds no information and does not sharpen the line (its width is fixed by the record length $T$), but
it evaluates the same spectrum on a grid eight times finer, interpolating between the original bins so
the peak is pinned to a small fraction of a bin. Fourth, take the magnitude of the transform and
locate its maximum inside the band $0.4\,(2E)<\omega<1.6\,(2E)$, which brackets the expected $2E$ and
rejects the residual drift and any harmonics; that peak position is $\omega_{\mathrm{ZB}}$.

\runin{Measuring the trembling amplitude.} The velocity amplitude $A_v$ of Eq.~\eqref{eq:zb-Av},
hence $R_{\mathrm{ZB}}=A_v/2E$ (the squares of Figure~\ref{fig:zb}), is measured by a least-squares
fit of $\langle\alpha_x(t)\rangle$ to the four functions $\{1,\,t,\,\cos 2Et,\,\sin 2Et\}$ over the
first trembling period $0\le t\le\pi/E$, from which $A_v=\sqrt{c_{\cos}^2+c_{\sin}^2}$ is the
magnitude of the fitted cosine and sine coefficients. A single period is used because the trembling
decays as the $+E$ and $-E$ halves of the packet, carried at opposite group velocities, separate in
space (the decay is visible in the left panel of Figure~\ref{fig:zb}); one period resolves the
amplitude at frequency $2E$ yet is short enough that this separation is still negligible, so the
measured $A_v$ matches the single-mode value Eq.~\eqref{eq:zb-Av} to within $2\%$.

\runin{Parameter sweep and reproduction.} The mass is swept over
$\mtil\in\{0,0.1,0.2,0.35,0.6,1.0,2.0\}$, spanning the crossover of Section~\ref{app:zb} at fixed
carrier $k_0$. The whole experiment, both solvers at every mass together with Table~\ref{tab:zb} and
Figure~\ref{fig:zb}, is produced by the single script
\texttt{qlb\_port/\allowbreak plot\_\allowbreak validation\_\allowbreak zitterbewegung.py}: its
routines \texttt{energy\_\allowbreak eigenmodes} (the same-block $\pm E$ pair), \texttt{initial\_\allowbreak state},
\texttt{run\_\allowbreak classical} (via \texttt{sweep\_\allowbreak operator}) and \texttt{run\_\allowbreak circuit} (via
\texttt{sweep\_\allowbreak circuit}) build and evolve the state, and \texttt{zb\_\allowbreak frequency} and
\texttt{zb\_\allowbreak velocity\_\allowbreak amplitude} carry out the two measurements just described.

\subsection{Results}
\begin{figure}[htb]
  \centering
  \includegraphics[width=0.98\linewidth]{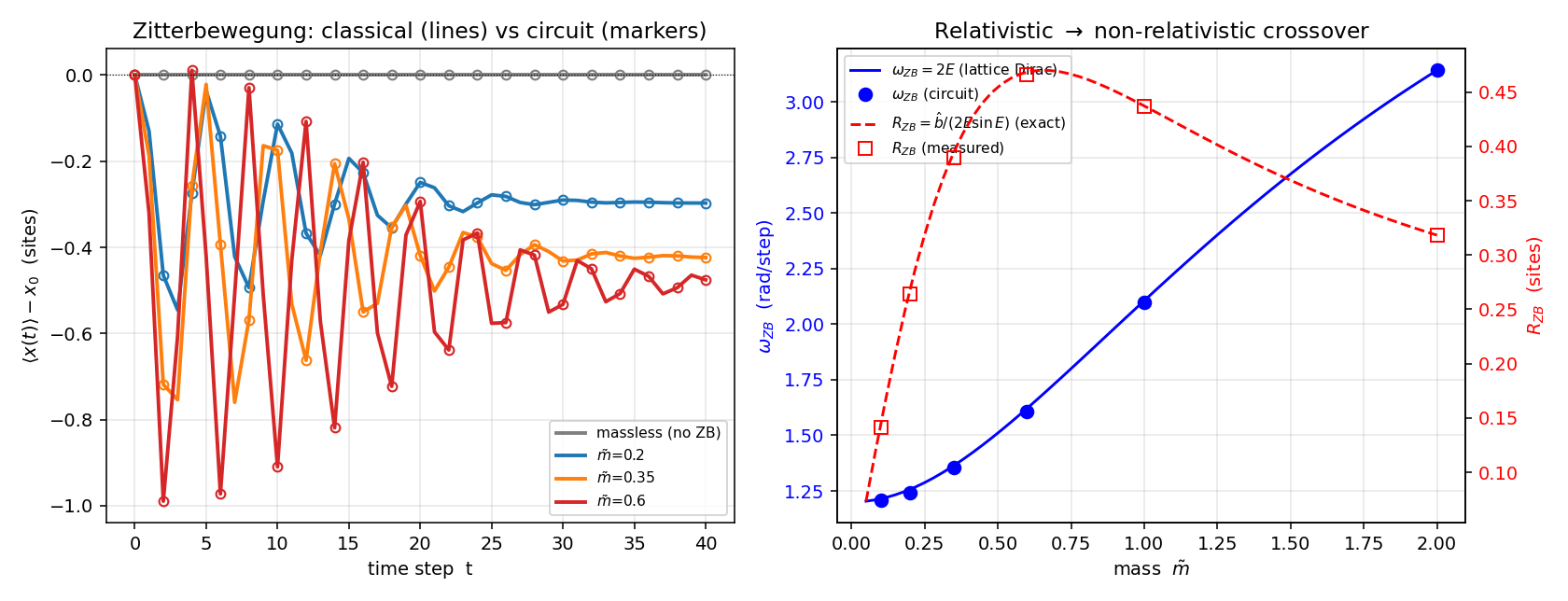}
  \caption{Zitterbewegung of the 1+1D Dirac packet in the QLB scheme, classical solver (lines)
    versus ported circuit (open markers). \emph{Left}: the mean position $\langle x(t)\rangle$
    for a massless packet (flat, no trembling) and three massive packets, which tremble and, as
    the $+E$ and $-E$ parts separate, relax to a drift; the frequency grows with mass and the
    circuit tracks the classical solver exactly. \emph{Right}: the crossover. The trembling
    frequency $\omega_{\mathrm{ZB}}$ from the circuit (filled points) follows $2E$ with the
    lattice dispersion Eq.~\eqref{eq:zb-disp} (solid line), while the position amplitude
    $R_{\mathrm{ZB}}$ measured from the trembling (open squares) tracks the exact
    $\bhat/(2E\sin E)$ (dashed, Eq.~\eqref{eq:zb-amp}) to within $2\%$, rising to a maximum near
    $\mtil\approx0.7$ and falling off in both the relativistic and non-relativistic limits.}
  \label{fig:zb}
\end{figure}

Table~\ref{tab:zb} and Figure~\ref{fig:zb} collect the measured Zitterbewegung, which recovers the
behavior Gerritsma et al.\ saw in their trapped ion. A massless packet does not tremble at all: with
$\mtil=0$ the velocity $\alpha_x$ is conserved, and the oscillation amplitude we measure sits at the
$10^{-16}$ level, numerically zero. Giving the packet a mass sets it trembling, and the trembling
quickens as the mass grows. Its frequency stays within about $1\%$ of $2E$ over the whole mass range
and follows the \emph{lattice} dispersion of Eq.~\eqref{eq:zb-disp} rather than the continuum form
$2\sqrt{(pc)^2+(mc^2)^2}$; the two agree while the packet still resolves the Compton scale
($\mtil\ll1$) and diverge once it does not, the measured $3.14$ at $\mtil=2$ falling well short of the
continuum $4.18$.

The amplitude of the trembling tells the same story as the experiment. Measured from the runs (the
squares in Figure~\ref{fig:zb}), the position amplitude $R_{\mathrm{ZB}}=\bhat/(2E\sin E)$ follows the
exact expression to within $2\%$: it rises from zero with the mass, peaks near $\mtil\approx0.7$, and
falls away again toward both extremes, just as Gerritsma et al.\ found. At large mass the two energy
branches drift apart before they can complete an oscillation, and at small mass the swing shrinks with
the Compton wavelength, so the trembling is strongest in the crossover between the two. Across every
mass the ported circuit reproduces the classical solver to $\max|\Delta\rho|\le2.3\times10^{-13}$ at
unit state fidelity, the same machine precision seen throughout Section~\ref{sec:verify}.

\begin{table}[htb]
  \centering
  \caption{Zitterbewegung on a $256$-site line ($10$ qubits), carrier $k_0=0.6$, from the QLB
    solver and the ported circuit. The trembling frequency $\omega_{\mathrm{ZB}}$ (the spectral
    peak of $\langle\alpha_x(t)\rangle$) is \emph{identical} from the circuit and the classical
    solver to the digits shown, and equals $2E$ with $E$ the lattice dispersion,
    Eq.~\eqref{eq:zb-disp}; the continuum value $2\sqrt{k_0^2+\mtil^2}$ is listed for
    comparison. The last column is the maximum circuit-vs-classical density deviation over all
    sites and the $90$ recorded steps; the state fidelity is $1$ to six digits throughout. The
    massless packet does not tremble, as $[\alpha_x,H]=0$.}
  \label{tab:zb}
  \begin{tabular}{@{}lccccc@{}}
    \toprule
    $\mtil$ & $E(k_0)$ & $\omega_{\mathrm{ZB}}$ & $2E$ & $2\sqrt{k_0^2+\mtil^2}$ & $\max|\Delta\rho|$ \\
    \midrule
    $0$    & $0.600$ & none    & ---     & ---     & $1.1\times10^{-13}$ \\
    $0.10$ & $0.607$ & $1.208$ & $1.215$ & $1.217$ & $1.0\times10^{-13}$ \\
    $0.20$ & $0.628$ & $1.243$ & $1.257$ & $1.265$ & $9.8\times10^{-14}$ \\
    $0.35$ & $0.682$ & $1.355$ & $1.364$ & $1.389$ & $9.8\times10^{-14}$ \\
    $0.60$ & $0.811$ & $1.605$ & $1.621$ & $1.697$ & $9.7\times10^{-14}$ \\
    $1.00$ & $1.053$ & $2.097$ & $2.105$ & $2.332$ & $9.5\times10^{-14}$ \\
    $2.00$ & $1.571$ & $3.142$ & $3.142$ & $4.176$ & $2.3\times10^{-13}$ \\
    \bottomrule
  \end{tabular}
\end{table}

Two connections close the comparison. Restoring physical units through the electron reference scale
of Eq.~\eqref{eq:refscale} (which fixes $\Delta x$ by taking the $\mtil=0.35$ packet to be an
electron), the trembling frequency $\omega_{\mathrm{ZB}}=2E/\hbar$ becomes $\sim10^{21}$~Hz and its
amplitude is of order the Compton wavelength, the free-electron Zitterbewegung that
\cite{gerritsma2010} identify as experimentally inaccessible in a real electron, and which is why
they simulate it. And whereas
the single-ion analog is intrinsically one-dimensional and free, the same circuit construction
extends unchanged to two and three dimensions and to arbitrary potentials, that is, to the Klein
tunneling of Section~\ref{sec:verify}.

\section{Execution on quantum hardware}
\label{app:hw}

The verification of Section~\ref{sec:verify} and Appendix~\ref{app:zb} is carried out on a
state-vector emulator, which writes the spinor field directly into the simulated register and
reads the full complex amplitude vector back out. A physical quantum processor does neither. Its
register begins in a fixed reference state, its output is available only as samples of a projective
measurement, and every gate must be expressed in the device's native instruction set on its fixed
qubit connectivity. This appendix sets out what those three constraints imply for the Dirac QLB
circuits, estimates the resulting gate cost, and shows that while the full Zitterbewegung run of
Appendix~\ref{app:zb} lies far beyond present hardware, a reduced instance of the same construction
is small enough to execute. The purpose of this Appendix is to demonstrate feasibility. Consistent 
with the scope of Section~\ref{sec:intro}, no claim of computational advantage is made or implied.

\subsection{Requirements for execution on quantum hardware}
\label{app:hw-req}

\runin{State preparation.} On the emulator the initial state, the length-$4N$ complex vector $\psi_0$
of Appendix~\ref{app:zb} on a lattice of $N=2^{\npos}$ sites, is loaded into the register in a single
operation. On hardware the $q=2+\npos$ qubits begin in $\ket{0}^{\otimes q}$, and $\psi_0$ must be
synthesized by a unitary $V$ with $V\ket{0}^{\otimes q}=\ket{\psi_0}$. We build $V$ \emph{exactly}, as
an amplitude encoding of the $2^{q}$ complex entries of the normalized field, using the standard
state-preparation routine of Qiskit \cite{qiskit2024,shende2006}. It synthesizes $V$ as the inverse
of a \emph{disentangling} circuit that would carry $\ket{\psi_0}$ back to $\ket{0}^{\otimes q}$: taking
the qubits one at a time, it applies to each a \emph{uniformly controlled} (multiplexed) pair of
single-qubit rotations, an $R_y$ that folds the two amplitude branches of that qubit into one and an
$R_z$ that cancels their relative phase, each conditioned on the qubits not yet processed, so that
after the last qubit every amplitude has been swept onto $\ket{0}^{\otimes q}$. Run in reverse with
negated angles, that cascade is $V$: it prepares $\ket{\psi_0}$ from $\ket{0}^{\otimes q}$ with no
approximation, and the compiled gate count is what the emulator cross-check of
Section~\ref{app:hw-demo} verifies. For a \emph{generic} $\psi_0$ the multiplexors expand to
$O(2^{q})$ two-qubit gates, the same exponential cost that makes arbitrary amplitude encoding expensive
and that we list among the open questions in Section~\ref{sec:discussion}. \emph{Structured} states
are far cheaper, which is why the reduced instance of Section~\ref{app:hw-demo} is chosen with
preparation in mind: on its three qubits both the localized density packet and the
mode-superposition velocity state factorize (a spinor state on the two spinor qubits times a position
state on the single position qubit) and prepare with \emph{no} two-qubit gates at all. The one
two-qubit gate that appears in the velocity circuit at $t=0$ is the $\Rmat^{-1}$ readout rotation of
Eq.~\eqref{eq:hw-alpha-deriv}, not the state preparation.

\runin{Readout.} The emulator returns the full amplitude vector; a device returns only
\emph{shots}. Each shot measures every qubit once, returns one bitstring of $0$s and $1$s, and
leaves the state collapsed, so the run is over. Two consequences follow. First, a single shot
reveals almost nothing: a probability or an average emerges only by repeating the \emph{identical}
circuit many times and histogramming the outcomes, which leaves a statistical (sampling) error of
order $1/\sqrt{S}$ after $S$ shots. Second, because the measurement ends the run, one cannot pause
the evolution, read a quantity, and resume; to follow an observable through time, each time $t$
needs its own circuit that prepares the state, applies $t$ steps, and measures at the end. The time
series in Appendix~\ref{app:zb} therefore becomes a separate circuit for every recorded step.

The two observables are read as follows. The position density $\rho(x)=\sum_{c}|\psi(x,c)|^{2}$ is
the probability of finding the packet at site $x$, summed over the four spinor components $c$; it is
obtained by keeping only the $\npos$ position bits of each measured bitstring and histogramming them.
The mean velocity $\langle\alpha_x\rangle$ cannot be extracted from that same density, and this is
not an oversight. The density records only \emph{populations}, how often each basis state is seen,
whereas $\alpha_x$ links different spinor components, so its value depends on the relative quantum
phases (the \emph{coherences}) between those components, which a straight measurement throws away. To
expose them one first rotates the spinor into the basis where $\alpha_x$ is diagonal: the
$x$-rotation $\Rmat$ does exactly this, $\Rmat^{-1}\alpha_x\Rmat=\diag(-1,-1,+1,+1)$
(Eq.~\eqref{eq:signs}), so appending $\Rmat^{-1}$ to the two spinor qubits and then measuring turns
$\langle\alpha_x\rangle$ into the population imbalance of a single spinor bit,
\begin{equation}
  \langle\alpha_x\rangle \;=\; P(q_1{=}1)-P(q_1{=}0),
  \label{eq:hw-alpha}
\end{equation}
where $q_1$ is the high spinor qubit. Explicitly, with the spinor component written $c=q_0+2q_1$
(Eq.~\eqref{eq:layout}), the diagonal entries of $\Rmat^{-1}\alpha_x\Rmat$ are $-1$ for $c\in\{0,1\}$
and $+1$ for $c\in\{2,3\}$, so the eigenvalue is fixed by the direction bit alone, $+1$ when
$q_1{=}1$ and $-1$ when $q_1{=}0$; whatever the value of the spin bit $q_0$ or of the position
register. Writing $\ket{\varphi}=\Rmat^{-1}\ket{\psi}$ for the rotated state that the circuit measures,
\begin{equation}
  \langle\alpha_x\rangle=\bra{\psi}\alpha_x\ket{\psi}
   =\bra{\varphi}\,\Rmat^{-1}\alpha_x\Rmat\,\ket{\varphi}
   =(+1)\,P(q_1{=}1)+(-1)\,P(q_1{=}0),
  \label{eq:hw-alpha-deriv}
\end{equation}
which is Eq.~\eqref{eq:hw-alpha}; here
$P(q_1{=}b)=\sum_{q_0,\,x}\lvert\braket{q_0,q_1{=}b,x}{\varphi}\rvert^{2}$ is the probability of the
direction bit, the sum over the spin bit $q_0$ and the position register being performed automatically
by the measurement. Reading the single qubit $q_1$ therefore returns the full velocity. That extra
rotation is a single two-qubit gate, a small price for information the density does not contain;
Figure~\ref{fig:hw-alpha-circuit} shows the complete measurement circuit.

\begin{figure}[h]
  \centering
  \begin{quantikz}[column sep=14pt,row sep=18pt]
    \lstick{$q_0$ (spin)}  & \gate[3]{\mathrm{Prep}(\psi_0)} & \gate[3]{\mathrm{Sweep}_x^{\,t}} & \gate[2]{\Rmat^{-1}} & \qw      & \qw       \\
    \lstick{$q_1$ (dir.)}  &                                 &                                  &                      & \qw      & \meter{}  \\
    \lstick{$\ket{x}$}     & \qwbundle{\npos}                &                                  & \qw                  & \qw      & \qw
  \end{quantikz}
  \caption{The circuit that measures the velocity $\langle\alpha_x\rangle$ on hardware. It prepares the
    initial spinor field $\psi_0$ (Section~\ref{app:hw-req}, \emph{State preparation}), applies $t$
    Dirac QLB sweeps $\mathrm{Sweep}_x$ (Figure~\ref{fig:sweep}), rotates the two spinor qubits by
    $\Rmat^{-1}$ into the $\alpha_x$ eigenbasis, and measures \emph{only} the direction qubit $q_1$;
    the population imbalance of that single bit is $\langle\alpha_x\rangle=P(q_1{=}1)-P(q_1{=}0)$
    (Eq.~\eqref{eq:hw-alpha-deriv}). The density circuit is identical up to the readout, where it drops
    $\Rmat^{-1}$ and instead measures all $q=2+\npos$ qubits. Compiled to the Rigetti Cepheus-1
    native set $\{R_x,R_z,\mathrm{iSWAP}\}$ and routed onto the square lattice, the per-step
    two-qubit-gate counts are those listed in Tables~\ref{tab:hw-position} and
    \ref{tab:hw-trembling}.}
  \label{fig:hw-alpha-circuit}
\end{figure}

\runin{Native gates and connectivity.} The unit operations of Section~\ref{sec:units} are written
with idealized multi-qubit gates: the fixed rotation $\Rmat$, the collision $\Qhat$, and the
streaming increment. A device realizes only a fixed native set, in practice a few single-qubit
rotations together with one two-qubit entangler, the $\mathrm{iSWAP}$ on the Rigetti processor used
here, and only between qubits that are physically coupled. Compiling to that set and routing to the
device's connectivity graph, a
two-dimensional grid for the superconducting processor used here, replaces each idealized gate by
native gates and inserts additional swap operations wherever the required interaction is not directly
available. Both steps raise the two-qubit-gate count and the \emph{depth} (the number of gate
layers that must run one after another) above the idealized figures of Section~\ref{sec:resources}.
A circuit with few layers and few two-qubit gates is called \emph{shallow}. Depth and gate count
are what limit a noisy device: every gate is slightly imperfect, and the longer a computation runs
the more the qubits lose their quantum state (they \emph{decohere}), so a shallow circuit finishes
before much error accumulates, whereas a deep one does not.

\begin{figure}[h]
  \centering
  \includegraphics[width=0.90\linewidth]{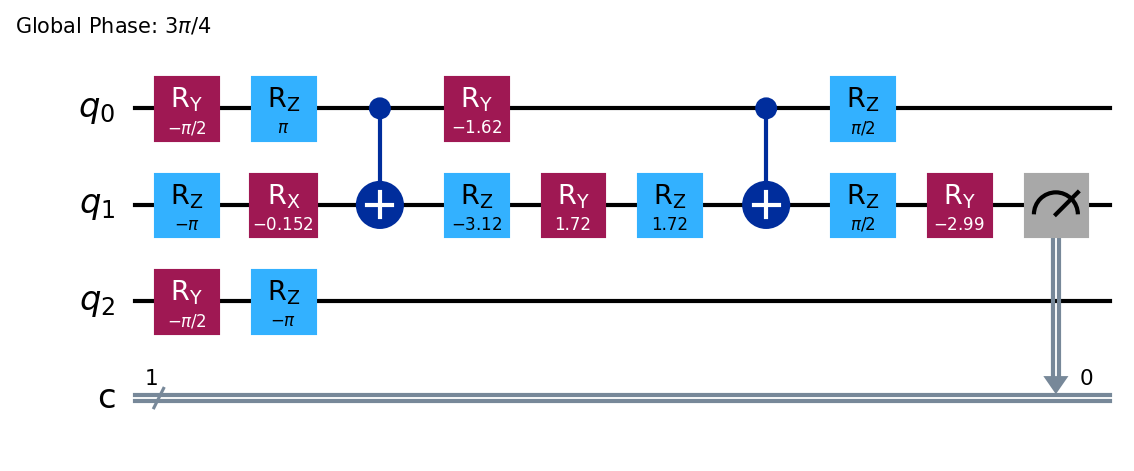}\\[10pt]
  \includegraphics[width=0.98\linewidth]{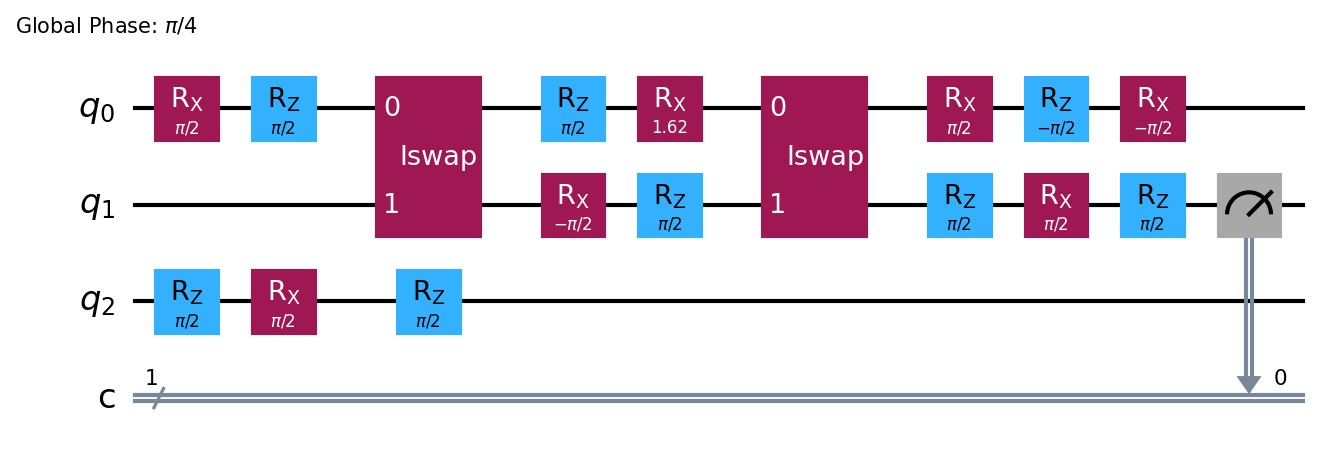}
  \caption{The two-site velocity circuit of Figure~\ref{fig:hw-alpha-circuit} at $t=1$, at the gate
    level in two gate sets, so the accommodations made for Rigetti are visible directly (qubits
    $q_0,q_1$ are the spinor, $q_2$ the position site). \emph{Top}: the circuit in the standard set
    $\{R_z,R_y,R_x,\mathrm{CX}\}$ used throughout the main text, two $\mathrm{CX}$ gates among
    single-qubit rotations. \emph{Bottom}: the same circuit recompiled to the Rigetti Cepheus-1
    native set $\{R_x,R_z,\mathrm{iSWAP}\}$; the entangling $\mathrm{CX}$ gates become
    $\mathrm{iSWAP}$, and because the device offers no $R_y$, the single-qubit rotations are rebuilt
    from $R_x$ and $R_z$ alone. The product-state preparation stays single-qubit, with no entangler,
    in both. Both panels are the automatic qiskit transpiler output at all-to-all connectivity;
    routing onto the physical square lattice adds swap gates, giving the per-step totals of
    Tables~\ref{tab:hw-position} and \ref{tab:hw-trembling}.}
  \label{fig:hw-native}
\end{figure}

\runin{Compiling the two-site circuit.} For the reduced instance the compilation is small enough to
display in full. Figure~\ref{fig:hw-native} shows the velocity circuit of
Figure~\ref{fig:hw-alpha-circuit}, at $t=1$, at the gate level in two gate sets, so the changes made
for the device can be read off directly. It is the \emph{complete} circuit, the state preparation, one
QLB sweep, and the $\Rmat^{-1}$ readout rotation compiled together; the transpiler fuses these three
stages, so they are not drawn as separate blocks, but all are present. In isolation the product-state
preparation is single-qubit only and the readout is a single $\Rmat^{-1}$ two-qubit gate, so the two
entangling gates shown here come from that readout gate merged with the sweep. The top row is the circuit in the
standard set $\{R_z,R_y,R_x,\mathrm{CX}\}$ used throughout the main text, two $\mathrm{CX}$ gates among
single-qubit rotations. The bottom row recompiles it to the Rigetti Cepheus-1 native set
$\{R_x,R_z,\mathrm{iSWAP}\}$: the entangling $\mathrm{CX}$ gates become $\mathrm{iSWAP}$, and because
the device offers no $R_y$, the single-qubit rotations are rebuilt from $R_x$ and $R_z$ alone. The
product-state preparation stays single-qubit in both, and the position qubit $q_2$ carries no
two-qubit gate at all.\footnote{On a single position qubit a $+1$ and a $-1$ shift coincide, both
exchanging the two sites, so the direction-controlled streaming increment collapses to an
unconditional bit-flip of $q_2$; the spinor therefore never entangles with the position register at
$\npos{=}1$, and a proper controlled increment, with genuine spinor-position entanglement, appears
only at $\npos\ge2$.} Both panels are the automatic output of the
qiskit transpiler at all-to-all connectivity; on the physical square lattice the router additionally
assigns the three logical qubits to three sites and inserts swap operations wherever an interacting
pair is not adjacent, which raises the per-step totals to the $1,4,10,12,18$ of
Table~\ref{tab:hw-trembling}.

\subsection{Resource estimate and the reach of present hardware}
\label{app:hw-cost}

The decisive quantity is the number of two-qubit gates, whose accumulated infidelity dominates the
error budget of a noisy device. Table~\ref{tab:hw-cost} gives the two-qubit-gate count of one QLB
sweep step as a function of the lattice size, obtained by compiling the sweep circuit of
Section~\ref{sec:assemble} to a native set (single-qubit rotations and the controlled-NOT) under
all-to-all connectivity; the count on a grid is higher, because routing adds swaps. The cost climbs
steeply with the position register, from $16$ gates at $\npos=2$ to about $2500$ at $\npos=6$, because
the streaming increment is a cascade of multi-controlled gates on the $\npos$ position qubits
(Section~\ref{sec:units}).

\begin{table}[htb]
  \centering
  \caption{Two-qubit-gate count of one Dirac QLB sweep step against the number of position qubits
    $\npos$ (lattice $N=2^{\npos}$ sites, $q=2+\npos$ qubits in all), from compiling the sweep circuit
    of Section~\ref{sec:assemble} to a native set (single-qubit rotations and the controlled-NOT)
    under all-to-all connectivity; this is a lower bound, as a grid adds routing swaps. These are
    \emph{un-optimized} per-step figures, counted block by block: in a fully compiled circuit
    adjacent rotations cancel and, at $\npos{=}1$, the two controlled increments collapse to a
    single-qubit flip, so the assembled velocity step of Figure~\ref{fig:hw-native} carries two
    entangling gates rather than the six tabulated. The last column is the survival probability
    $f^{\,n_2}$ that no two-qubit gate errs in one step at a representative fidelity $f=0.99$. The
    Zitterbewegung run of Appendix~\ref{app:zb} is $\npos=8$ over $90$ steps.}
  \label{tab:hw-cost}
  \begin{tabular}{@{}ccccc@{}}
    \toprule
    $\npos$ & $q=2+\npos$ & $N=2^{\npos}$ & two-qubit gates / step & $f^{\,n_2}$ at $f=0.99$ \\
    \midrule
    $1$ & $3$ & $2$  & $6$    & $0.94$ \\
    $2$ & $4$ & $4$  & $16$   & $0.85$ \\
    $3$ & $5$ & $8$  & $120$  & $0.30$ \\
    $4$ & $6$ & $16$ & $352$  & $0.029$ \\
    $6$ & $8$ & $64$ & $2480$ & $1.5\times10^{-11}$ \\
    \bottomrule
  \end{tabular}
\end{table}

The Zitterbewegung run of Appendix~\ref{app:zb} uses $\npos=8$ (a $256$-site line, ten qubits) over
$T=90$ steps. Extrapolating Table~\ref{tab:hw-cost}, a single step is already of order $10^{4}$
two-qubit gates, so the full evolution exceeds $10^{6}$ two-qubit gates, before the state
preparation and before grid routing. On a device whose two-qubit gates have fidelity $f\approx0.99$,
the probability that not one of them errs across the circuit is about $f^{10^{6}}\approx0$: the output
would be indistinguishable from noise. Even a single step at $\npos=6$, at $f^{2500}\sim10^{-11}$, is
already hopeless. The full experiment is therefore intrinsically an emulator result, and its
reproduction on hardware awaits substantially lower error rates or quantum error correction.

What is within reach today is a reduced instance at the smallest nontrivial size. The top row of
Table~\ref{tab:hw-cost} shows that at $\npos=1$, a two-site line on three qubits, one sweep step
compiles to about six two-qubit gates under all-to-all connectivity, for which $f^{6}\approx0.94$
at $f=0.99$ keeps the signal well clear of the noise floor even after the extra gates that grid
routing inserts. A second point matters just as much when an observable is followed in time: by
Section~\ref{app:hw-req} each recorded step is a \emph{separate} circuit that prepares the state,
applies that many sweeps, and measures, and at $\npos=1$ the early steps that carry the largest
signal are only one or two two-qubit gates before routing, so a whole trembling history can be
traced step by step rather than collapsing after the first. The following two subsections describe
this instance and its execution on hardware.

\subsection{A reduced instance that runs on hardware}
\label{app:hw-demo}

The reduced instance keeps the full Dirac QLB step but takes the smallest nontrivial lattice,
$\npos=1$: a line of $N=2$ sites, encoded in $q=2+\npos=3$ qubits, namely the two shared spinor
qubits and a single position qubit. Every operation of the scheme is retained, the fixed rotation
$\Rmat$, the two-qubit collision $\Qhat$, and the streaming increment, but on one position qubit the
increment is a single controlled operation, so one sweep is as shallow as the scheme allows. We use
this instance to follow the QLB dynamics in time through two observables, each on its own tailored
state: the mean position of a localized packet, which sloshes between the two sites under streaming,
and the mean velocity of the $\pm E$ mode superposition, which trembles at the Zitterbewegung
frequency.

\runin{Lattice momenta and the two states.} A periodic line of $N$ sites carries the discrete momenta
$k_j=2\pi j/N$ for $j=0,\dots,N-1$. On $N=2$ sites only two survive, $k_0=0$ and $k_0=\pi$, and both
are \emph{band edges}. The \emph{group velocity} of a wave packet is the speed at which its center of
mass drifts, and equals the slope of the dispersion curve $E(k)$, $v_g=\mathrm dE/\mathrm dk$ (in
lattice units the momentum is the wavenumber, $p\leftrightarrow k$, so
$\mathrm dE/\mathrm dp=\mathrm dE/\mathrm dk$; Appendix~\ref{app:zb}). Differentiating the lattice
dispersion $\cos E(k)=\ahat\cos k$ of Eq.~\eqref{eq:zb-disp}, that is
$-\sin E\,\mathrm dE=-\ahat\sin k\,\mathrm dk$, gives
\[
  v_g=\frac{\mathrm dE}{\mathrm dk}=\frac{\ahat\sin k}{\sin E}
     =\frac{\ahat\sin k}{\sqrt{1-\ahat^2\cos^2 k}} ,
\]
which is proportional to $\sin k$ and so vanishes at \emph{both} $k=0$ and $k=\pi$, the bottom and top
of the energy band where $E(k)$ is stationary. (These are the only two momenta $N=2$ allows, so every
carrier the lattice can hold is a band edge.) Zero group velocity is the property we want. In the full
run of Appendix~\ref{app:zb} the carrier is $k_0=0.6$, where $v_g\neq0$; there the positive- and
negative-energy modes $u_\pm$ of the trembling state travel at the opposite speeds $\pm v_g$ and slowly
move apart, and once they no longer overlap in space they can no longer interfere, so the trembling dies
away. At a band edge both speeds are zero, the two modes stay on top of each other for all time, and the
trembling persists undamped; this is what lets us trace a full period on only two sites. We therefore
take $k_0=0$.

We prepare two initial states. Both carry the same four-component spinor, the equal superposition
$\tfrac{1}{\sqrt2}(u_++u_-)$ of the two energy modes whose interference is the trembling
(Eq.~\eqref{eq:zb-split}); the states differ only in how that spinor is spread over the two sites, and
it is this spatial profile that decides which observable moves. The \emph{velocity state} puts the same
spinor, with equal amplitude, on both sites: it is the plane wave $e^{ik_0x}$ of
Eq.~\eqref{eq:initcond} at $k_0=0$, where $e^{i0\cdot x}=1$ makes every site identical. Its probability
is then shared equally between the sites, $\rho(0)=\rho(1)=\tfrac12$, and the one-site streaming shift
only swaps these two equal halves, so the mean position $\langle x\rangle=\rho(1)$ never leaves $\tfrac12$
and the sole quantity that changes in time is the mean velocity $\langle\alpha_x\rangle$; this state
isolates the velocity trembling. The \emph{position state} instead weights the sites unequally, by
multiplying the spinor by a Gaussian centered on one site, so most of the probability sits there,
$\rho(0)\neq\rho(1)$, and the packet is localized rather than uniform. Now the streaming shift carries the
peak from one site to the other, so it is the mean position $\langle x\rangle$ that moves; this state
isolates the position sloshing. Both are structured states, so their preparation is cheap: on three qubits the localized packet
compiles to single-qubit rotations with \emph{no} two-qubit gates at all, and the velocity state,
together with the rotation into the $\alpha_x$ basis, to a single two-qubit gate at $t=0$.

The two observables are read by the recipes of Section~\ref{app:hw-req}. The position
density $\rho(x)=\sum_c|\psi(x,c)|^2$ is the histogram of the single position bit over the shots. The
mean velocity $\langle\alpha_x\rangle$ follows from Eq.~\eqref{eq:hw-alpha}: appending the inverse
rotation $\Rmat^{-1}$ to the two spinor qubits diagonalizes $\alpha_x$, after which
$\langle\alpha_x\rangle=P(q_1{=}1)-P(q_1{=}0)$ is the population imbalance of the high spinor bit.

The trembling is full-scale here. The velocity trembles with amplitude
$A_v=\bhat/\sin E$ (Eq.~\eqref{eq:zb-Av}). At a band edge this is not merely large but \emph{exactly
one}, for any mass. With $k_0=0$ the dispersion Eq.~\eqref{eq:zb-disp} gives $\cos E=\ahat$, and since
the collision amplitudes obey $\ahat^2+\bhat^2=1$ (Eq.~\eqref{eq:collision}) it follows that
$\sin E=\sqrt{1-\ahat^2}=\bhat$, whence $A_v=\bhat/\bhat=1$. The mean velocity therefore swings across
the entire physical interval $[-1,+1]$. By the same band-edge symmetry the steady part of
Eq.~\eqref{eq:zb-split} vanishes, so the exact signal is the pure sinusoid
$\langle\alpha_x\rangle(t)=\sin(2Et)$: it starts at zero, rises to $+1$ near $t=1$, and reverses to
$-1$ near $t=3$. At the chosen mass $\mtil=0.8$ the amplitudes are $\ahat=0.724$ and $\bhat=0.690$, so
$E=\arccos(0.724)=0.761$, the trembling frequency is $2E=1.522$ radians per step, and the period is
$2\pi/2E\approx4.1$ steps. This resolvable, full-amplitude oscillation is something that another
choice of a four-site carrier at $k_0=\pi/2$ could not provide, where $E=\pi/2$ collapses the 
trembling into a bare two-step sign alternation.

The packet sloshes with period two. The position signal has an entirely different origin
from the velocity. Under the streaming increment the two spinor halves of the packet are shifted by
one site in opposite directions, and on a two-site ring a one-site shift merely exchanges the two
sites, so the density returns to its start after two steps and the mean position
$\langle x\rangle=\rho(1)$ oscillates with period two. This sloshing is present already for a
\emph{single} energy branch, for which $\langle\alpha_x\rangle=0$ at every step, so it is produced by
the streaming shift and not by the $\pm E$ interference: the two observables therefore probe
complementary halves of the step, the streaming through the position (a test of populations) and the
collision-induced $\pm E$ beat through the velocity (a test of coherences).

One circuit is needed per recorded step. Because a projective measurement ends the run, an observable as
a function of time is not one circuit but a family: the circuit for step $t$ prepares the chosen
state, applies $t$ sweeps, appends the readout rotation where one is needed, and measures. 
Its two-qubit gate count grows with $t$, but from a very low base, so the early steps that carry the largest
signal are also the shallowest. This is the structural reason the entire evolution can survive on
hardware, whereas the single deep circuit of the full Zitterbewegung run of Appendix~\ref{app:zb}, of
order $10^6$ two-qubit gates, cannot.

\runin{Emulator cross-check.} Before submission the sampling and readout pipeline is verified against
the exact result on the state-vector emulator at the same shot count. The one-step quantities are
reproduced to the sampling error: the $t=0$ density to a \emph{total-variation distance} of a few
parts in $10^3$ (the total-variation distance between two probability distributions $p$ and $q$ is
$\tfrac12\sum_x|p(x)-q(x)|$, zero for identical distributions and one for disjoint ones, so a small
value means close agreement), and each $\langle\alpha_x\rangle(t)$ to within a few percent, confirming
that the state preparation, the rotation into the $\alpha_x$ basis, and the histogramming are correct.
The identical compiled circuits are then submitted to the device. The whole procedure, the state
preparations, the density and velocity circuits, the emulator cross-check, and the hardware
submission, is carried out by the scripts \texttt{qlb\_port/\allowbreak run\_2site\_density.py} (the
position sloshing) and \texttt{qlb\_port/\allowbreak run\_2site\_trembling.py} (the velocity period),
with \texttt{qlb\_port/\allowbreak plot\_2site\_results.py} assembling Figure~\ref{fig:hw-results}, in
the accompanying library (Section~\ref{sec:repro}).

\subsection{Results on a superconducting processor}
\label{app:hw-results}

\begin{figure}[htb]
  \centering
  \includegraphics[width=0.98\linewidth]{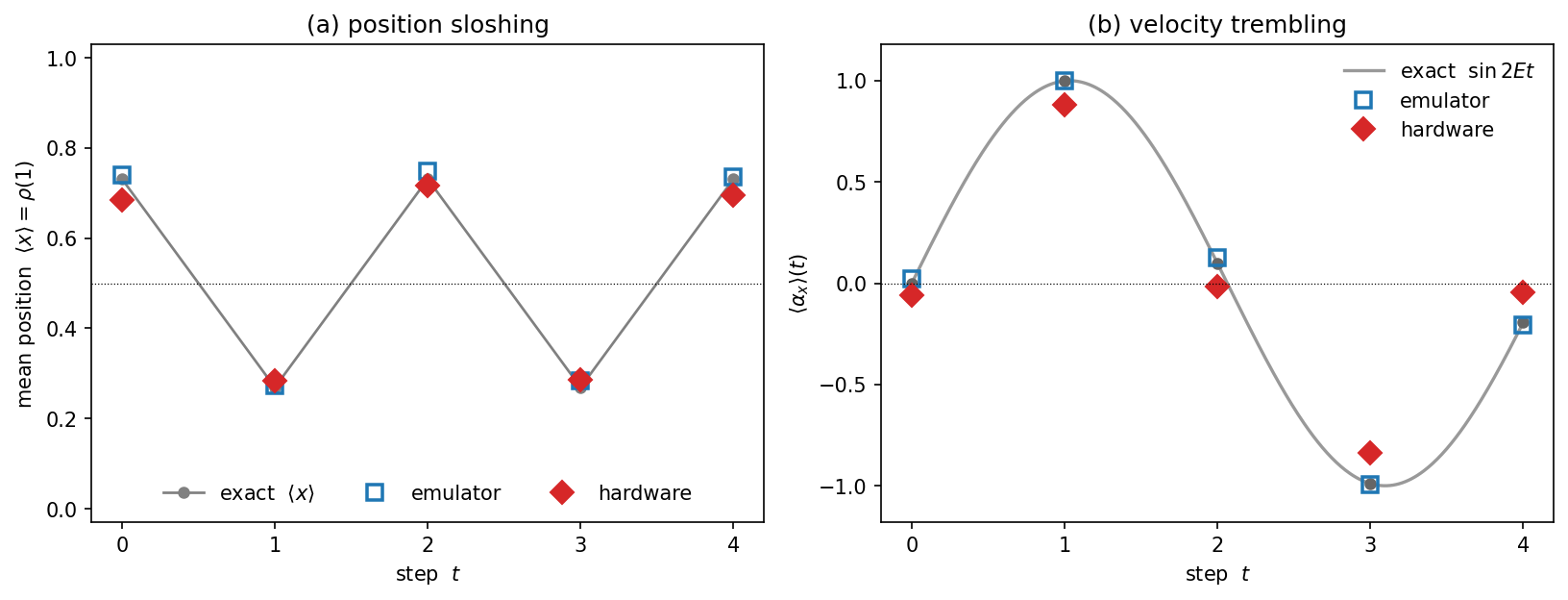}
  \caption{Reduced Dirac QLB instance on the two-site lattice (three qubits): exact (grey),
    state-vector emulator (blue, $2000$ shots), and Rigetti Cepheus-1 hardware (red).
    \emph{(a)}~Mean position $\langle x\rangle=\rho(1)$ of a localized packet, which the streaming
    shift sends sloshing between the two sites with period two; the hardware retains about $90\%$ of
    the swing through $t=4$. \emph{(b)}~Mean velocity $\langle\alpha_x\rangle(t)$, the Zitterbewegung
    beat of the $\pm E$ superposition over one full period. The exact values are the pure sinusoid
    $\sin(2Et)$ (band-edge carrier $k_0=0$, mass $\mtil=0.8$, period $\approx4.1$ steps); the hardware
    traces the entire oscillation, recovering $85$--$88\%$ of the full swing at the extrema ($t=1,3$)
    and returning to the readout bias at the small values ($t=0,2,4$). Each time point is an
    independent shallow circuit rather than one deep one.}
  \label{fig:hw-results}
\end{figure}

The reduced instance was executed on the Rigetti Cepheus-1 processor ($108$ superconducting qubits,
square-lattice connectivity) through the Open Quantum cloud~\cite{openquantum}, at $2000$ shots per circuit. As a
reference for the device's combined two-qubit and readout quality, a Bell pair prepared and measured
on the processor returned the correct correlated outcomes $00$ and $11$ in about $92\%$ of the shots.

\begin{table}[htb]
  \centering
  \caption{Mean position $\langle x\rangle=\rho(1)$ of the localized two-site packet over
    $t=0,\dots,4$, from the exact operator, the state-vector emulator, and the Rigetti Cepheus-1
    processor ($2000$ shots each). The packet exchanges the two sites each step, so $\langle x\rangle$
    alternates with period two between $0.731$ and $0.269$. The last column is the two-qubit-gate
    count of each circuit after routing onto the square lattice; at $t=0$ the state compiles to
    single-qubit rotations only, isolating the preparation-and-readout floor. The hardware retains
    about $90\%$ of the swing out to $t=4$.}
  \label{tab:hw-position}
  \begin{tabular}{@{}ccccc@{}}
    \toprule
    step $t$ & exact & emulator & hardware & two-qubit gates \\
    \midrule
    $0$ & $0.731$ & $0.740$ & $0.684$ & $0$  \\
    $1$ & $0.269$ & $0.274$ & $0.284$ & $5$  \\
    $2$ & $0.731$ & $0.749$ & $0.716$ & $10$ \\
    $3$ & $0.269$ & $0.284$ & $0.286$ & $13$ \\
    $4$ & $0.731$ & $0.736$ & $0.696$ & $18$ \\
    \bottomrule
  \end{tabular}
\end{table}

\runin{The position sloshing.} Table~\ref{tab:hw-position} and Figure~\ref{fig:hw-results}(a) follow
the mean position $\langle x\rangle=\rho(1)$ of the localized packet over $t=0,\dots,4$. Exactly, the
packet exchanges the two sites at every step, so $\langle x\rangle$ alternates between $0.731$ and
$0.269$ with period two. The hardware reproduces this cleanly, tracing
$0.684,\,0.284,\,0.716,\,0.286,\,0.696$: the full alternation is intact out to $t=4$ ($18$ two-qubit
gates), the swing damped only to about $90\%$ of the ideal. At $t=0$, where the packet compiles to
single-qubit rotations with no two-qubit gates, the per-site density measures the pure
preparation-and-readout floor, exact $(0.269,0.731)$ against the hardware $(0.316,0.684)$, a
total-variation distance of $0.047$ set essentially by the residual readout asymmetry of the device.

\begin{table}[htb]
  \centering
  \caption{Mean velocity $\langle\alpha_x\rangle(t)$ over one Zitterbewegung period on the two-site
    lattice ($\npos=1$, mass $\mtil=0.8$, band-edge carrier $k_0=0$), from the exact operator, the
    state-vector emulator, and the Rigetti Cepheus-1 processor ($2000$ shots each). The exact values
    are the pure sinusoid $\sin(2Et)$ with $2E=1.522$ (period $\approx4.1$ steps). The last column is
    the two-qubit-gate count of each circuit after routing onto the square lattice. The hardware
    recovers $88\%$ and $85\%$ of the ideal swing at the two extrema ($t=1,3$) and returns to the
    readout bias at the small values ($t=0,2,4$).}
  \label{tab:hw-trembling}
  \begin{tabular}{@{}ccccc@{}}
    \toprule
    step $t$ & exact & emulator & hardware & two-qubit gates \\
    \midrule
    $0$ & $+0.000$ & $+0.020$ & $-0.059$ & $1$  \\
    $1$ & $+0.999$ & $+0.999$ & $+0.883$ & $4$  \\
    $2$ & $+0.097$ & $+0.128$ & $-0.018$ & $10$ \\
    $3$ & $-0.989$ & $-0.993$ & $-0.836$ & $12$ \\
    $4$ & $-0.194$ & $-0.206$ & $-0.047$ & $18$ \\
    \bottomrule
  \end{tabular}
\end{table}

\runin{The velocity trembling.} Table~\ref{tab:hw-trembling} and Figure~\ref{fig:hw-results}(b) give
the mean velocity $\langle\alpha_x\rangle(t)$ over one full period, $t=0,\dots,4$. Each point is a separate
circuit whose two-qubit count after routing onto the square lattice is listed in the last column of
the table; routing raises the idealized one or two gates per step to $1,4,10,12,18$, the overhead
growing with the number of streaming increments. Against the exact sinusoid $\sin(2Et)$, the hardware
reproduces the trembling in full. The two extrema, $+0.999$ at $t=1$ and $-0.989$ at $t=3$, are
measured at $+0.883$ and $-0.836$, recovering $88\%$ and $85\%$ of the ideal swing, in line with the
shallow gate counts and the per-step survival of Table~\ref{tab:hw-cost}; the smaller values at
$t=0,2,4$ are all measured within about $0.06$ of zero, at the level of the readout bias set by the
$t=0$ point. The emulator, plotted alongside, sits on the exact curve to the sampling error.

Two features are worth drawing out. First, the noise damps the amplitude but preserves the shape: the
hardware points still fall on a sinusoid of the same frequency, scaled down by roughly a constant
factor. This is the signature of an approximately depolarizing error, which multiplies the expectation
of a traceless observable such as $\alpha_x$ by an overall factor while leaving its sign structure and
frequency intact, so the trembling remains unambiguous even where it is attenuated. Second, the two
observables test complementary halves of the step: the position sloshing is driven by the streaming
shift and survives as a nearly undamped population signal ($\sim90\%$), whereas the velocity trembling
is the collision-induced $\pm E$ beat, a coherence that the same noise attenuates somewhat more
($\sim85\%$). Together they certify that both the streaming and the collision of the Dirac QLB step run
on the device, over a full period, not merely that a state can be prepared and read.


\subsection{Qubit selection on shallow circuits: automatic routing versus a pinned triple}
\label{app:hw-pinned}

\begin{figure}[htb]
  \centering
  \includegraphics[width=0.98\linewidth]{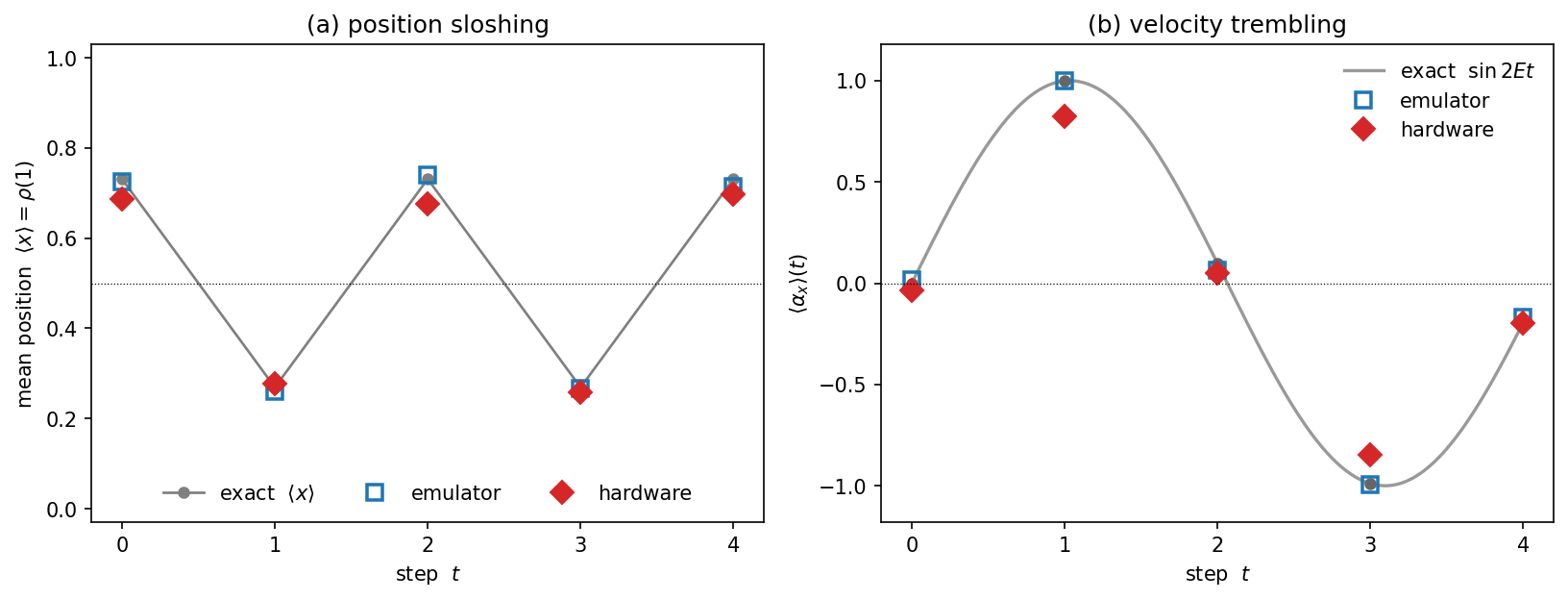}
  \caption{The two-site instance of Figure~\ref{fig:hw-results} run directly on Rigetti Cepheus-1
    through Quantum Cloud Services, with the three qubits pinned by hand to a high-fidelity triple
    drawn from the live calibration: the position register sits on the highest-readout qubit of the
    device, and the two spinor qubits on the highest-fidelity coupled pair. Exact values are grey,
    the state-vector emulator blue, and the hardware red. \emph{(a)}~Position sloshing
    $\langle x\rangle=\rho(1)$; \emph{(b)}~velocity trembling $\langle\alpha_x\rangle(t)$, each over
    one period. Against the automatic, connectivity-only placement of Table~\ref{tab:hw-pinned},
    pinning lifts the preparation-and-readout floor at $t=0$ from a total-variation distance of
    $0.319$ to $0.044$, while the coherent dynamics are unchanged.}
  \label{fig:hw-pinned}
\end{figure}

The run of Section~\ref{app:hw-results} was carried out through the Open Quantum cloud, which chose the
physical qubits on the reduced instance's behalf. Repeating it directly on the same processor through
Rigetti's Quantum Cloud Services, and taking that choice into our own hands, exposes a feature of
shallow circuits that the first run had left implicit: when a circuit is only a few gates deep, its
accuracy is governed less by the gates than by the quality of the particular qubits it is placed on,
and most of all by the fidelity with which those qubits can be measured.

\runin{From qiskit to Quil.} The circuits of this paper are built and verified with qiskit, whereas
Quantum Cloud Services accepts programs in Quil, the instruction language of the Rigetti stack, and no
ready-made converter links the two. We therefore wrote a small exporter, but it is glue code rather
than a second implementation: it rebuilds no circuit and repeats none of the physics of
Section~\ref{sec:units}. It takes each finished qiskit circuit, already transpiled to the universal
basis $\{R_x,R_y,R_z,\mathrm{CX}\}$ used throughout the main text, reads through its list of gates in
order, and writes each one out as the Quil instruction of the same name: an $R_x$ rotation becomes
\texttt{RX}, a $\mathrm{CX}$ becomes \texttt{CNOT}, and a measurement becomes \texttt{MEASURE}. No
angle is recomputed and no matrix is re-synthesized, because the two languages define these gates
identically, so a qiskit $R_x(\theta)$ is the Quil \texttt{RX} rotation by the same $\theta$ and a
qiskit $\mathrm{CX}$ is a Quil \texttt{CNOT}; the transcription is a literal, gate-for-gate
relabeling. The only quantity dropped is the circuit's overall phase, which qiskit records but Quil
does not represent and which is in any case physically unobservable. From that generic Quil the native
compiler quilc takes over automatically, rewriting the program into the processor's native gates and
routing it onto the lattice, the same role the qiskit transpiler filled for the Open Quantum run. The
two paths therefore differ only in the compiler that performs the final native mapping and routing, not
in the logical circuit itself, which is what makes their comparison meaningful.

On a circuit this shallow, automatic placement is limited by readout rather than by depth. Left to
choose for itself, the compiler
assigns the three logical qubits to physical ones so as to satisfy the circuit's connectivity, without
weighing the wide spread in readout quality across the device's $108$ qubits. On the direct run this
gave a poor home to the position register, and the two-site density made the consequence plain. The
mean position $\langle x\rangle=\rho(1)$ is read straight from the measured bit of the position qubit,
so an error in that qubit's readout passes undiminished into $\langle x\rangle$; the velocity, by
contrast, is the difference $P(q_1{=}1)-P(q_1{=}0)$ of two spinor-bit populations, which a symmetric
readout error only scales rather than displaces, making it the more forgiving of the two observables.
The symptom is sharpest at $t=0$, where the circuit is a single-qubit state preparation followed by a
measurement, with no entangling gate at all: the automatic run returned $\rho=(0.588,0.412)$ where both
the exact operator and the noiseless emulator give $(0.269,0.731)$, a total-variation distance of
$0.319$ from the ideal (Table~\ref{tab:hw-pinned}). Since the emulator reproduces the exact value in
every case, nothing in the dynamics has failed; the whole of the discrepancy is the readout of a badly
chosen qubit. The error runs largest at the even steps $t=0,2,4$, where the packet rests mainly on the
far site and the position qubit is therefore mostly in $\ket1$. Readout on a superconducting qubit is
asymmetric: a qubit in $\ket1$ is misreported as $\ket0$ more often than the reverse, because it
relaxes toward its ground state during the finite measurement, so a large $\rho(1)$ is pulled down
harder than a small one. This is exactly the even-step sag of the automatic column of
Table~\ref{tab:hw-pinned}.

\begin{table}[htb]
  \centering
  \caption{Direct execution on Rigetti Cepheus-1 through Quantum Cloud Services: the compiler's
    automatic (connectivity-only) qubit placement against a placement pinned by hand to a
    high-fidelity triple read from the live calibration, for the mean position
    $\langle x\rangle=\rho(1)$ and the mean velocity $\langle\alpha_x\rangle$ over $t=0,\dots,4$
    ($2000$ shots; exact values shown for reference). Automatic placement degrades the position most at
    the even steps, where $\langle x\rangle$ is large and the position qubit sits mostly in $\ket1$;
    pinning that register to the best-readout qubit removes the bias. The velocity, a difference of
    two populations, is the less sensitive observable and shifts little either way.}
  \label{tab:hw-pinned}
  \begin{tabular}{@{}cccccccc@{}}
    \toprule
    & \multicolumn{3}{c}{position $\langle x\rangle$} & & \multicolumn{3}{c}{velocity $\langle\alpha_x\rangle$}\\
    \cmidrule(lr){2-4}\cmidrule(lr){6-8}
    step $t$ & exact & automatic & pinned & & exact & automatic & pinned \\
    \midrule
    $0$ & $0.731$ & $0.412$ & $0.688$ & & $+0.000$ & $+0.021$ & $-0.035$ \\
    $1$ & $0.269$ & $0.264$ & $0.278$ & & $+0.999$ & $+0.773$ & $+0.825$ \\
    $2$ & $0.731$ & $0.497$ & $0.676$ & & $+0.097$ & $+0.176$ & $+0.049$ \\
    $3$ & $0.269$ & $0.293$ & $0.258$ & & $-0.989$ & $-0.681$ & $-0.847$ \\
    $4$ & $0.731$ & $0.418$ & $0.698$ & & $-0.194$ & $-0.111$ & $-0.199$ \\
    \bottomrule
  \end{tabular}
\end{table}

A better placement can be read from the calibration. The processor publishes a live calibration, a per-qubit
and per-edge characterization measured shortly before use, which reports among other quantities the
single-qubit readout fidelity of every qubit and the two-qubit gate fidelity of every coupled pair.
We read this table and fixed the three physical qubits by hand rather than leaving them to the router.
The position register, on whose readout the density depends most directly, went to the single
highest-readout qubit on the device, of fidelity $0.988$. The two spinor qubits, which carry the
collision and the readout rotation between them and are themselves measured, went to the coupled pair
of highest two-qubit gate fidelity whose two members also read out well, an edge of gate fidelity
$0.989$ with member readouts $0.983$ and $0.984$. Nothing else was touched: the circuits, the shot
count, and the compilation to native gates were identical to the automatic run, and the chosen map was
enforced through the compiler's initial-rewiring directive, which places logical qubit $i$ on the
requested physical qubit and suppresses the router's own search. No reservation, recalibration, or
pulse-level control entered; the sole change was the three-number assignment from logical to physical
qubits.

With the qubits pinned, the run returns to the quality of the
Open Quantum measurement of Section~\ref{app:hw-results}, as Figure~\ref{fig:hw-pinned} and the pinned
columns of Table~\ref{tab:hw-pinned} show. The state-preparation floor at $t=0$ recovers from
$\rho=(0.588,0.412)$ to $(0.312,0.688)$, so its total-variation distance from the ideal falls from
$0.319$ to $0.044$, a sevenfold reduction, and the even-step sag disappears: the position now
alternates $0.688,\,0.278,\,0.676,\,0.258,\,0.698$ about the exact $0.731$ and $0.269$, in place of the
$0.412,\,0.264,\,0.497,\,0.293,\,0.418$ of the automatic placement. The velocity, already the sturdier
observable, tightens as well, its two extrema climbing from $0.77$ and $0.69$ of the ideal unit swing
to $0.83$ and $0.86$. The device benchmark moves in step: a Bell pair prepared on the pinned edge
returns the correct correlated outcomes in $93\%$ of its shots, against $82\%$ on the pair the router
had chosen. On a circuit this shallow, the strongest handle on the result is not the gate sequence but
the qubit assignment. The algorithm fixes the gates; what is left free is the map from logical to
physical qubits, and above all the readout fidelity of the qubit that carries the measured observable.
Reading that map from the calibration recovers essentially all of the reachable accuracy at no cost in
circuit depth. This handle is
peculiar to the shallow regime. As the position register grows and the streaming increment becomes the
deep cascade of entangling gates counted in Table~\ref{tab:hw-cost}, the accumulated two-qubit gate
error overtakes readout as the limiting factor, and qubit selection, though still worthwhile, can no
longer on its own bring the signal back.

The scripts for this direct run accompany the library. \texttt{qlb\_port/\allowbreak export\_quil.py}
transcribes the qiskit circuits into Quil and writes their manifest, and
\texttt{qlb\_port/\allowbreak qcs\_quil/\allowbreak run\_qcs.py} compiles them through quilc, pins the
qubits through its \texttt{-{}-qubits} option, and submits them to the processor.
Figure~\ref{fig:hw-pinned} is produced by the same \texttt{qlb\_port/\allowbreak plot\_2site\_results.py}
used for Figure~\ref{fig:hw-results}, while \texttt{qlb\_port/\allowbreak HARDWARE.md} documents the
credential setup and the calibration query from which the pinned qubits were chosen.

\end{document}